\documentclass[%
 aps,
 prb,
 superscriptaddress,
 amsmath,amssymb,
 twocolumn,
 reprint
]{revtex4-2}

\usepackage{graphicx}
\usepackage{dcolumn}
\usepackage{bm}
\usepackage{physics}
\usepackage{newtxtext}
\usepackage{siunitx}
\usepackage{xcolor}
\usepackage{bm}
\usepackage{braket}
\usepackage[legacycolonsymbols]{mathtools}

\usepackage[
    colorlinks = true,
    linkcolor = blue,
    citecolor=blue,
    urlcolor = blue
]{hyperref}

\graphicspath{{figure/}} 

\begin{document}

\title{Wideband wide-field imaging of spin-wave propagation using diamond quantum sensors}

\author{Kensuke Ogawa} 
\email{kensuke.ogawa@phys.s.u-tokyo.ac.jp}
\affiliation{Department of Physics, The University of Tokyo, Bunkyo-ku, Tokyo, 113-0033, Japan}
\author{Moeta Tsukamoto}
\affiliation{Department of Physics, The University of Tokyo, Bunkyo-ku, Tokyo, 113-0033, Japan}
\author{Yusuke Mori} 
\affiliation{Department of Physics, Osaka University, Toyonaka, Osaka 560-0043, Japan}
\author{Daigo Takafuji} 
\affiliation{Department of Physics, Osaka University, Toyonaka, Osaka 560-0043, Japan}
\author{Junichi Shiogai}
\affiliation{Department of Physics, Osaka University, Toyonaka, Osaka 560-0043, Japan}
\affiliation{Division of Spintronics Research Network, Institute for Open and Transdisciplinary Research Initiatives, Osaka University, Suita,
Osaka 565-0871, Japan}
\author{Kohei Ueda}
\affiliation{Department of Physics, Osaka University, Toyonaka, Osaka 560-0043, Japan}
\affiliation{Division of Spintronics Research Network, Institute for Open and Transdisciplinary Research Initiatives, Osaka University, Suita,
Osaka 565-0871, Japan}
\author{Jobu Matsuno}
\affiliation{Department of Physics, Osaka University, Toyonaka, Osaka 560-0043, Japan}
\affiliation{Division of Spintronics Research Network, Institute for Open and Transdisciplinary Research Initiatives, Osaka University, Suita,
Osaka 565-0871, Japan}
\author{Kento Sasaki}
\affiliation{Department of Physics, The University of Tokyo, Bunkyo-ku, Tokyo, 113-0033, Japan}
\author{Kensuke Kobayashi}
\affiliation{Department of Physics, The University of Tokyo, Bunkyo-ku, Tokyo, 113-0033, Japan}
\affiliation{Institute for Physics of Intelligence, The University of Tokyo, Bunkyo-ku, Tokyo, 113-0033, Japan}
\affiliation{Trans-Scale Quantum Science Institute, The University of Tokyo, Bunkyo-ku, Tokyo, 113-0033, Japan}

\begin{abstract}
Imaging spin-wave propagation in magnetic materials in a wide frequency range is crucial for understanding and applying spin-wave dynamics.
Recently, nitrogen-vacancy (NV) centers in diamond have attracted attention as sensors capable of quantitatively measuring the amplitude and phase of coherent spin waves.
However, the conventional sensing protocol has been limited to detecting spin waves whose frequencies match the resonance frequency of the NV spins.
We demonstrate that by utilizing the AC Zeeman effect, it is possible to image spin waves propagating in a yttrium iron garnet (YIG) thin film over a wide frequency range up to a maximum detuning of $567 \, \mathrm{MHz}$ without changing the external magnetic field. 
Our results expand the applicability of NV centers for spin-wave sensing and pave the way for quantitative investigations of the dynamics in various magnetic materials, such as metallic ferromagnets and van der Waals magnets.
\end{abstract}

\maketitle
\date{\today}

\section{Introduction}
In magnetic materials, spin-wave dynamics show fascinating physics by reflecting various spin-spin interactions and anisotropies \cite{Pirro2021}.
Rich dispersion relationships are observed depending on the parameters and shape of magnetic materials. 
Nonlinear phenomena such as parametric instability \cite{bryant1988spin}, soliton formation \cite{Sulymenko2018}, and Bose-Einstein condensation \cite{Demokritov2006} are also interesting ones observed in spin-wave dynamics.
Additionally, coherent spin waves are expected for applications in next-generation information devices, such as logic circuits and transistors, because they avoid the heat generation associated with conventional charge-based devices \cite{Flebus2024}.
Therefore, quantitative visualization of spin-wave dynamics is essential for understanding and applying spin-wave dynamics. 
\par
Recently, nitrogen-vacancy (NV) centers in diamond \cite{doherty2013nitrogen} have attracted significant attention as sensors for microwave magnetic fields.
NV spins have resonance frequencies around $3 \, \mathrm{GHz}$ in a low magnetic field and they can detect microwave fields with the frequencies with high sensitivity \cite{Wang2015, Horsley2018}.
In particular, microwave sensing using NV centers is well-suited for probing spin-wave dynamics, which typically have frequencies in the range of a few $\mathrm{GHz}$. 
Previous studies have demonstrated quantitative imaging of coherent spin waves \cite{van2015nanometre, Andrich2017, kikuchi2017long, bertelli2020magnetic,zhou2021magnon}. 
\par
Microwave sensing using NV centers is conventionally based on measuring the Rabi oscillation of NV spins \cite{Jelezko2004}. In this protocol, it is necessary to match the resonance frequency of NV spins with the target microwave frequency. If we want to change the target microwave frequency, we must tune the static magnetic field strength.
It implies that NV centers can detect only spin waves whose frequency matches the resonance frequency of NV spins, which limits the range of applicable materials and wavenumbers.
\par
To overcome this limitation, a previous study observes the ferromagnetic resonance in a ferromagnetic micro disc whose frequency is far from the resonance frequencies of the NV spin by detecting the reduction of the the time-averaged longitudinal magnetization \cite{van2015nanometre}. However, the off-resonant detection of the microwave from the spin waves is not performed. Also, recent studies utilize spin-wave mixing for frequency conversion of off-resonant spin waves \cite{Carmiggelt2023, Wu2024}.
They convert the target off-resonant spin waves to the spin waves whose frequency matches the resonant frequency of NV spins by tuning the frequency of the additional pump spin waves.
This approach is practical, but because it relies on complex nonlinear processes of the spin waves, observing directly the propagation of the target spin waves is challenging.
If it is possible to directly detect the propagation of off-resonant spin waves, we can understand the properties of the target spin waves more straightforwardly.
\par
In this paper, we focus on the AC Zeeman effect. The AC Zeeman effect refers to a phenomenon in which the resonance frequency of the NV spins slightly shifts when an off-resonant microwave is applied \cite{autler1955stark,ramsey1955resonance,wei1997strongly}.
The effect where the energy levels of an atom shift due to an alternating electric field is called the AC Stark effect. This effect has been applied in optical dipole traps for atomic cooling \cite{grimm2000optical} and microwave electric field sensing using Rydberg atoms \cite{meyer2020assessment,simons2016simultaneous,simons2016using}.
\par
By detecting the AC Zeeman effect, detection of off-resonant microwaves with a maximum detuning of approximately $1 \, \mathrm{GHz}$ using a single NV center \cite{li2019wideband} and visualization of the frequency characteristics of resonator antennas \cite{ogawa2023demonstration} have been demonstrated.
Here, we demonstrate an imaging of surface spin waves propagating through a magnetic thin film with a detuning of more than 500~MHz without changing a bias magnetic field.
Furthermore, we validate the future applicability of this protocol based on the sensitivity evaluation and show that this protocol has a potential to detect spin waves with a frequency of several tens of GHz.
\par
This paper is organized as follows. We describe the experimental setup and principle in Sec.~\ref{sec:exp}. In Sec.~\ref{sec:rabi}, we present the results of sensing resonant spin waves by the conventional Rabi-based protocol. In Sec.~\ref{sec:ac_zeeman}, we show the results of wideband spin-wave sensing using the AC Zeeman effect, which are key results of this paper. In Sec.~\ref{sec:power}, we investigate the spin-wave excitation power dependence of spin-wave amplitude. In Sec.~\ref{sec:sensitivity}, we estimate the sensitivity of our measurement and discuss the future applicability. Finally, we summarize the paper in Sec.~\ref{sec:summary}.
\par

\begin{figure}[t]
\centering
    \includegraphics[width=\linewidth]{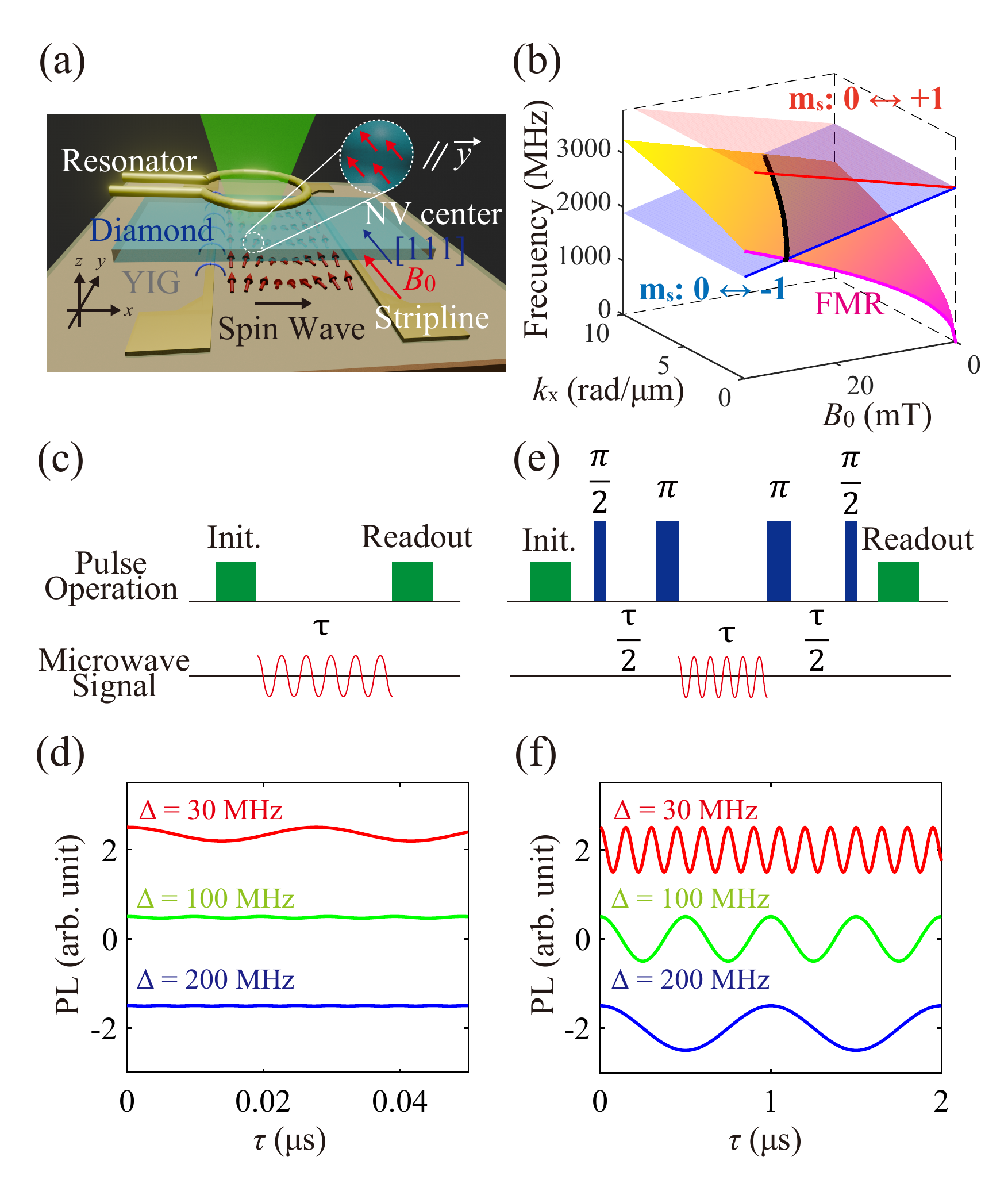}
    \caption{
    (a) Schematic of the experimental setup.
    (b) The dispersion relationship of surface spin waves in the YIG thin film with an effective magnetization $\mu_{0} M_{\mathrm{eff}} = 170 \, \mathrm{mT}$ and the resonance frequencies of the NV spins under various magnetic fields. The black line corresponds to the overlap between the resonance frequency of the NV spins and the dispersion relationship.
    (c) Pulse sequence used to detect the Rabi oscillation.
    (d) Numerical simulation results of off-resonant microwave detection using the Rabi oscillation with detunings $\Delta = 30 \, \mathrm{MHz}, 100 \, \mathrm{MHz}$, and $200 \, \mathrm{MHz}$. Microwave amplitude is fixed at $f_{\mathrm{rabi}} = 20 \, \mathrm{MHz}$ regarding Rabi frequency.
    (e) Pulse sequence used to detect the AC Zeeman effect.
    (f) Numerical simulation results of off-resonant microwave detection using the AC Zeeman effect with detunings $\Delta = 30 \, \mathrm{MHz}, 100 \, \mathrm{MHz}$, and $200 \, \mathrm{MHz}$. Microwave amplitude is fixed at $f_{\mathrm{rabi}} = 20 \, \mathrm{MHz}$ regarding Rabi frequency.
   }
    \label{fig:setup_principle}
\end{figure}

\begin{figure*}[t]
    \centering
    \includegraphics[width=\linewidth]{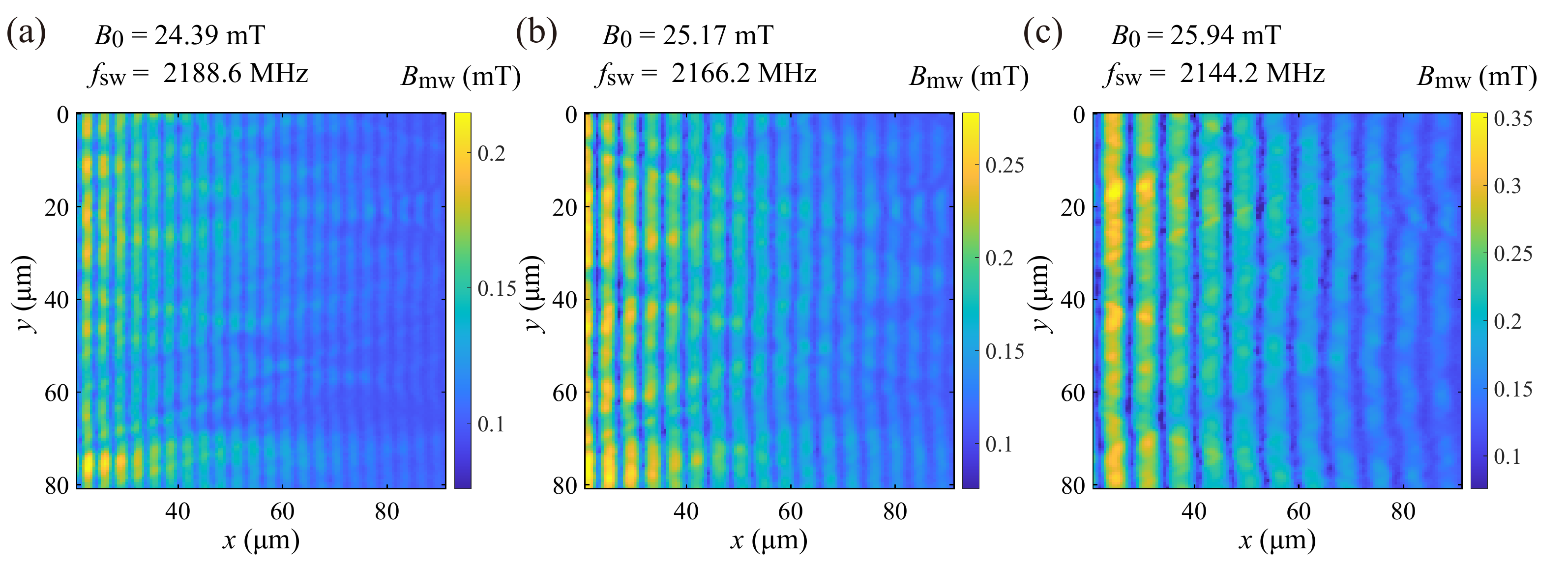}
    \caption{Results of imaging resonant spin-wave propagation using the conventional protocol based on the Rabi oscillation at external magnetic fields $B_{0} = 24.39 \, \mathrm{mT}, 25.17 \, \mathrm{mT}$, and $25.94 \, \mathrm{mT}$. The position $x = 0 \, \mathrm{\mu m}$ corresponds to the location of the right edge of the stripline, with the spin waves propagating in the right direction.
    }
    \label{fig:res_2d}
\end{figure*}

\section{Experimental Setup and Principle}
\label{sec:exp}
The experimental setup in this paper is shown in Fig.~\ref{fig:setup_principle}(a). 
For the measurement system, we employ a widefield NV microscopy, where a green laser with a wavelength of $515 \, \mathrm{nm}$ is irradiated to ensemble NV centers, and the red fluorescence is detected with a CMOS camera \cite{ogawa2023demonstration}. All the measurements are conducted at ambient conditions.
\par
The diamond chip used in this study is fabricated by cutting a (111) oriented Ib-type diamond single crystal (Element Six, MD111) into a substrate with a thickness of $38 \, \mathrm{\mu m}$ in the (110) orientation. 
An NV ensemble layer distributed over $30 \, \mathrm{nm}$ at a depth of $100 \, \mathrm{nm}$ from the surface is formed by carbon ion implantation ($30 \, \mathrm{keV}$, $1 \times 10^{12} \, \mathrm{cm^{-2}}$) and subsequent anneals \cite{tetienne2018spin}.
We utilize the NV centers oriented in the in-plane direction ($y$-axis) [see Fig.~\ref{fig:setup_principle}(a)].
Using in-plane NV centers is experimentally advantageous for spin-wave sensing. Applying a bias magnetic field $B_{0}$ parallel to the NV centers makes it possible to magnetize the target material in the in-plane direction. Several fundamental spin-wave modes, such as a backward volume mode and a surface mode are excited when a magnetic film is magnetized in the in-plane direction. Therefore, using in-plane NV centers simultaneously satisfies the condition of applying an external magnetic field along the NV axis and the condition of magnetizing the target material in the in-plane direction.
The diamond chip is attached to the magnetic film with the NV surface facing down. The distance $z_{\mathrm{d}}$ between the NV layer and the surface of the magnetic film is estimated to be $(878 \pm 20) \, \mathrm{nm}$, determined by evaluating the distribution of the stray magnetic field from the stripline on the YIG film surface under current application (see Appendix~\hyperref[app:distance]{D}).
\par
As the magnetic material, we utilize a yttrium iron garnet (YIG) thin film with a thickness of $d = 54$~nm, epitaxially grown on (111)-oriented gadolinium gallium garnet (GGG) substrate using RF sputtering \cite{fukushima2022spin}. We measured x-ray diffraction to confirm that the YIG film is coherently strained to the GGG substrate; this is consistent with a small lattice mismatch ($-0.06$\%) between bulk YIG and GGG.
Striplines with a width of $20 \, \mathrm{\mu m}$ are deposited on the YIG film surface, and spin waves are excited by irradiating microwave to the stripline on the left side.
A static magnetic field is applied in the in-plane direction along the stripline using a Helmholtz coil. In this geometry, surface spin waves propagating perpendicular to the magnetic field are observed \cite{Serga2010}.
We define the edge of the $+x$ side of the stripline as the origin of the $x$-axis ($x=0$).

Additionally, as shown in Fig.~\ref{fig:setup_principle}(a), a microwave resonator \cite{Sasaki2016} is hung above the diamond chip, from which a uniform microwave can be irradiated over the entire field of view (see Appendix~\hyperref[app:resonator]{E}).
By applying microwaves to this resonator, which is inductively coupled to the stripline \cite{Mariani2020}, it is possible to excite the surface spin waves indirectly.
NV centers feel the microwaves emitted from both spin waves and the antennas (resonator and stripline).
Hereafter, we call the microwaves from the antennas as the reference microwave.
\par
Next, we explain the principles and protocols of spin-wave sensing using NV centers, including resonant spin-wave sensing based on the Rabi oscillation and wideband spin-wave sensing using the AC Zeeman effect.
\par
Figure~\ref{fig:setup_principle}(b) shows the resonance frequencies of NV spins and the dispersion relationship of the surface spin waves in the YIG film (see Appendix~\hyperref[app:dispersion]{B}) under various magnetic fields $B_{0}$ \cite{zhou2021magnon}.
The dispersion relationship of the surface spin waves covers a wide range of frequencies, depending on the magnetic field and wavenumber $k_{x}$. 
In spin-wave sensing based on the Rabi oscillation, NV centers are sensitive to only spin waves whose frequency matches their resonance frequency [black line in Fig.~\ref{fig:setup_principle}(b)]. Therefore, when the external magnetic field is fixed, the detectable wavenumber of the spin waves is specified. Utilizing the protocol with the AC Zeeman effect enables the detection of spin waves across a wide frequency range centered around the resonance frequency for each magnetic field.
\par
Figure~\ref{fig:setup_principle}(c) shows the sensing protocol based on the Rabi oscillation. After initializing the spin state with a green laser pulse, the signal microwave pulse is irradiated for a period of $\tau$, and finally, the spin state is read out. The resulting photoluminescence (PL) intensity can be expressed as \cite{alsid2023solid}
\begin{align}
    S(\tau) =& 1 - \frac{C}{2} \notag \\
    & + \qty[\frac{1}{2} - \frac{f_{\mathrm{rabi}}^2}{f_{\mathrm{rabi}}^2 + \Delta^2} \sin (2 \pi \sqrt{f_{\mathrm{rabi}}^2 + \Delta^2} \tau)] C e^{-\frac{\tau}{T_{2 \mathrm{\rho}}}},
    \label{rabi_eq}
\end{align}
where $\Delta = f_{\mathrm{nv}} - f_{\mathrm{mw}}$ represents the detuning between the microwave frequency $f_{\mathrm{mw}}$ and the resonance frequency of the NV spins $f_{\mathrm{nv}}$.
$f_{\mathrm{rabi}} = \sqrt{2} \gamma_{e} B_\text{mw}$ corresponds to the Rabi frequency of the NV spins, which is proportional to amplitude $B_{\mathrm{mw}}$ of the microwave with circular polarization perpendicular to the NV axis, and $\gamma_{e}$ is the gyromagnetic ratio of an electron spin. $C$ is the PL contrast between the spin $m_{s} = 0, -1$ states, and $T_{2 \mathrm{\rho}}$ is the relaxation time during the Rabi oscillation.
This conventional protocol has high sensitivity when the detuning is small; however, sensitivity dramatically decreases as the detuning $\Delta$ increases.
Figure~\ref{fig:setup_principle}(d) shows the numerical simulation results of the PL intensity. The microwave amplitude is fixed at $f_{\mathrm{rabi}} = 20 \, \mathrm{MHz}$ regarding Rabi frequency, and the detunings $\Delta$ are varied from $30 \, \mathrm{MHz}$ to $200 \, \mathrm{MHz}$. At $\Delta = 200 \, \mathrm{MHz}$, the signal contrast is lost, and estimation of microwave amplitude is almost impossible.
\par
Figure~\ref{fig:setup_principle}(e) presents the measurement protocol based on the AC Zeeman effect \cite{li2019wideband, ogawa2023demonstration}. 
After initializing the spin state with a green laser pulse, the Carr-Purcell sequence with two $\pi$ pulses (CP-2) is performed by resonant microwave pulses followed by the reading out of the spin state. The signal microwave is applied only for the time $\tau$ between the two $\pi$ pulses to accumulate the phase due to the AC Zeeman effect.
The resulting normalized PL intensity when the detuning is sufficiently larger than Rabi frequency ($\Delta \gg f_{\mathrm{rabi}}$), and yet sufficiently small compared to the resonance frequencies of the NV spins can be expressed as
\begin{equation}
\begin{split}
    S(\tau) = 1 &- \frac{1 - \cos \qty(2 \pi f_{\mathrm{ACZ}} \tau) e^{-\frac{2\tau}{T_{2}}}}{2}, \\
    f_{\mathrm{ACZ}} &= \frac{B_{\mathrm{mw}}^2}{\Delta},
    \label{eqsig}
\end{split}
\end{equation}
where $T_{2}$ represents the coherence time during the CP-2 sequence. A more rigorous expression when the detuning becomes larger will be discussed in Sec.~\ref{sec:sensitivity}.
Figure~\ref{fig:setup_principle}(f) presents the numerical simulation results under the same conditions in Fig.~\ref{fig:setup_principle}(d). Signal oscillations can be clearly observed even at large detunings, compared to the results based on the Rabi oscillation [Fig.~\ref{fig:setup_principle}(d)]. As the detuning increases, the signal frequency decreases, and the sensitivity degrades, but the contrast does not decrease as in the conventional method. Therefore, estimation of microwave amplitude is possible as long as the oscillation cycle is not much longer than the coherence time $T_{2}$.
\par

\section{Sensing resonant spin waves using the Rabi oscillation}
\label{sec:rabi}
This section presents the results of imaging resonant spin waves using the Rabi oscillation.
We set the field of view as $71 \, \mathrm{\mu m}$ in the $x$-direction and $80 \, \mathrm{\mu m}$ in the $y$-direction. The external magnetic field $B_{0}$ is swept from $24.39 \, \mathrm{mT}$ to $26.14 \, \mathrm{mT}$. At each magnetic field, the spin-wave frequency is set to the resonance frequency of the NV spins determined by the optically detected magnetic resonance (ODMR) spectrum. We input a microwave with the frequency to the resonator, and spin waves are excited by the inductively coupled stripline.
\par
Figures~\ref{fig:res_2d}(a), \ref{fig:res_2d}(b), and \ref{fig:res_2d}(c) show the two-dimensional images of the microwave amplitude $B_{\mathrm{mw}}$ for three external magnetic fields, $B_{0} = 24.39 \, \mathrm{mT}, \, 25.17 \, \mathrm{mT}$, and $25.94 \, \mathrm{mT}$, respectively.
In the analysis, signals from each $520 \, \mathrm{nm}$ in the $x$- and $y$-directions are integrated and treated as a single pixel.
The microwave amplitude oscillates periodically in the $x$-direction, which is due to the interference between the microwave generated by the spin waves that propagate in the $x$-direction and the reference microwave whose phase is spatially uniform in the entire field of view \cite{bertelli2020magnetic,zhou2021magnon}.
The phase of the oscillations corresponds to that of the spin waves. 
The period of the oscillations, which is equal to the wavelength of the spin waves, decreases as the magnetic field increases.
It can be explained by the dispersion relationship shown in Fig.~\ref{fig:setup_principle}(b). 
As the magnetic field increases, the resonance frequency of the NV spins between $m_{s} = 0$ and $-1$ states decreases, and the dispersion relationship of the spin waves shifts to higher frequencies. As a result, the wavenumber of the spin waves at the resonance frequency of NV spins becomes smaller.
In addition to the spin-wave propagation in the $x$-direction, microwave amplitude increases at a specific angle in the diagonal direction.
This phenomenon is known as “caustics” and arises from the anisotropy of the spin-wave dispersion relationship \cite{Schneider2010}.
The stripline used in this study is long enough, and ideally, only spin waves in the $x$-direction should be excited, and caustics should not be observed.
The caustics pattern can occur due to two magnon scattering caused by surface roughness or defects in the YIG thin film \cite{hurben1998, Bertelli2021} or non-uniformities in the stripline.
\par

\begin{figure}[!ht]
    \centering
    \includegraphics[width=\linewidth]{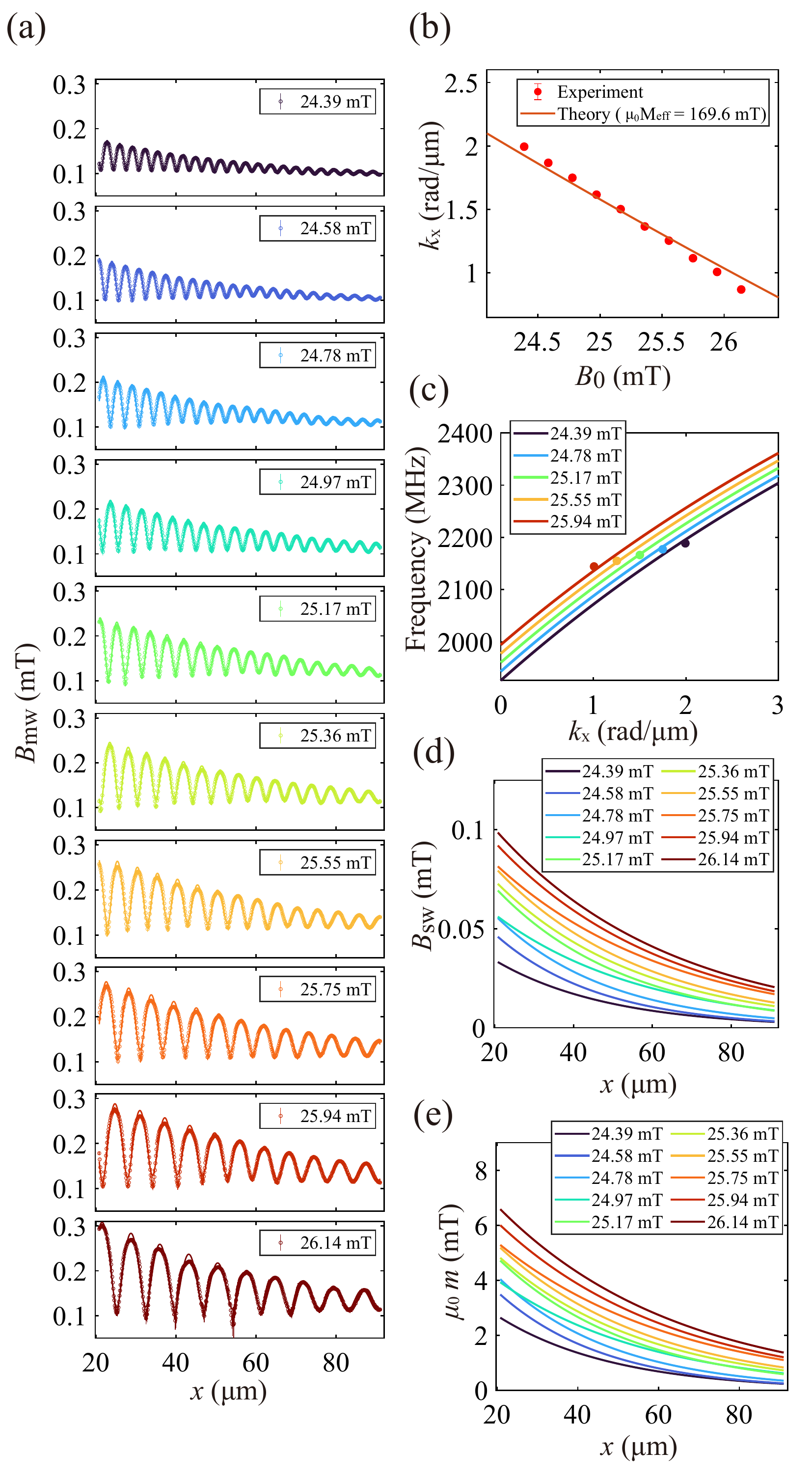}
    \caption{
    One-dimensional analysis results of sensing resonant spin-wave propagation based on the conventional protocol using the Rabi oscillation (a) One-dimensional microwave amplitude distribution $B_{\mathrm{mw}}(x)$ for each external magnetic field from $B_{0} = 24.39 \, \mathrm{mT}$ to $B_{0} = 26.14 \, \mathrm{mT}$. The points with error bars represent experimental data, and the solid lines correspond to the fitting results.
    (b) Relationship between the external magnetic fields and the experimentally obtained spin-wave wavenumbers. The solid line represents the theoretical line at the effective magnetization $\mu_{0} M_{\mathrm{eff}} = 169.6 \, \mathrm{mT}$ determined by the fitting.
    (c) Plots of the results of (b) in the form of the dispersion relationship between the spin-wave wavenumber and spin-wave frequency.
    (d) Microwave amplitude distribution from the spin waves $B_{\mathrm{sw}}(x)$ at each magnetic field, calculated from the fitting result.
    (e) Spin-wave amplitude distribution $m(x)$ multiplied by the permeability of vacuum $\mu_{0}$, converted from the result of (d).
    }
    \label{fig:res_1d}
\end{figure}

We analyze the obtained microwave amplitude in one dimension along the $x$-direction to extract quantitative information about the spin waves.
Figure~\ref{fig:res_1d}(a) shows the microwave amplitude distribution $B_{\mathrm{mw}}(x)$ for each magnetic field $B_{0}$. Here, at each magnetic field, we integrate signals over a width of $13 \, \mathrm{\mu m}$ from $y = 31.87 \, \mathrm{\mu m}$ to $y = 44.87 \, \mathrm{\mu m}$ in the $y$-direction.
At each magnetic field, the microwave amplitude oscillates with the wavelength of the spin waves. In this measurement, $B_{\mathrm{ref}}(x) \gg B_{\mathrm{sw}}(x)$ holds over the entire field of view (see Appendix~\hyperref[app:rabi]{F}). Therefore, the oscillation amplitude corresponds to the microwave amplitude from the spin wave $B_{\mathrm{sw}}(x)$, and the envelope of the microwave amplitude corresponds to the reference microwave amplitude $B_{\mathrm{ref}}(x)$.
To extract information on the wavenumber and amplitude of the spin waves, we fit the obtained microwave amplitude distribution using the following equation:
\begin{equation}
\begin{split}
    \label{rabi:fittng}
    B_{\mathrm{mw}}(x) &= B_{\mathrm{ref}}(x) + B_{\mathrm{sw}}(x)\cos \qty(k_{x}(x-x_{0}) + \theta_{0}), \\
    B_{\mathrm{sw}}(x) &= B^{0}_{\mathrm{sw}} e^{-\frac{x-x_{0}}{l_{\mathrm{d}}}}, \\ 
    B_{\mathrm{ref}}(x) &= B^{0}_{\mathrm{ref}} e^{-\qty(\frac{x-x_{0}}{l_{\mathrm{ref}}})^{p_{\mathrm{ref}}}} + B^{c}_{\mathrm{ref}},
\end{split}
\end{equation}
where $x_{0}$ represents the distance from the end of the stripline to the left edge of the field of view ($x_{0} = 20 \, \mathrm{\mu m}$ in this measurement), $k_{x}$ is the wavenumber of the spin waves, and $\theta_{0}$ corresponds to the phase of the spin waves at $x = x_{0}$.
For the microwave amplitude distribution from spin waves $B_{\mathrm{sw}}(x)$, we take into account the Gilbert damping and assume it to be an exponential decay, characterized by the amplitude at the left edge of the field of view $B^{0}_{\mathrm{sw}} = B_{\mathrm{sw}}(x_0)$ and the decay length $l_{\mathrm{d}}$.
For the reference microwave amplitude distribution $B_{\mathrm{ref}}(x)$, due to the difficulty in obtaining an analytical expression, we adopt a function that is the sum of an exponential decay characterized by the amplitude $B^{0}_{\mathrm{ref}}$ at the left edge of the field of view, the decay length $l_{\mathrm{ref}}$, and the stretch factor $p_{\mathrm{ref}}$ and the constant term $B^{c}_{\mathrm{ref}}$.
The fitting results accurately reproduce the obtained microwave amplitude at each magnetic field. 
\par
Figure~\ref{fig:res_1d}(b) shows the relationship between the external magnetic field and the wavenumber of the spin waves obtained from the fitting results.
The solid line represents the fitting line. For the fitting, we utilize the following theoretical dispersion relationship of the surface spin waves (see Appendix~\hyperref[app:dispersion]{B}):
\begin{equation}
\begin{split}
    f_{\mathrm{sw}} = \gamma_{e} \mu_{0}  &\sqrt{\qty(H_{0} + M_{\mathrm{eff}}(1-g(k_{x}d))) \qty(H_{0} + M_{\mathrm{eff}} g(k_{x}d))}, \\
    g(k_{x}d) &= 1 - \frac{1-e^{-k_{x}d}}{k_{x}d},
    \label{eq:dispersion}
\end{split}
\end{equation}
where $H_{0}$ is the external magnetic field ($B_{0} = \mu_{0} H_{0}$), $\mu_0$ is the magnetic permeability in vacuum, $M_{\mathrm{eff}}$ is the effective magnetization, and $d$ is the thickness of the YIG film.
We set the external magnetic field $B_{0}$, the obtained wavenumber of the spin waves $k_{x}$, and the spin-wave frequency $f_{\mathrm{sw}}$ as input variables, and the fitting parameter is only the effective magnetization $M_{\mathrm{eff}}$. The fitting line agrees well with the experimental result, and the obtained effective magnetization $\mu_{0} M_{\mathrm{eff}} = (169.6 \pm 0.7) \, \mathrm{mT}$ is consistent with a value in a previous study Ref.~\cite{chang2014nano}.
\par
Figure~\ref{fig:res_1d}(c) shows the relationship between the obtained spin-wave wavenumber and frequency for each magnetic field (markers) with dispersion relationships calculated with the effective magnetization $\mu_{0} M_{\mathrm{eff}} = 169.6 \, \mathrm{mT}$ (solid lines).
As evident from the figure, for the conventional spin-wave sensing using the Rabi oscillation, when we fix the external magnetic field, the wavenumber of the detectable spin waves is also specified. While changing the magnetic field makes it possible to detect spin waves with different wavenumbers, it is invasive for the target material because it also alters the dispersion relationship of the material.
\par
Next, we extract quantitative information about the amplitude of the spin waves.
Figure~\ref{fig:res_1d}(d) shows the microwave amplitude distribution from the spin waves $B_{\mathrm{sw}}(x)$, calculated from the fitting result. The microwave amplitude increases as the bias magnetic field becomes larger and the wavenumber decreases.
Also, at each magnetic field, the microwave amplitude from the spin waves decays on the scale of tens of micrometers. In particular, for the data at $B_{0} = 25.36 \, \mathrm{mT}$, the obtained decay length is $l_{\mathrm{d}} = 37 \, \mathrm{\mu m}$. From this decay length, we can calculate the Gilbert damping constant $\alpha$ at this condition as $\alpha = 1.1 \times 10^{-3}$ (see Appendix~\hyperref[app:dispersion]{B}). 
\par
Theoretically, the microwave amplitude from the spin waves at the position of the NV layer can be represented by the product of the spin-wave amplitude $m(x)$ and a wavenumber-dependent transfer function $D(k,z)$ \cite{zhou2021magnon,bertelli2020magnetic}:
\begin{equation}
\begin{split}
    B_{\mathrm{sw}}(x) &= \frac{\mu_{0}}{2} D(k_{x},z) (1  + \eta_{k_{x}}) m(x), \\
    D(k,z) &= e^{-kz} (1-e^{-kd}),\\
    \label{eq:m_Bmw}
\end{split}
\end{equation}
where $\eta_{k_{x}}$ is the ellipticity of the spin waves (see Appendix~\hyperref[app:dispersion]{B}).
Figure \ref{fig:res_1d}(e) shows the results of converting the microwave amplitude $B_{\mathrm{sw}}(x)$ [Fig.~\ref{fig:res_1d}(d)] into the spin-wave amplitude $m(x)$ using Eq.~(\ref{eq:m_Bmw}).
As the bias magnetic field increases, the spin-wave amplitude increases the same way as the microwave amplitude [Fig.~\ref{fig:res_1d}(d)].
The main reason for the observed behavior would be related to the wavenumber dependence of the excitation efficiency of the spin waves, which is determined by the dimension of the stripline.
The excitation efficiency of the spin waves is proportional to the Fourier amplitude of the microwave from the stripline at their wavenumber. In our measurement, the wavenumber decreases as the bias magnetic field becomes larger. It is generally known that the Fourier amplitude becomes smaller as the wavenumber increases \cite{Qin2018} (see Appendix~\hyperref[app:stripline_mw]{H}), which is consistent with the experimental result.
\par 
Although the impact is minor, in addition to the stripline's excitation efficiency, the resonator's frequency characteristics also affect the amplitude of the spin waves. The resonator has a resonance frequency of around $2100 \, \mathrm{MHz}$ (see Appendix~\hyperref[app:resonator]{E}). As the magnetic field increases, the frequency of the spin waves approaches this resonance frequency, resulting in a greater microwave intensity for spin-wave excitation. From the magnitude of the reference microwave, contributions from this effect are estimated to account for a maximum of about $30\%$
\par
As mentioned above, the ability to quantitatively detect the phase and amplitude of spin waves is a unique property of NV centers. However, as repeatedly emphasized, the conventional protocol based on the Rabi oscillation can detect only one wavenumber of spin waves for a given bias magnetic field. Moreover, the protocol cannot be applied to materials whose spin-wave frequency does not match the resonance frequency of the NV spins. This limitation restricts the future applicability of NV centers in spin-wave sensing. To address this issue, the following section demonstrates wideband spin-wave sensing using the AC Zeeman effect.
\par

\begin{figure*}[t]
    \centering
    \includegraphics[width=\linewidth]{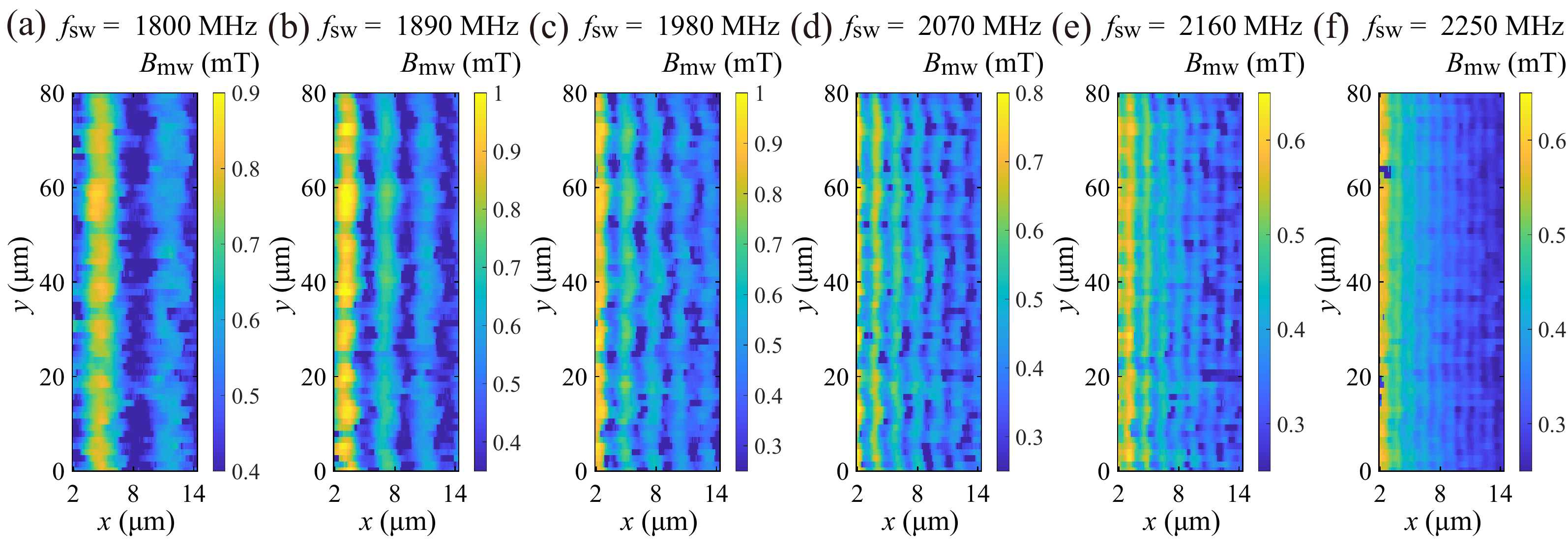}
    \caption{Results of imaging off-resonant spin-wave propagation using the AC Zeeman effect at spin-wave frequencies $f_{\mathrm{sw}}$ ranging from $1800 \, \mathrm{MHz}$ to $2250 \, \mathrm{MHz}$ at intervals of $90 \, \mathrm{MHz}$. The external magnetic field $B_{0}$ is fixed at $18.37 \, \mathrm{mT}$.}
    \label{fig:nres_2d}
\end{figure*}

\section{sensing wideband spin waves using the AC Zeeman effect}
\label{sec:ac_zeeman}

This section presents the results of wideband spin-wave sensing using the AC Zeeman effect. As shown in Fig.~\ref{fig:setup_principle}(e), wideband spin-wave sensing using the AC Zeeman effect requires pulse operations of NV spins. For pulse operations, it is necessary to manipulate NV spins fast enough compared to the dephasing rate. To meet this condition, we directly input microwave pulses for both pulse operations of the NV spins and spin-wave excitation (the resonator is not used in the following measurements).
In all measurements, the external magnetic field is fixed at $B_{0} = 18.4 \, \mathrm{mT}$ (the resonance frequency of the NV centers is $2356.7 \, \mathrm{MHz}$).
At this bias magnetic field, the wavenumber of spin waves whose frequency matches the resonance frequency of the NV spins is sufficiently large ($k_{x} \approx 6 \, \mathrm{rad}/\mathrm{\mu m}$), and the microwave amplitude from the spin waves at the frequency is small (see Appendix~\hyperref[app:wideband_setup]{G}). Thus, when we input microwaves with the same frequency as the resonance frequency of the NV spins into the stripline, most of the microwave field generated at the position of NV centers originates from the stripline. Therefore, we utilize this microwave field for pulse operations.
For imaging measurement, it is also necessary to manipulate the NV spins uniformly within the field of view. 
We adopt the SCROFULOUS composite pulse sequence \cite{cummins2003tackling, nomura2021composite, ogawa2023demonstration} to suppress pulse errors caused by spatial non-uniformity of the microwave field amplitude generated from the stripline (see Appendix~\hyperref[app:wideband_setup]{G}).
This composite pulse sequence makes it possible to adequately estimate the microwave amplitude, even when an approximately $50 \, \%$ pulse amplitude error exists.
With this technique, we image the area from $x = 1.95 \, \mathrm{\mu m}$ to $x = 14.3 \, \mathrm{\mu m}$ where the microwave field for pulse operations is sufficiently strong and pulse errors can be compensated. 
Under the conditions, we sweep the spin-wave frequency $f_{\mathrm{sw}}$ from $1800 \, \mathrm{MHz} \, (\Delta = 556.7 \, \mathrm{MHz})$ to $2310 \, \mathrm{MHz} \, (\Delta = 46.7 \, \mathrm{MHz})$. At all frequencies, the input microwave power $P_{\mathrm{mw}}$ is fixed at $22.9 \, \mathrm{dBm}$.
\par

\begin{figure}[t]
    \centering
    \includegraphics[width=\linewidth]{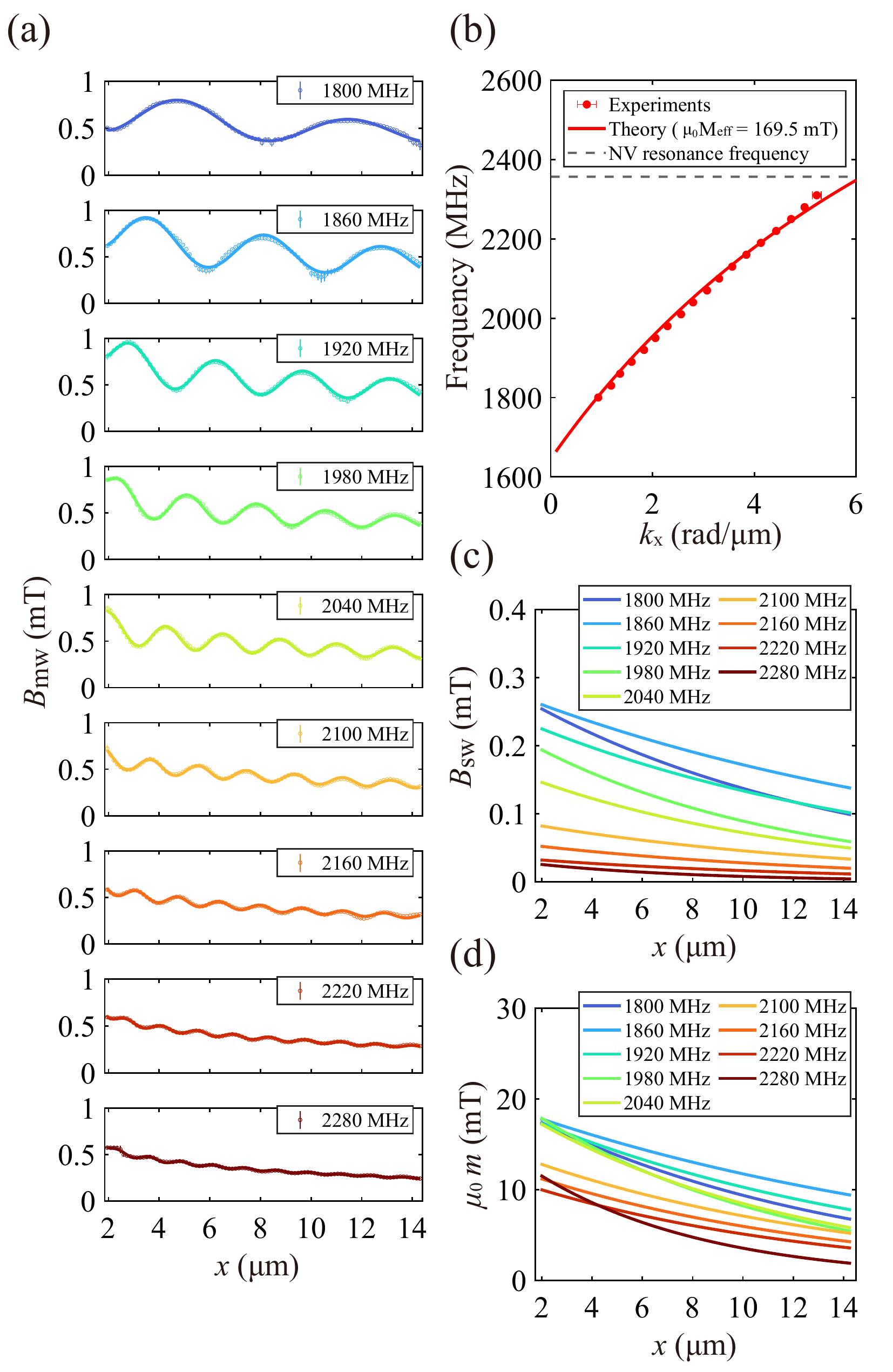}
    \caption{
    One-dimensional analysis of sensing off-resonant spin-wave propagation using the AC Zeeman effect.
    (a) One-dimensional microwave amplitude distribution $B_{\mathrm{mw}}(x)$ at each spin-wave frequency from $f_{\mathrm{sw}} = 1800 \, \mathrm{MHz}$ to $f_{\mathrm{sw}} = 2280 \, \mathrm{MHz}$. The points with error bars represent the experimental data, and the solid lines correspond to the fitting results.
    (b) Relationship between the experimentally obtained spin-wave wavenumbers and spin-wave frequencies. The solid line represents the theoretical line at the effective magnetization of $\mu_{0} M_{\mathrm{eff}} = 169.5 \, \mathrm{mT}$ determined by the fitting.
    (c) Microwave amplitude distribution from spin waves $B_{\mathrm{sw}}(x)$ at each spin-wave frequency, calculated from the fitting result.
    (d) Spin-wave amplitude distribution $m(x)$ multiplied by the permeability of vacuum $\mu_{0}$, converted from the result of (c).
    }
    \label{fig:nres_1d}
\end{figure}

Figure~\ref{fig:nres_2d} displays the two-dimensional images of the microwave amplitude $B_{\mathrm{mw}}$ at frequency intervals of $90 \, \mathrm{MHz}$ from $f_{\mathrm{sw}} = 1800 \, \mathrm{MHz}$ to $f_{\mathrm{sw}} = 2250 \, \mathrm{MHz}$. In the analysis, signals from $130 \, \mathrm{nm}$ in the $x$-direction and $1.3 \, \mathrm{\mu m}$ in the $y$-direction are integrated and treated as a single pixel. The interference patterns are clearly observed, similar to the results of imaging resonant spin-waves (Fig.~\ref{fig:res_2d}). Notably, at $f_{\mathrm{sw}} = 1800 \, \mathrm{MHz}$, two interference peaks are visible within the field of view, whereas at $f_{\mathrm{sw}} = 2250 \, \mathrm{MHz}$, nine interference peaks are observed. 
Additionally, variations in the two-dimensional microwave amplitude in the $y$ direction are noticeable at frequencies such as $f_{\mathrm{sw}} = 1980 \, \mathrm{MHz}$. This variation can be attributed to the influence of other spin wave modes similar to that observed in Fig.~\ref{fig:res_2d}.
\par
These results are obtained without changing the bias magnetic field and underscore the utility of wideband spin-wave sensing using the AC Zeeman effect.
\par

Next, to extract the quantitative information of the spin waves, we analyze the obtained microwave amplitude distribution in one dimension along the $x$-direction. Figure~\ref{fig:nres_1d}(a) shows the one-dimensional microwave amplitude distribution along the $x$-direction for each spin-wave frequency. For the analysis, we integrate signals over a width of $26 \, \mathrm{\mu m}$ from $y = 31.9 \, \mathrm{\mu m}$ to $y = 57.9 \, \mathrm{\mu m}$ in the $y$-direction. As with the previous section, the fitting is performed using Eq.~(\ref{rabi:fittng}) to extract the contributions from the spin waves. Since the reference microwave originates only from the stripline and becomes negligible at a sufficient distance from the stripline, we set $B^{c}_{\mathrm{ref}} = 0$. The fitting curve accurately reproduces the experimentally obtained microwave amplitude distribution at every spin-wave frequency.
\par
Figure~\ref{fig:nres_1d}(b) shows the experimentally obtained dispersion relationship. The solid curve represents the dispersion relationship obtained by the fitting, using Eq.~(\ref{eq:dispersion}), with the effective magnetization $M_{\mathrm{eff}}$ as the fitting parameter. The fitting curve accurately reproduces the experimental results. The obtained effective magnetization $\mu_{0} M_{\mathrm{eff}} = (169.5 \pm 0.7) \, \mathrm{mT}$ is consistent with the effective magnetization derived in the previous section $\mu_{0} M_{\mathrm{eff}} = (169.6 \pm 0.7) \, \mathrm{mT}$. This consistency guarantees the validity of the obtained results. 
\par
Figure~\ref{fig:nres_1d}(c) displays the microwave amplitude distribution $B_{\mathrm{sw}}(x)$ from the spin waves calculated from the fitting result. A decrease in microwave amplitude is observed as the distance from the stripline increases, similar to the conventional resonant spin-wave sensing [Fig.~\ref{fig:res_1d}(d)]. The decay lengths appear to vary among some results (e.g., $f_{\mathrm{sw}} = 2280 \, \mathrm{MHz}$), which is likely due to large spatial variations in the reference microwave distribution. The discrepancy between the actual distribution of the reference microwave and the fitting result can affect the microwave distribution from the spin waves.
\par

The microwave amplitude distribution from the spin waves is represented as the product of the spin-wave amplitude and a wavenumber-dependent transfer function, as expressed by Eq.~(\ref{eq:m_Bmw}). 
Figure~\ref{fig:nres_1d}(d) illustrates the results of converting the microwave amplitude distribution from the spin waves in Fig.~\ref{fig:nres_1d}(c) into the spin-wave amplitude distribution $m(x)$ using Eq.~(\ref{eq:m_Bmw}). The spin-wave amplitude decreases as the spin-wave frequency increases. In this measurement, unlike the results in the previous section, the spin waves are excited by inputting microwaves into the stripline directly. Since the stripline has no specific frequency characteristics, the reduction of the spin-wave amplitude is considered to derive primely from the wavenumber dependence of the spin-wave excitation efficiency of the stripline. A more detailed analysis of this wavenumber dependence of the spin-wave amplitude is provided in Appendix~\hyperref[app:sw_amplitude]{I}.
Note that at regions with large wavenumbers, the wavelength of the spin waves approaches as small as approximately twice the optical diffraction limit, and as a result, the oscillation amplitude of the microwave from the spin waves can be reduced, resulting in the spin-wave amplitude being underestimated by 10 to 20\%. It is possible to obtain a more accurate estimation of the spin-wave amplitude by measuring the point spread function (PSF) of the optical system, calculating how much the PSF reduces the microwave oscillation amplitude, and applying corrections to the value of spin-wave amplitude \cite{Kevin2024practical,nishimura2024investigations}. However, since this effect does not alter the essence of our results, we have analyzed the data without making any corrections.
\par
The above results demonstrate the possibility of obtaining quantitative information on the wavenumber and spin-wave amplitude of the surface spin waves with up to a maximum detuning of $556.7 \, \mathrm{MHz}$ using the AC Zeeman effect.
\par

\section{Power dependence of spin-wave amplitude measured by AC Zeeman effect}
\label{sec:power}
\begin{figure}[t]
    \centering
    \includegraphics[width=\linewidth]{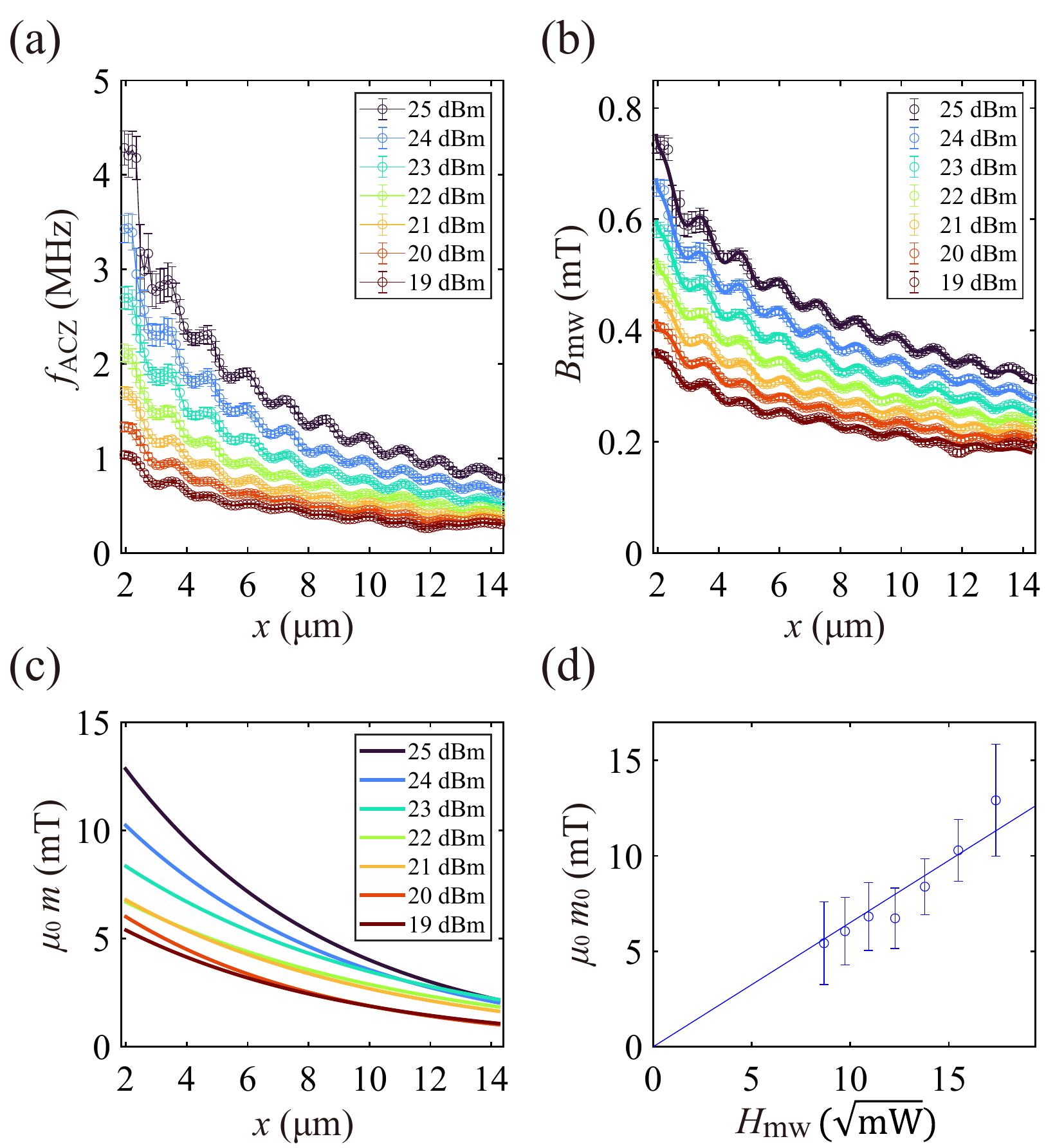}
    \caption{
    Off-resonant spin-wave amplitude dependence on the microwave input power measured using the AC Zeeman effect.
    (a) Signal frequency distribution $f_{\mathrm{ACZ}}(x)$ at each microwave input power with the spin-wave frequency $f_{\mathrm{sw}} = 2260 \, \mathrm{MHz} \, (\Delta = 96.7 \, \mathrm{MHz})$.
    (b) Microwave amplitude distribution $B_{\mathrm{mw}}(x)$ at each microwave input power.
    (c) Spin-wave amplitude distribution $m(x)$ multiplied by the permeability of vacuum $\mu_{0}$ at each microwave input power, converted from the result of (b).
    (d) The relationship between the input microwave amplitude $H_{\mathrm{mw}}$ and the spin-wave amplitude $m_{0}$ multiplied by the permeability of vacuum $\mu_{0}$ at the left edge of the field of view ($x = 1.95 \, \mathrm{\mu m}$). The solid line represents a linear fit through the origin.
    }
    \label{fig:nres_pow}
\end{figure}
In this section, we examine the excitation power dependence of the spin-wave amplitude to validate our quantitative method further. Here, we set the spin-wave frequency as $f_{\mathrm{sw}} = 2260 \, \mathrm{MHz} \, (\Delta = 96.7 \, \mathrm{MHz})$ and sweep the input microwave power $P_{\mathrm{mw}}$.
\par
Figure~\ref{fig:nres_pow}(a) displays the distribution of the signal frequency $f_{\mathrm{ACZ}}(x)$ at each microwave power, analyzing the same area as in the previous section. The signal frequency increases as the input microwave power is raised. As Eq.~(\ref{eqsig}) indicates, the signal frequency is proportional to the square of the input microwave amplitude. Consequently, significant changes in the signal frequency are observed as the input power varies. 
\par
Figure~\ref{fig:nres_pow}(b) presents the results of converting $f_{\mathrm{ACZ}}(x)$ [Fig.~\ref{fig:nres_pow}(a)] into the microwave amplitude distribution $B_{\mathrm{mw}}(x)$. The solid lines represent the fitting results obtained using the same procedure as in the previous section. The fitting result is consistent with the experimental result for each microwave power.
\par
Figure~\ref{fig:nres_pow}(c) displays the results of converting the microwave amplitude distribution from the spin waves, extracted from the fitting result, into the spin-wave amplitude distribution $m(x)$ using Eq.~(\ref{eq:m_Bmw}). The spin-wave amplitude increases with input microwave power, and a similar decay curve is observed across all microwave input powers.
\par
We discuss the relationship between the input microwave amplitude and the obtained spin-wave amplitude.
Figure~\ref{fig:nres_pow}(d) illustrates the dependence of the spin-wave amplitude $m_{\mathrm{0}}$ at the left edge of the field of view on the input microwave amplitude $H_{\mathrm{mw}} \, (\coloneqq \sqrt{P_{\mathrm{mw}}})$. The solid line represents a fitting line obtained through linear regression that passes through the origin. The fitting result demonstrates that the obtained spin-wave amplitude is proportional to the input microwave amplitude. In this measurement, the maximum spin-wave amplitude is about $12 \, \mathrm{mT}$, considerably smaller than the saturation magnetization. Therefore, the observed linear relationship is consistent with the expectation.
\par
Previous studies have demonstrated that as the amplitude of spin waves increases, nonlinear effects such as effective static magnetization reduction and higher-order spin-wave scattering processes become apparent, leading to phase modulation and nonlinear decay of propagating spin waves \cite{Scott2004, Hula2020, lake2022interplay}. Our method holds a potential for quantitatively investigating such nonlinear spin-wave dynamics across a broad range of frequencies and wavenumbers.

\section{Sensitivity estimation and considerations of future applicability}
\label{sec:sensitivity}
\begin{figure}[t]
    \centering
    \includegraphics[width=\linewidth]{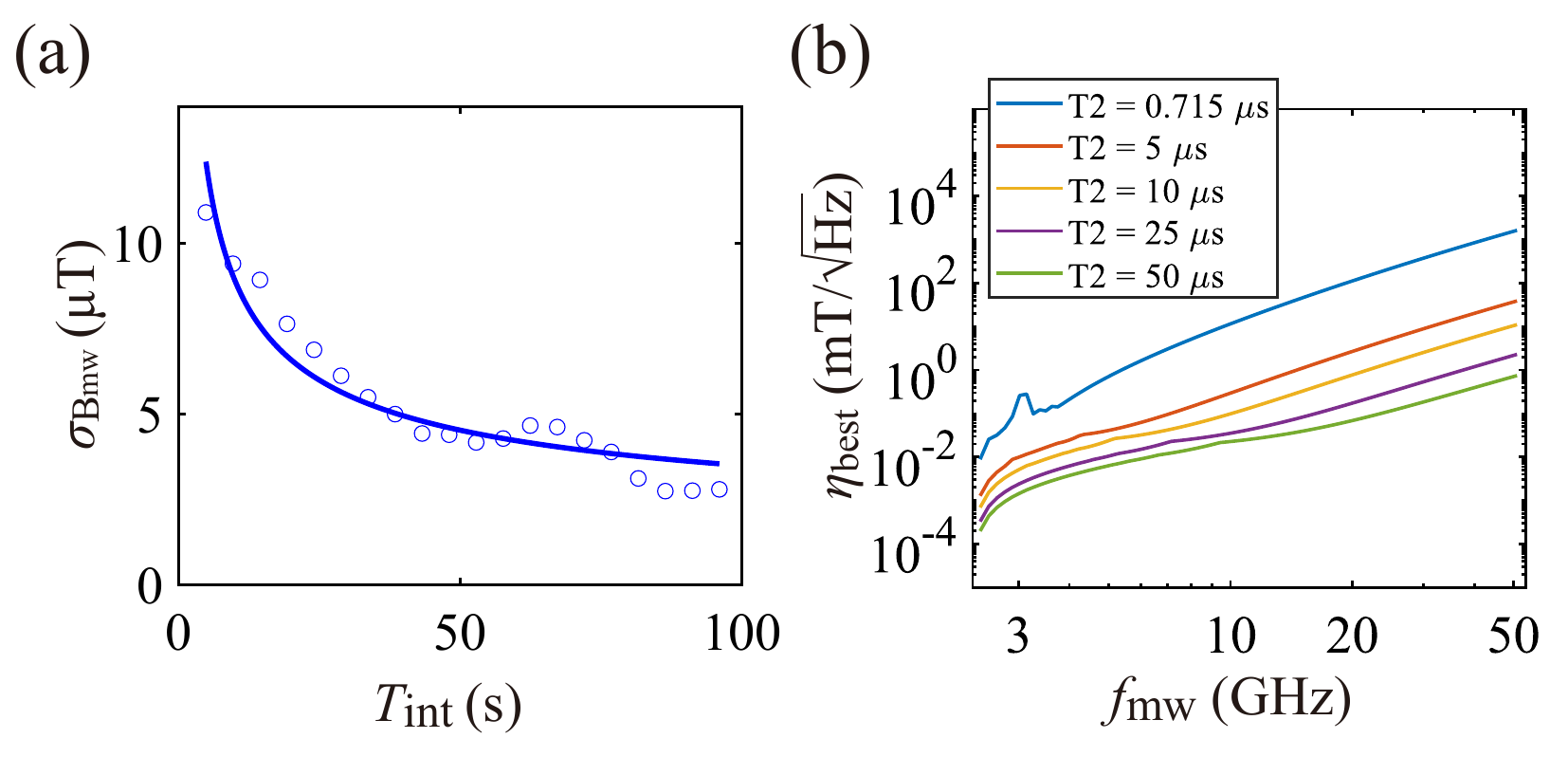}
    \caption{
    Evaluation of the sensitivity of off-resonant spin-wave sensing using the AC Zeeman effect.
    (a) The relationship between integration time $T_{\mathrm{int}}$ and the standard error of microwave amplitude $\sigma_{\mathrm{B_{mw}}}$ at the position $x = 5.85 \, \mathrm{\mu m}$ with a spin-wave frequency $f_{\mathrm{sw}} = 2256.7 \, \mathrm{MHz} \, (\Delta = 100 \, \mathrm{MHz})$ and microwave amplitude $B_{\mathrm{mw}} = 0.42 \, \mathrm{mT}$.
    (b) Estimation of the sensitivity $\eta_{\mathrm{best}}$ of spin-wave microwave amplitude $B_{\mathrm{sw}}$ at various spin-wave frequencies under different coherence time $T_{2}$. The reference microwave amplitude and the spin-wave microwave amplitude are assumed to be $B_{\mathrm{ref}} = 0.505 \, \mathrm{mT}$ and $B_{\mathrm{sw}} = 75.8 \, \mathrm{\mu T}$, respectively.
    }
    \label{fig:nres_sensitivity}
\end{figure}

Finally, we evaluate the sensitivity of the present off-resonant spin-wave sensing.
To evaluate the sensitivity, we conduct a measurement at the spin-wave frequency $f_{\mathrm{sw}} = 2256.7 \, \mathrm{MHz} \, (\Delta = 100 \, \mathrm{MHz})$ under the same conditions in the frequency sweep measurement (Sec.~\ref{sec:ac_zeeman}).
Like the one-dimensional analysis in Fig.~\ref{fig:nres_1d}, we integrate signals from $130 \, \mathrm{nm}$ in the $x$-direction and $26 \, \mathrm{\mu m}$ in the $y$-direction, and analyze the data at the position $x = 5.85 \, \mathrm{\mu m}$. At this position, the microwave amplitude is estimated to be $B_{\mathrm{mw}} = 0.42 \, \mathrm{mT}$.
We follow the same procedure described in Ref.~\cite{ogawa2023demonstration} for the derivation of the sensitivity.
We calculate the standard error of the estimated microwave amplitude for each integration time, and sensitivity is determined by fitting the calculation result.
\par
Figure~\ref{fig:nres_sensitivity}(a) displays the standard error $\sigma_{\mathrm{B_{mw}}}$ of the microwave amplitude for each integration time $T_{\mathrm{int}}$. The standard error decreases with a longer integration time. We categorize the error into the statistical error, which arises from the optical shot noise and is inversely proportional to the square root of the integration time, and contributions from other systematic errors $\sigma_{0}$. The time dependence of the standard error can be expressed by the following equation:
\begin{equation}
    \sigma_{\mathrm{B_{mw}}}(T) = \frac{\eta_{\mathrm{B_{mw}}}}{\sqrt{T_{\mathrm{int}}}} + \sigma_{0},
\end{equation}
where $\eta_{\mathrm{B_{mw}}}$ corresponds to the sensitivity.
From the fitting, the sensitivity $\eta_{\mathrm{B_{mw}}}$ is estimated to be $25 \, \mathrm{\mu T}/\sqrt{\mathrm{Hz}}$. This value indicates that for a microwave signal with an amplitude of $0.42 \, \mathrm{mT}$ and a detuning of $\Delta = 100 \, \mathrm{MHz}$, the standard error of the measurement over a one-second integration period is $25 \, \mathrm{\mu T}$. This sensitivity demonstrates that our method can sufficiently sense off-resonant spin waves with high sensitivity.
\par
Based on the sensitivity, we examine the capability of this protocol to detect far-off-resonant spin waves. In this situation, we have to take care of the effects of the NV spin $m_{s} = 1$ state and the breakdown of the rotating wave approximation. A more rigorous expression of the resonance frequency shift $f_{\mathrm{ACZ}}$ due to the AC Zeeman effect that includes these effects can be derived using Floquet perturbation theory as \cite{van2015nanometre, li2019wideband, sambe1973steady, beloy2009theory}
\begin{equation}
    f_{\mathrm{ACZ}} = \frac{(\gamma_{e} B_{-})^{2}}{f_{-} - f_{\mathrm{mw}}} + \frac{(\gamma_{e} B_{-})^{2}}{2(f_{+} + f_{\mathrm{mw}})} + \frac{(\gamma_{e} B_{+})^{2}}{f_{-} + f_{\mathrm{mw}}} + \frac{(\gamma_{e} B_{+})^{2}}{2(f_{+} - f_{\mathrm{mw}})},
    \label{eq:ac_zeeman_all}
\end{equation}
where $f_{\pm}$ are the resonance frequencies of the NV spins between $m_{\mathrm{s}} = 0$ state and $m_{\mathrm{s}} = \pm 1$ state, $B_{\pm}$ correspond to the amplitudes of the circularly polarized microwave rotating counterclockwise (clockwise), and $f_{\mathrm{mw}}$ is the microwave frequency. A detailed derivation is provided in Appendix~\hyperref[app:AC_Zeeman_rigorous]{J}.
\par
Using the derived expression, we estimate how sensitivity scales with the detuning. We assume surface spin-wave interference measurements where the microwave from the stripline is linearly polarized while the microwave from the surface spin waves is circularly polarized. In this case, the circular polarization components of the microwave amplitude can be expressed as 
\begin{equation}
    \begin{split}
        B_{-} &= B_{\mathrm{ref}} + B_{\mathrm{sw}},  \\
        B_{+} &= B_{\mathrm{ref}}, \\
    \end{split}
\end{equation}
where $B_{\mathrm{ref}}$ and $B_{\mathrm{sw}}$ represent the reference microwave amplitude and the microwave amplitude from the spin waves, respectively.
We estimate the sensitivity of the microwave amplitude from spin waves $B_{\mathrm{sw}}$ under conditions with an external magnetic field $B_{0} = 20 \, \mathrm{mT}$, the reference microwave amplitude $B_{\mathrm{ref}} = 0.505 \, \mathrm{mT}$, and the microwave amplitude from spin waves $B_{\mathrm{sw}} = 75.8 \, \mathrm{\mu T}$. Following the process in Ref.~\cite{ogawa2023demonstration}, we calculate the best sensitivity $\eta_{\mathrm{best}}$ at a single microwave irradiation time $\tau_{\mathrm{best}}$.
Figure~\ref{fig:nres_sensitivity}(b) illustrates the dependence of sensitivity on the microwave frequency $f_{\mathrm{mw}}$. The blue line represents the estimation of the sensitivity for the diamond sample used in this study, characterized by a coherence time $T_{2} = 0.72 \, \mathrm{\mu s}$. The other lines correspond to the estimations obtained by varying coherence time $T_{2}$ up to $50 \, \mathrm{\mu s}$ while keeping the PL contrast and the shot noise constant.
The sensitivity degrades as the microwave frequency increases and the detuning becomes larger. Notably, there is a peak-like feature at $f_{\mathrm{mw}} = 3 \, \mathrm{GHz}$. It occurs because the microwave frequency approaches the resonant frequency of the NV spins' $m_{s} = 1$ state. In this condition, perturbation theory does not hold, and conducting microwave sensing with the conventional protocol using the Rabi oscillation is reasonable.
Regarding the sensitivity with a much longer coherence time where $T_{2} = 50 \, \mathrm{\mu s}$, the sensitivity at a microwave frequency $f_{\mathrm{mw}} = 50 \, \mathrm{GHz}$ is estimated to be $\eta_{\mathrm{best}} = 708 \, \mathrm{\mu T}/\sqrt{\mathrm{Hz}}$. This sensitivity means that the standard error can be reduced to $\sigma_{\mathrm{B_{mw}}} = 70.8 \, \mathrm{\mu T}$ through one hundred seconds of integration and gets less than the microwave amplitude from spin waves $B_{\mathrm{sw}} = 75.8 \, \mathrm{\mu T}$. This sensitivity implies that the protocol based on the AC Zeeman effect has a sufficient potential to image spin-wave propagation at the frequency of several tens of $\mathrm{GHz}$.
\par
We show some strategies to enhance the sensitivity. The diamond sample used in this study is fabricated from an Ib substrate with a high nitrogen concentration, and it has a short coherence time of $T_{2} = 0.72 \, \mathrm{\mu s}$ and a relatively small PL contrast of $C = 0.01$. By adjusting the nitrogen concentration and ion irradiation conditions, it should be possible to optimize the sensitivity concerning PL contrast, PL intensity, and coherence time \cite{healey2020comparison}. Moreover, extending the coherence time could also be possible by replacing the CP-2 sequence in the protocol with sequences such as XY8, which can cancel out the effects of magnetic noise sources more effectively \cite{ogawa2023demonstration}. By enhancing the quality of the diamond sample and employing advanced pulse sequences, achieving coherence times up to $50 \, \mathrm{\mu s}$ could be possible while maintaining the sensitivity determined by PL contrast and PL intensity, enabling the detection of wideband spin-wave dynamics up to $50 \, \mathrm{GHz}$.

\section{Conclusion}
\label{sec:summary}
In this study, we demonstrated wideband wide-field quantitative imaging of the spin-wave propagation using the AC Zeeman effect.
In particular, we quantitatively visualized surface spin waves, both in phase and amplitude, over a wide range of frequencies (up to a maximum detuning of $567 \, \mathrm{MHz}$). 
Furthermore, our sensitivity estimation showed the potential to image spin-wave propagation with detunings up to $50 \, \mathrm{GHz}$ effectively.
Our results expand the applicability of NV centers to visualizing spin-wave dynamics. This includes exploring ferromagnetic metals like permalloy \cite{Stigloher2016}, and CoFeB \cite{Heussner2020}, which exhibit spin wave frequencies around $10 \, \mathrm{GHz}$. Additionally, our approach paves the way to explore two-dimensional Van der Waals ferromagnets \cite{Schulz2023, Xu2023}, whose spin wave frequencies often range from several $\mathrm{GHz}$ to tens of $\mathrm{GHz}$.
Quantitative investigation of the spin-wave dynamics of these materials through NV centers is essential and valuable, not only for understanding the dynamics themselves but also for their engineering applications.

\section*{Acknowledgements}
This work was partially supported by JST, CREST Grant No. JPMJCR23I2, Japan; 
Grants-in-Aid for Scientific Research (Nos. JP22K03524, JP23H01103, JP22KJ1058, JP22KJ1059, and JP24H01666);
the Mitsubishi Foundation (Grant No. 202310021);
Kondo Memorial Foundation; Daikin Industry Ltd; 
the Cooperative Research Project of RIEC, Tohoku University; 
“Advanced Research Infrastructure for Materials and Nanotechnology in Japan (ARIM)” (No. JPMXP1222UT1131) of the Ministry of Education, Culture, Sports, Science and Technology of Japan (MEXT).
K.O. and M.T. acknowledge supports from FoPM, WINGS Program, The University of Tokyo. \\

\renewcommand{\thefigure}{A\arabic{figure}}
\renewcommand{\theequation}{A\arabic{equation}}
\setcounter{equation}{0}
\setcounter{figure}{0}

\section*{Appendix A: Details of experimental setup}
\begin{figure}[t]
    \centering
    \includegraphics[width=\linewidth]{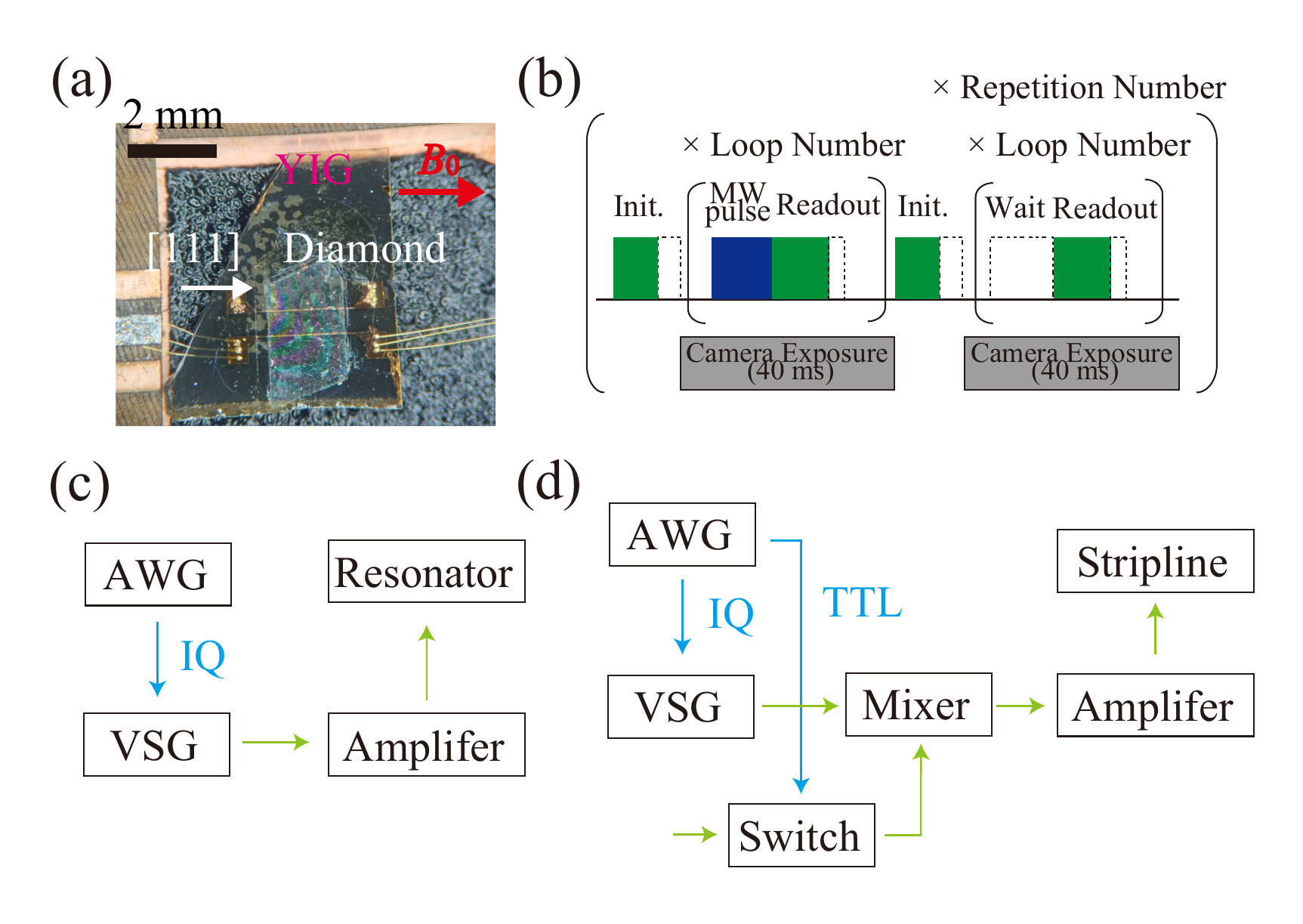}
    \caption{Experiment details.
    (a) The optical image of the measurement device.
    (b) The signal acquisition protocol for the wide-field measurements.
    (c) The microwave circuit for the resonant spin-wave sensing using the Rabi oscillation.
    (d) The microwave circuit for the off-resonant spin-wave sensing using the AC Zeeman effect.
    }
    \label{fig:exp_detail}
\end{figure}

An optical image of the measurement device is shown in Fig.~\ref{fig:exp_detail}(a). The stripline on the YIG film consists of Ti/Cu/Au ($10 \, \mathrm{nm}/190 \, \mathrm{nm}/9 \, \mathrm{nm}$).
The diamond chip and the YIG film are bonded by varnish. 
\par
Figure~\ref{fig:exp_detail}(b) shows the signal acquisition protocol for the the wide-field measurements. The protocol is the same as in Refs.~\cite{Horsley2018,ogawa2023demonstration}.
After initializing the NV spins with a green laser pulse ($16.384 \, \mathrm{\mu s}$ laser on + $16.384 \, \mathrm{\mu s}$ wait), the exposure of the sCMOS camera (Andor Zyla 5.5) starts. During the exposure, we repeat a sequence that consists of microwave pulse operations, and reading out and re-initialization with a green laser pulse ($20 \, \mathrm{\mu s}$ laser on + $2 \, \mathrm{\mu s}$ wait). Subsequently, a frame without the microwave pulse operations is acquired as a reference; we use the ratio of the two frames as a signal. The exposure time is fixed at $40 \, \mathrm{ms}$, and the laser power is set to $600 \, \mathrm{mW}$ in all measurements.
\par
The microwave circuits used for the resonant spin-wave sensing and the wideband off-resonant spin-wave sensing are shown in Figs.~\ref{fig:exp_detail}(c) and (d), respectively.
For the resonant spin-wave sensing, the microwave pulse that is controlled in amplitude and phase via IQ modulation is output by a vector signal generator (Anritsu MG3700A), amplified by an amplifier, and finally input into a microwave resonator, which is hung above the diamond chip.
For the wideband off-resonant spin wave sensing, resonant microwaves for pulse operations of NV spins are output from a vector signal generator, similar to the measurement based on the Rabi oscillation. The off-resonant microwave is output constantly from a signal generator (ROHDE \& SCHWARZ SMU200A) and controlled by a microwave switch (Mini-Circuits ZYSWA-2-50DR+). The two microwaves are mixed in a microwave mixer (Mini-Circuits ZFRSC-42-S+) and amplified before input into the stripline.

\renewcommand{\thefigure}{B\arabic{figure}}
\renewcommand{\theequation}{B\arabic{equation}}
\setcounter{equation}{0}
\setcounter{figure}{0}
\section*{Appendix B: Derivation of spin wave dispersion relationship}
\label{app:dispersion}
\begin{figure}[t]
    \centering
    \includegraphics[width=\linewidth]{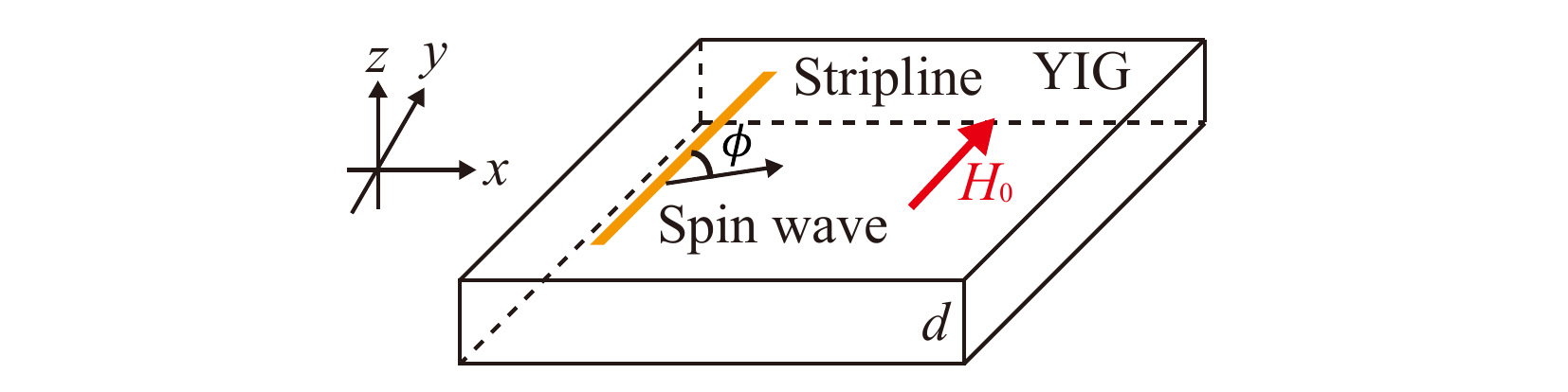}
    \caption{Schematic of spin-wave propagation.}
    \label{fig:disp_setup}
\end{figure}

In this section, we derive the dispersion relationship of the spin waves used in this study \cite{kalinikos1986theory,Yu2019,Bertelli2021}.
The magnetization $\bm{M}(\bm{r},t)$ in a magnetic film evolves over time according to the Landau-Lifshitz-Gilbert (LLG) equation
\begin{equation}
\label{LL-equation}
    \frac{d \bm{M}}{dt} = - \gamma^{\prime}_{e} \mu_{0} \bm{M} \times \bm{H}_{\mathrm{eff}} + \frac{\alpha}{M_{\mathrm{s}}} \bm{M} \times \frac{d \bm{M}}{dt},
\end{equation}
where $\bm{H}_{\mathrm{eff}}$ is the effective magnetic field, $M_{\mathrm{s}}$ is the saturation magnetization of the magnetic film, $\alpha$ is the Gilbert damping constant, and $\gamma^{\prime}_{e} (= 2\pi \gamma_{e})$ is the gyromagnetic ratio of an electron spin.
For the effective magnetic field, we consider the external magnetic field $H_{0}$, the dipolar magnetic field $H_{\mathrm{dip}}$, the exchange magnetic field $H_{\mathrm{ex}}$, and the surface anisotropy field $H_{\mathrm{s}}$. The dipolar and exchange magnetic fields can be expressed as
\begin{equation}
    \bm{H}_{\mathrm{dip}}(\bm{r},t) = \frac{1}{4\pi} \nabla \int \frac{\nabla^{\prime} \cdot \bm{M}(\bm{r}^{\prime},t)}{|\bm{r}-\bm{r}'|} d\bm{r}^{\prime},
\end{equation}
\begin{equation}
    \bm{H}_{\mathrm{ex}}(\bm{r},t) = \alpha_{\mathrm{ex}} \qty(\frac{\partial^2 \bm{M}}{\partial x^2} + \frac{\partial^2 \bm{M}}{\partial y^2} + \frac{\partial^2 \bm{M}}{\partial z^2}),
\end{equation}
where $\alpha_{\mathrm{ex}}$ is exchange stiffness. 
\par
Regarding the surface anisotropy field, we assume the sueface anisortopy energy $E_{\mathrm{s}}$ per unit volume as
\begin{equation}
    E_{\mathrm{s}} = -\frac{K_{s}}{M^{2}_{\mathrm{s}}d} M^{2}_{z},
\end{equation}
where $K_{\mathrm{s}}$ is the surface anisotropy constant \cite{Rodrigo1999Extrinsic}. From the derivative of $E_{\mathrm{S}}$, the surface anisotropy field has only the $z$-component $H_{\mathrm{s},z}(\bm{r},t)$ non-zero, which can be calculated as
\begin{equation}
\begin{split}
    H_{\mathrm{s},z}(\bm{r},t) &= \frac{\partial E_{\mathrm{s}}}{\partial M_{z}} = \frac{2 K_{s}}{M^{2}_{\mathrm{s}}d} M_{z}(\bm{r},t) = \frac{H_{\mathrm{ani}}}{M_{\mathrm{s}}} M_{z}(\bm{r},t),
\end{split}
\end{equation}
where $H_{\mathrm{ani}} = \frac{2 K_{\mathrm{s}}}{M_{\mathrm{s}}d}$ is the anisortopy field.

We consider the above equations in wavenumber and frequency space to derive the dispersion relationship.
Regarding the wavenumber space, considering the propagation of spin waves in a magnetic thin film, the two-dimensional $k_{x}$-$k_{y}$ plane is considered.
\par
First, we calculate the Fourier transformation of the dipolar field and the exchange field in the wavenumber space.
\par
The dipolar field is expressed in the form of a convolution
\begin{equation}
\label{eq-dipole}
\begin{split}
    H_{\mathrm{dip},p}(\bm{r},t) &= \int d\bm{r}' \, D_{pq}(\bm{r}-\bm{r'}) M_{q}(\bm{r'},t), \\
    D_{pq}(\bm{r}) &= \frac{1}{4\pi} \frac{\partial}{\partial p} \frac{\partial}{\partial q} \frac{1}{|\bm{r}|},
\end{split}
\end{equation}
where $H_{\mathrm{dip},p}$ is the $p$-component ($p = x, \, y, \, z$) of the dipole magnetic field $\bm{H}_{\mathrm{dip}}$. \\
Therefore, it can be expressed in the wavenumber space as
\begin{equation}
\label{eq-dipole-conv}
\begin{split}
    H_{\mathrm{dip},p}(\bm{k}, z, t) &= \int d \bm{\rho} \, H_{\mathrm{dip},p}(\bm{\rho}, z, t) e^{-i \bm{k} \cdot \bm{\rho}} \\
    &= \int dz' \, D_{p q}(\bm{k}, z-z') M_{q} (\bm{k}, z', t),
\end{split}
\end{equation}
where $\bm{\rho} = (x,y)$ represents the two-dimensional coordinate and
\begin{equation}
\begin{split}
    D_{p q}(\bm{k}, z) = \frac{1}{4\pi} \int \qty(\frac{\partial}{\partial p} \frac{\partial}{\partial q} \frac{1}{|\bm{r}|}) e^{-i \bm{k} \cdot \bm{\rho}} d \bm{\rho}. \\
\end{split}
\end{equation}
Each component of $D_{p q}(\bm{k}, z)$ can be calculated as
\begin{equation}
\begin{split}
    &D(\bm{k}, z) = \\
    &\mqty(-\frac{k_{x}^2}{2k} e^{-k|z|} & -\frac{k_{x}k_{y}}{2k} e^{-k|z|} & -ik_{x} \mathrm{sign}(z) e^{-k|z|} \\
    -\frac{k_{x}k_{y}}{2k} e^{-k|z|} & -\frac{k_{y}^2}{2k} e^{-k|z|} & -ik_{y} \mathrm{sign}(z) e^{-k|z|} \\
    -ik_{x} \mathrm{sign}(z) e^{-k|z|} & -ik_{y} \mathrm{sign}(z) e^{-k|z|} & \frac{1}{2} k e^{-k|z|} - \delta (z)  
    ),
\end{split}
\end{equation}
where $k = \sqrt{k^{2}_{x} + k^{2}_{y}}$. \\
Regarding the integration in the $z$-direction of Eq.~(\ref{eq-dipole-conv}), we consider the condition where the wavelength of the spin waves is sufficiently longer than the thickness of the film, and the magnetization is uniform across the $z$-direction. Also, we approximate that the dipolar field takes the average value across the thickness direction and does not depend on $z$:
\begin{equation}
    H_{\mathrm{dip},p}(\bm{k},t) = \frac{1}{d} \int_{-d}^{0} dz \int_{-d}^{0} dz' D_{p q}(\bm{k}, z-z') M_{q} (\bm{k},t). \\
\end{equation}
Under these conditions, the dipolar field in the wavenumber space can be calculated as
\begin{equation}
\begin{split}
    &\bm{H}_{\mathrm{dip}}(\bm{k},t) \\
    &= \mqty(-\frac{k_{x}^2}{k^2} g(kd) & -\frac{k_{x}k_{y}}{k^2} g(kd) & 0 \\
    -\frac{k_{x}k_{y}}{k^2} g(kd) & -\frac{k_{y}^2}{k^2} g(kd) & 0 \\
    0 & 0 & g(kd)-1
    )
    \bm{M}(\bm{k},t) \\
    &= \mqty(- g(kd)\sin^{2}\phi & - \frac{g(kd) \sin 2\phi}{2} & 0 \\
    - \frac{g(kd) \sin 2\phi}{2} & - g(kd) \cos^{2}\phi & 0 \\
    0 & 0 & g(kd)-1
    )
    \bm{M}(\bm{k},t),
\end{split}
\end{equation}
where
\begin{equation}
    \bm{M}(\bm{k},t) = \mqty(M_{x}(\bm{k},t) \\  M_{y}(\bm{k},t) \\ M_{z}(\bm{k},t)), \\
\end{equation}
\begin{equation}
    g(kd) = 1 - \frac{1-e^{-kd}}{kd}, \\
\end{equation}
and $\phi$ is the angle between the direction of the external magnet and that of spin-wave propagation [Fig.\ref{fig:disp_setup}].
\par
The Fourier transformation of the exchange field in the wavenumber space can be calculated as
\begin{equation}
    \bm{H}_{\mathrm{ex}}(\bm{k},t) = -\alpha_{\mathrm{ex}} k^{2} \bm{M}(\bm{k},t).
\end{equation}
\par
The Fourier transformation of the surface anisotropy field in the wavenumber space can be calculated as
\begin{equation}
    H_{\mathrm{s},z}(\bm{k},t) = \frac{H_{\mathrm{ani}}}{M_{\mathrm{s}}} M_{z}(\bm{k},t).
\end{equation}
\par
We assume the external magnetic field is applied in the $y$-direction, and the magnetic film is magnetized in this in-plane direction.
Also, we consider the condition where the precession amplitude of the spin waves is sufficiently small compared to the saturation magnetization:
\begin{equation}
\begin{split}
    M_{x}(\bm{r},t), \, &M_{z}(\bm{r},t) \ll M_{s}, \\
    M_{y}(\bm{r},t) &= \sqrt{M_{s} - M^2_{x}(\bm{r},t) - M^{2}_{z}(\bm{r},t)} \\ 
                    &\simeq M_{s}.
\end{split}
\end{equation}
\par
Based on the above preparations, we consider the LLG equation (\ref{LL-equation}) in the wavenumber and the frequency space. Also, we ignore second-order terms of small quantities. 
The LLG equation (\ref{LL-equation}) can be expressed as
\begin{equation}
\label{m_res_eq}
\begin{split}
    -i \omega m_{x,\bm{k},\omega} &= \qty(\omega_{1} - i \alpha \omega) m_{z,\bm{k},\omega}, \\
    -i \omega m_{z,\bm{k},\omega} &= \qty(-\omega_{2} + i \alpha \omega) m_{x,\bm{k},\omega}, \\
\end{split}
\end{equation}
where $m_{x,\bm{k},\omega}$, $m_{z,\bm{k},\omega}$ are the Fourier transformations of magnetization in the $x$- and $z$-direction respectively:
\begin{equation}
    \begin{split}
        m_{x,\bm{k},\omega} &= \int d \bm{\rho} \int dt \, M_{x}(\bm{r},t) e^{-i\bm{k} \cdot \rho } e^{-i\omega t}, \\
        m_{z,\bm{k},\omega} &= \int d \bm{\rho} \int dt \, M_{z}(\bm{r},t) e^{-i\bm{k} \cdot \rho } e^{-i\omega t}, \\
    \end{split}
\end{equation}
and
\begin{equation}
\begin{split}
    \omega_{1} &= \gamma^{\prime}_{e} \mu_{0} \qty(H_{0} + M_{\mathrm{s}}(1-g(kd)) - H_{\mathrm{ani}} + \alpha_{\mathrm{ex}} M_{s} k^{2}), \\
    \omega_{2} &= \gamma^{\prime}_{e} \mu_{0} \qty(H_{0} + M_{\mathrm{s}}g(kd)\sin^2 \phi + \alpha_{\mathrm{ex}} M_{s} k^{2}). \\
\end{split}
\label{eq:omega_12}
\end{equation}
From Eq.~(\ref{m_res_eq}), the resonance frequency of the spin-waves can be calculated as
\begin{equation}
\label{eq:disp_complex}
    \omega = \sqrt{\omega_{1} \omega_{2}} - i\alpha \frac{\omega_{1} + \omega_{2}}{2}.
\end{equation}
From the real part of Eq.~(\ref{eq:disp_complex}), the following dispersion relationship of the spin waves is obtained:
\begin{equation}
\begin{split}
    \omega = \gamma^{\prime}_{e} \mu_{0} &\sqrt{\qty(H_{0} + M_{\mathrm{s}}(1-g(kd)) - H_{\mathrm{ani}} + \alpha_{\mathrm{ex}} M_{s} k^{2})} \\
        & \times \sqrt{\qty(H_{0} + M_{\mathrm{s}}g(kd)\sin^2 \phi + \alpha_{\mathrm{ex}} M_{s} k^{2})}
\end{split}
\label{eq:dispersion_all}
\end{equation}
For the analysis of the experimental results in this paper, since we examine only the regions where the wavenumber is small and the effects of the external magnetic field and the dipolar magnetic field dominant, we use the dispersion relationship that ignores the terms of the exchange field. Furthermore, when the wavelength is sufficiently longer than the film thickness ($kd \ll 1$) and the anisotropy field is not large, $H_{\mathrm{ani}} g(kd) \ll H_{0}, M_{\mathrm{s}}$ holds. In this situation, by defining the effective magnetization $M_{\mathrm{eff}}$ as $M_{\mathrm{eff}} = M_{\mathrm{s}} - H_{\mathrm{ani}}$, $\omega_{1}, \omega_{2}$ in Eq.~(\ref{eq:omega_12}) can be approximated as
\begin{equation}
\begin{split}
    \omega_{1} &\simeq \gamma^{\prime}_{e} \mu_{0} \qty(H_{0} + M_{\mathrm{eff}}(1-g(kd))), \\
    \omega_{2} &\simeq \gamma^{\prime}_{e} \mu_{0} \qty(H_{0} + M_{\mathrm{eff}}g(kd)\sin^2 \phi). \\
\end{split}
\end{equation}
The dispersion relationship in Eq.~(\ref{eq:dispersion_all}) can be therefore rewritten as \cite{Qin2018}
\begin{equation}
    \omega \simeq \gamma^{\prime}_{e} \mu_{0} \sqrt{\qty(H_{0} + M_{\mathrm{eff}}(1-g(kd))) \qty(H_{0} + M_{\mathrm{eff}}g(kd)\sin^2 \phi)},
\end{equation}
which is equivalent to Eq.~(\ref{eq:dispersion}) in the main text.
\par
The imaginary part of the resonance frequency $\Delta \omega = \alpha \frac{\omega_{1} + \omega_{2}}{2}$ corresponds to the resonance linewidth. Also, the linewidth in the corresponding wavenumber $\Delta \bm{k}$ is linked to the decay length of the spin waves propagating on the magnetic film.
The relationship between the linewidth of resonance frequency and that of the wavenumber is 
\begin{equation}
    \Delta \omega = \frac{\partial \omega}{\partial k} \Delta k  = v_{\mathrm{g}} \Delta k,  
    \label{eq:domega}
\end{equation}
where $v_{\mathrm{g}}$ is the group velocity of the spin waves.
From Eq.~(\ref{eq:domega}), the decay length of the spin waves $l_{\mathrm{d}}$ can be calculated as
\begin{equation}
    l_{\mathrm{d}} = \frac{2 v_{\mathrm{g}}}{\alpha (\omega_{1} + \omega_{2})}.
\end{equation}
Also, the ellipticity $\eta_{\bm{k}}$ of the spin wave precession can be calculated from Eq. (\ref{m_res_eq}) as 
\begin{equation}
\begin{split}
    \eta_{\bm{k}} &= \left| \frac{m_{z,\bm{k},\omega}}{m_{x,\bm{k},\omega}} \right| \\
         &= \sqrt{\frac{\omega_{2}}{\omega_{1}}}. 
\end{split}
\end{equation}

\renewcommand{\thefigure}{C\arabic{figure}}
\renewcommand{\theequation}{C\arabic{equation}}
\setcounter{equation}{0}
\setcounter{figure}{0}
\section*{Appendix C: Calculation of microwave amplitude from spin waves sensed by NV centers}

\subsection{Microwave amplitude from spin waves}
In this section, we derive the microwave amplitude from the spin waves at the position of the NV centers \cite{bertelli2020magnetic,zhou2021magnon}.
\par
We consider the spin waves propagating in a magnetic thin film and assume the temporal precession of the spin waves at a position $\rho$ as
\begin{equation}
\begin{split}
    m_{x}(\bm{\rho}, t) &= m_{0} e^{i (\bm{k} \cdot \bm{\rho} - \omega t)},  \\
    m_{z}(\bm{\rho}, t) &= - i \eta_{\bm{k}} m_{0} e^{-i  (\bm{k} \cdot \bm{\rho} - \omega t)}.
\end{split}
\end{equation}
Using Eq.~(\ref{eq-dipole-conv}), the microwave dipolar magnetic field in the wavenumber space at a position $z > 0$ can be expressed as 
\begin{equation}
\begin{split}
    B_{\mathrm{sw},p}(\bm{k},t) &= \mu_{0} \int_{-d}^{0} dz^{\prime} D_{p q}(\bm{k}, z-z^{\prime}) m_{q}(\bm{k},t). \\
\end{split}
\end{equation}
By calculating the above equation and converting it back into the coordinate representation, the microwave amplitude at the position $(\bm{\rho}, z)$ from the spin waves can be expressed as
\begin{equation}
\begin{split}
    &B_{\mathrm{sw}, x}(\bm{\rho}, t) \\
    &= - \frac{\mu_{0}}{2} e^{-kz} (1-e^{-kd}) \sin \phi (\sin \phi  + \eta_{\bm{k}})  m_{0} e^{i (\bm{k} \cdot \bm{\rho} - \omega t)}, \\
    &B_{\mathrm{sw}, y}(\bm{\rho}, t) \\
    &= - \frac{\mu_{0}}{2} e^{-kz} (1-e^{-kd}) \cos \phi (\sin \phi + \eta_{\bm{k}}) m_{0} e^{i (\bm{k} \cdot \bm{\rho} - \omega t)}, \\
    &B_{\mathrm{sw}, z}(\bm{\rho}, t) \\
    &= - i \frac{\mu_{0}}{2} e^{-kz} (1-e^{-kd}) (\sin \phi  + \eta_{\bm{k}}) m_{0} e^{i (\bm{k} \cdot \bm{\rho} - \omega t)}.
\end{split}
\end{equation}
In this study, we detect the surface spin waves corresponding to $\phi = \frac{\pi}{2}$. Also, the NV centers used in the measurements have their quantization axis along the $y$-direction, and they are sensitive to microwave magnetic fields perpendicular to this direction. Therefore, the microwave magnetic field sensed by the NV centers is
\begin{equation}
\label{eq:mw_sw_vec}
\begin{split}
    &B_{\mathrm{sw}, x}(\bm{\rho}, t) \\
    &= - \frac{\mu_{0}}{2} e^{-kz} (1-e^{-kd}) (1  + \eta_{\bm{k}}) m_{0} e^{i (\bm{k} \cdot \bm{\rho} - \omega t)}, \\
    &B_{\mathrm{sw}, z}(\bm{\rho}, t) \\
    &= - i \frac{\mu_{0}}{2} e^{-kz} (1-e^{-kd}) (1  + \eta_{\bm{k}}) m_{0} e^{i (\bm{k} \cdot \bm{\rho} - \omega t)}.
\end{split}
\end{equation}
This microwave is circularly polarized and induces transitions between the $m_{\mathrm{s}} = 0 \leftrightarrow m_{\mathrm{s}} = -1$ states of the NV spins. Finally, from Eq.~(\ref{eq:mw_sw_vec}), we can obtain the expression for the circularly polarized microwave amplitude from the spin waves as
\begin{equation}
    B_{\mathrm{sw}} = \frac{\mu_{0}}{2} e^{-kz} (1-e^{-kd}) (1  + \eta_{\bm{k}}) m_{0}. \\
\end{equation}

\subsection{Interference between microwave field from spin waves and reference microwave field}
In our measurements, the microwave sensed by the NV centers results from the interference between the microwave from the spin waves and the reference microwave radiated from the antenna. 
Since the wavelength of the reference microwave is on the order of centimeters, which is sufficiently longer than the length of the field of view, the phase of the reference microwave can be regarded as uniform. 
On the other hand, the wavelength of the spin waves is on the order of several micrometers and their phase significantly changes within the field of view. 
We define the microwave from the spin waves $B_{\mathrm{sw}}(\bm{\rho}, t)$ and the reference microwave $B_{\mathrm{ref}}(\bm{\rho}, t)$ as
\begin{equation}
\begin{split}
    B_{\mathrm{sw}}(\bm{\rho}, t) &=  B_{\mathrm{sw}}(\bm{\rho}) e^{i (\bm{k} \cdot \bm{\rho} - \omega t + \theta)}, \\
    B_{\mathrm{ref}}(\bm{\rho}, t) &=  B_{\mathrm{ref}}(\bm{\rho}) e^{-i\omega t}. \\
\end{split}
\end{equation}
Under this condition, the microwave amplitude $B_{\mathrm{mw}}(\bm{\rho}, t)$ sensed by NV centers is
\begin{equation}
\begin{split}
    B_{\mathrm{mw}}(\bm{\rho}, t) &= \Re \qty[B_{\mathrm{sw}}(\bm{\rho}, t) + B_{\mathrm{ref}}(\bm{\rho}, t)] \\
                                  &= B_{\mathrm{tot}}(\bm{\rho}) \cos(\omega t + \theta), \\
    \label{eq:Btot}
\end{split}
\end{equation}
where
\begin{equation}
\begin{split}
    &B_{\mathrm{tot}}(\bm{\rho}) \\
    &\, = \sqrt{B^2_{\mathrm{sw}}(\bm{\rho}) + 2  B_{\mathrm{sw}}(\bm{\rho}) B_{\mathrm{ref}}(\bm{\rho}) \cos (\bm{k} \cdot \bm{\rho} + \theta) + B^2_{\mathrm{ref}}(\bm{\rho})}.
\end{split}
\end{equation}
When the reference microwave amplitude is sufficiently larger than the microwave amplitude from the spin waves $\qty(B_{\mathrm{ref}}(\bm{\rho}) \gg B_{\mathrm{sw}}(\bm{\rho}))$, the total microwave amplitude $B_{\mathrm{tot}}(\bm{\rho})$ can be approximated as
\begin{equation}
    B_{\mathrm{tot}}(\bm{\rho}) \simeq B_{\mathrm{ref}}(\bm{\rho}) + B_{\mathrm{sw}}(\bm{\rho}) \cos (\bm{k} \cdot \bm{\rho} + \theta).
    \label{eq:Btot_approx}
\end{equation}

\renewcommand{\thefigure}{D\arabic{figure}}
\renewcommand{\theequation}{D\arabic{equation}}
\setcounter{equation}{0}
\setcounter{figure}{0}
\section*{Appendix D: Distance calibration between NV layer and YIG film surface}
\label{app:distance}
\begin{figure}[b]
    \centering
    \includegraphics[width=\linewidth]{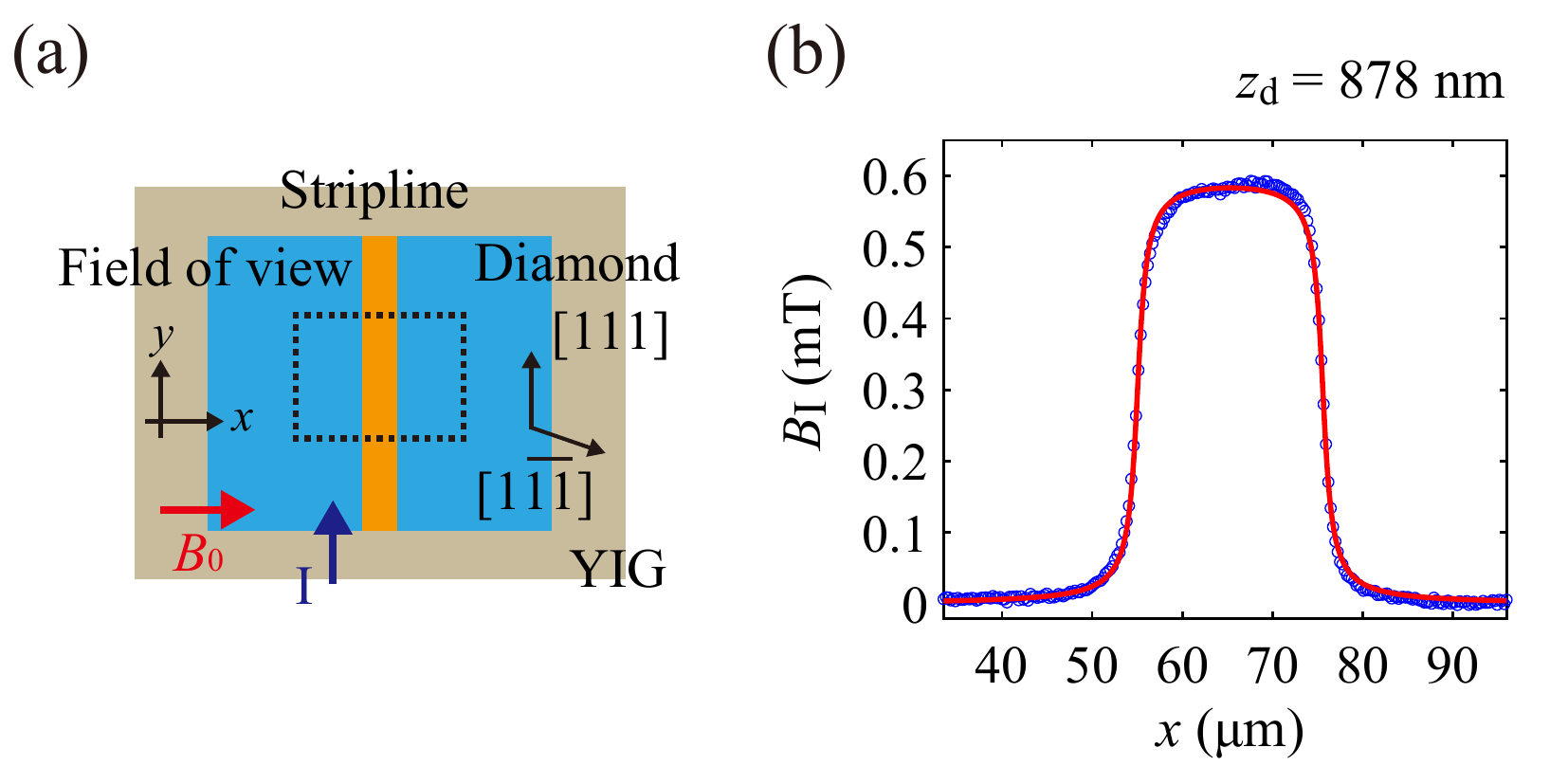}
    \caption{
        Distance calibration between the NV layer and the YIG film surface.
        (a) Schematic of the measurement.
        (b) Magnetic field distribution from the stripline current. (blue) Experimental results. (red) The fitting result based on the analytical model.
    }
    \label{fig:distance_calib}
\end{figure}

The distance between the NV layer and the YIG film surface is estimated by applying dc current to the stripline on the YIG film surface and fitting the distribution of the stray magnetic field from the stripline current using an analytical model \cite{bertelli2020magnetic}. 
\par
The measurement setup is shown in Figure~\ref{fig:distance_calib}(a). The $x$-component dominates the stray magnetic field from the stripline current except near the edge of the stripline. To detect the magnetic field in this direction, we utilize NV centers oriented in the $[1\overline{11}]$ direction within the plane, which is $\ang{19.47}$ tilted from the $x$-direction.
An external magnetic field $B_{0} = 9.62 \, \mathrm{mT}$ is applied in the $x$-direction, and dc current $I = \pm \, 20 \, \mathrm{mA}$ is applied to the stripline using a DC voltage source (Yokogawa GS200). We measure the optically detected magnetic resonance (ODMR) spectrum of NV centers to obtain the magnetic field distribution. We integrate signals in the $y$-direction within the field of view, and the magnetic field distribution is calculated by estimating the two resonance frequencies of the NV centers at each position and converting them to the magnetic field \cite{Tsukamoto2021}.
The magnetic field distribution from the dc current $B_{\mathrm{I}}(x)$ is estimated by taking the difference between the distributions obtained with positive and negative currents. We fit the obtained magnetic field distribution using Amp\`{e}re's law:
\begin{equation}
    B_{\mathrm{I}}(x) = \int_{0}^{t} dz \frac{\mu_{0} I}{2\pi w t} \arctan(\frac{w(z_{\mathrm{d}}-z)}{(x-x_{0})^{2} + (z_{\mathrm{d}}-z)^{2} - (\frac{w}{2})^2}), 
\end{equation}
where $z_{\mathrm{d}}$ represents the distance between the NV layer and the YIG film surface, $w$ is the width of the stripline, $x_{0}$ is the $x$-coordinate of the central position of the stripline, and $I$ corresponds to the current value.
The integration in the $z$-direction refers to the integration across the thickness of the stripline. In this measurement, the thickness of the stripline $t$ is estimated to be $t = 209 \, \mathrm{nm}$ from the calibration values of the coating thickness gauge.
In this configuration, the $z$-component of the magnetic field from the stripline current can be ignored because it is perpendicular to the NV axis, and the expected maximum magnitude of $0.5 \, \mathrm{mT}$ is sufficiently small compared to the bias external magnetic field in the $x$-direction $B_{0} = 9.62 \, \mathrm{mT}$. 
For the fitting, only parameters $w$, $x_{0}$, and $z_{\mathrm{d}}$ are used as fitting parameters; all other parameters are fixed.

Figure~\ref{fig:distance_calib}(b) shows the experimentally obtained magnetic field distribution (blue points) and the fitting result (red line). The model accurately reproduces the experimental results, and the distance between the NV layer and the YIG film surface is estimated to be $z_{\mathrm{d}} = (878 \pm 20) \, \mathrm{nm}$.

\renewcommand{\thefigure}{E\arabic{figure}}
\renewcommand{\theequation}{E\arabic{equation}}
\setcounter{equation}{0}
\setcounter{figure}{0}

\section*{Appendix E: Frequency characteristics of the microwave resonator}
\label{app:resonator}
\begin{figure}[b]
    \centering
    \includegraphics[width=\linewidth]{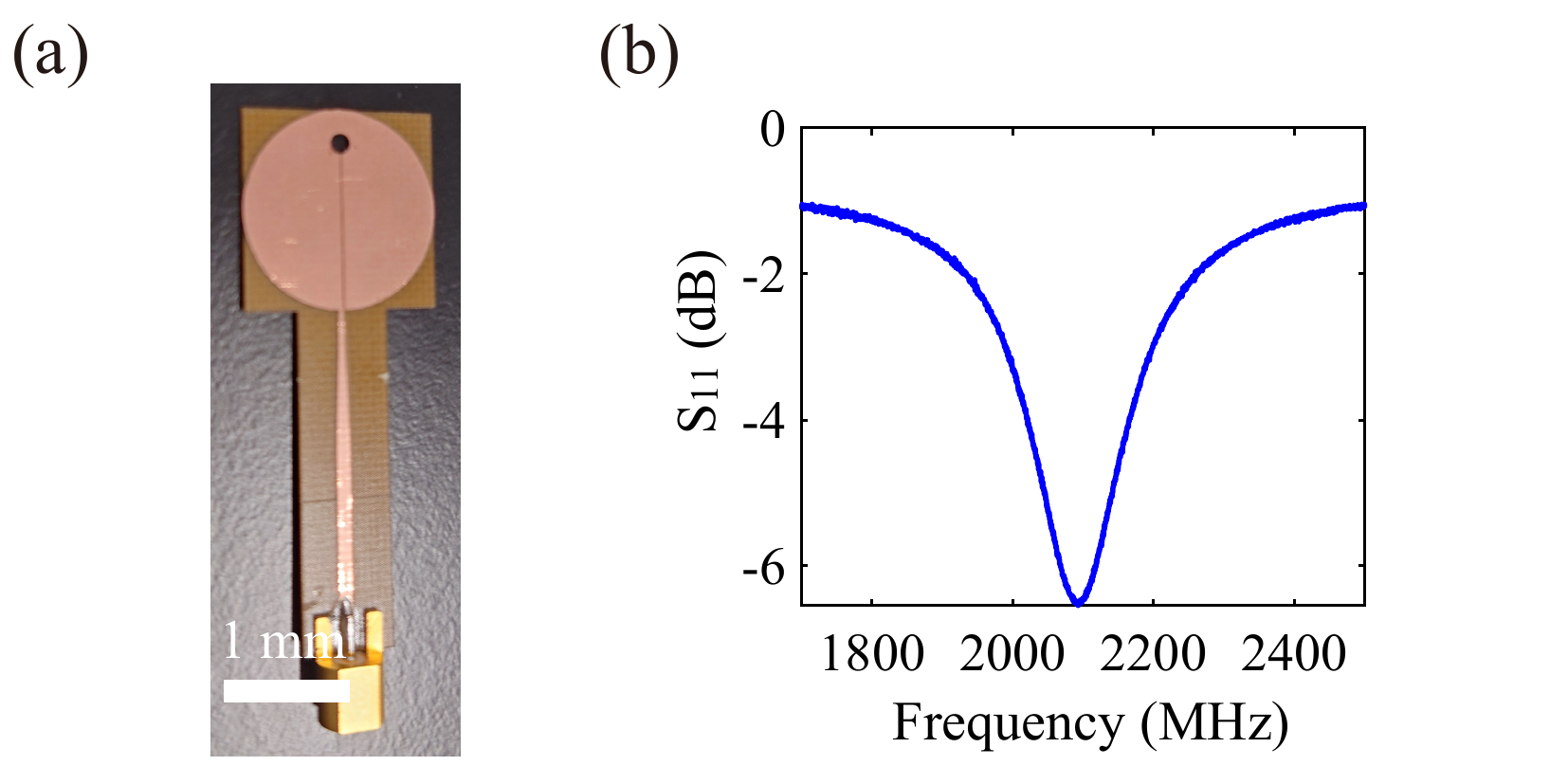}
    \caption{The resonator antenna used in the resonant spin-wave sensing. (a) An optical image of the resonator. (b) $S_{\mathrm{11}}$ characteristics of the resonator obtained with a network analyzer.}
    \label{fig:resonator_chara}
\end{figure}

In the resonant spin-wave detection based on the protocol using the Rabi oscillation, a microwave resonator is hung above the diamond chip. The microwave is input to this resonator to excite spin waves and radiate uniform microwaves in the field of view.
Figure~\ref{fig:resonator_chara}(a) shows an optical image of the resonator.
The design is based on Ref.~\cite{Sasaki2016}. It is manufactured by laser processing a PCB (PCB: FR-4.0) board of a thickness of $300 \, \mathrm{\mu m}$ with $18 \, \mathrm{\mu m}$ copper foil.
\par 
Figure~\ref{fig:resonator_chara}(b) shows the $S_{11}$ characteristics of the resonator. The figure displays that it has a resonance frequency of around $2100 \, \mathrm{MHz}$.

\renewcommand{\thefigure}{F\arabic{figure}}
\renewcommand{\theequation}{F\arabic{equation}}
\setcounter{equation}{0}
\setcounter{figure}{0}
\section*{Appendix F: Analysis of microwave amplitude in resonant spin-wave sensing based on the Rabi oscillation}
\label{app:rabi}

In this section, we explain the details of the analysis of the microwave amplitude in resonant spin-wave sensing based on the protocol using the Rabi oscillation.
The resulting microwave distribution in the measurements is represented by Eq.~(\ref{eq:Btot}). 
Information about the reference microwave distribution is necessary to extract the microwave amplitude from the spin waves.
\par
\begin{figure}[t]
    \centering
    \includegraphics[width=\linewidth]{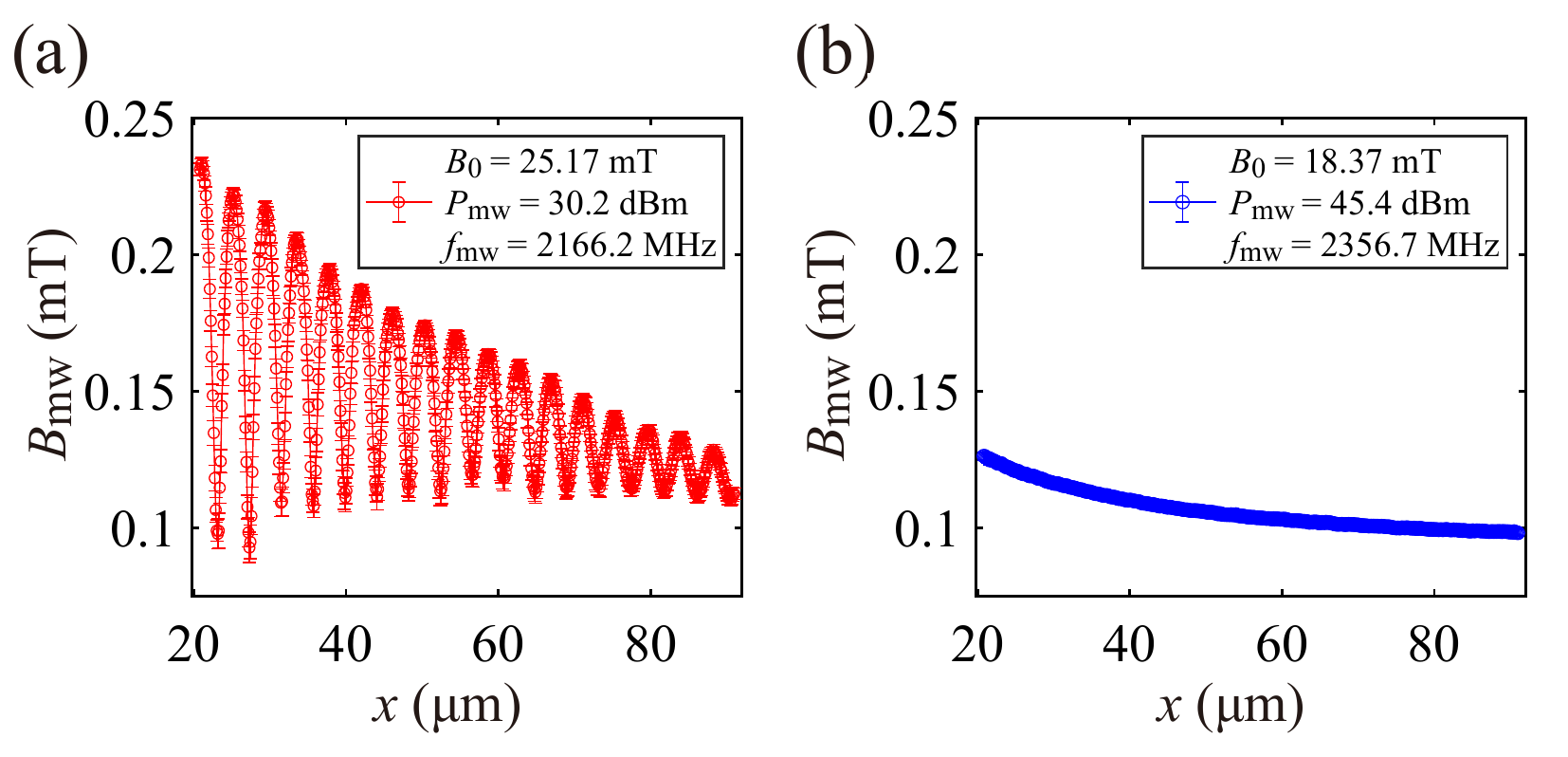}
    \caption{
        (a) Microwave amplitude distribution at an external magnetic field $B_{0} = 25.17 \, \mathrm{mT}$ and a microwave frequency $f_{\mathrm{mw}} = 2166.2 \, \mathrm{MHz}$.
        (b) Microwave amplitude distribution at an external magnetic field of $B_{0} = 18.37 \, \mathrm{mT}$ and a microwave frequency $f_{\mathrm{mw}} = 2256.7 \, \mathrm{MHz}$.
    }
    \label{fig:rabi_ref}
\end{figure}

Figure~\ref{fig:rabi_ref}(a) shows the microwave amplitude distribution at an external magnetic field of $B_{0} = 25.17 \, \mathrm{mT}$ and a microwave frequency of $f_{\mathrm{mw}} = 2166.2 \, \mathrm{MHz}$. Clear interference between the microwave from the spin waves and the reference microwave is observed.
The envelope of the obtained microwave amplitude distribution is greater than the amplitude of the oscillations at each position, and the microwave amplitude of the envelope only changes by about $30 \, \%$ from the left to the right edge of the field of view. On the other hand, the amplitude of the oscillations decreases by a factor of about ten from the left to the right edge of the field of view.
\par
Figure~\ref{fig:rabi_ref}(b) shows the microwave amplitude distribution at an external magnetic field of $B_{0} = 18.37 \, \mathrm{mT}$ and a microwave frequency of $f_{\mathrm{mw}} = 2356.7 \, \mathrm{MHz}$. At this magnetic field, the microwave amplitude from the spin waves is small because their wavenumber at the frequency is large, making the contribution from the reference microwave dominant.
Since the microwave frequency is far from the resonator's resonance frequency, the microwave amplitude of the envelope is smaller than that at $B_{0} = 25.17 \, \mathrm{mT}$. However, the change in the microwave amplitude from the left to the right edge of the field of view is about $30 \, \%$, consistent with the envelope trend at $B_{0} = 25.17 \, \mathrm{mT}$.
Therefore, we conclude that the envelope represents the reference microwave amplitude distribution, and the oscillations derive from the spin waves. Since the amplitude of the reference microwave is larger than that from the spin waves, we use Eq.~(\ref{eq:Btot_approx}) for the analysis.

\renewcommand{\thefigure}{G\arabic{figure}}
\renewcommand{\theequation}{G\arabic{equation}}
\setcounter{equation}{0}
\setcounter{figure}{0}

In wideband spin-wave sensing based on the AC Zeeman effect, pulse operations of NV spins are necessary to read out the resonance frequency shifts due to the AC Zeeman effect using the CP-2 sequence.
Since we simultaneously image a wide field of view, we must manipulate the NV spins uniformly and sufficiently fast compared to the coherence time across the field of view.
As described in the main text, we manipulate the NV spins by the microwave radiated from the stripline.
Figure~\ref{fig:composite_pulse}(a) shows the Rabi frequency distribution at an external magnetic field $B_{0} = 18.37 \, \mathrm{mT}$, a microwave frequency $f_{\mathrm{mw}} = 2356.7 \, \mathrm{MHz}$, and an input microwave power $P_{\mathrm{mw}} = 30.8 \, \mathrm{dBm}$.
Though the Rabi frequency is sufficiently large ($\sim 60 \, \mathrm{MHz}$) around the edge of the stripline, it also exhibits significant inhomogeneity. We employ SCROFULOUS composite pulses to compensate for the amplitude error \cite{cummins2003tackling,nomura2021composite,ogawa2023demonstration}.
SCROFULOUS composite pulses, shown in Fig.~\ref{fig:composite_pulse}(b), can reduce pulse amplitude errors by replacing the $\frac{\pi}{2}$ pulse and $\pi$ pulse with three composite pulses with specific phases and rotation angles. 

\section*{Appendix G: Details of wideband spin-wave sensing}
\label{app:wideband_setup}
\begin{figure}[b]
    \centering
    \includegraphics[width=\linewidth]{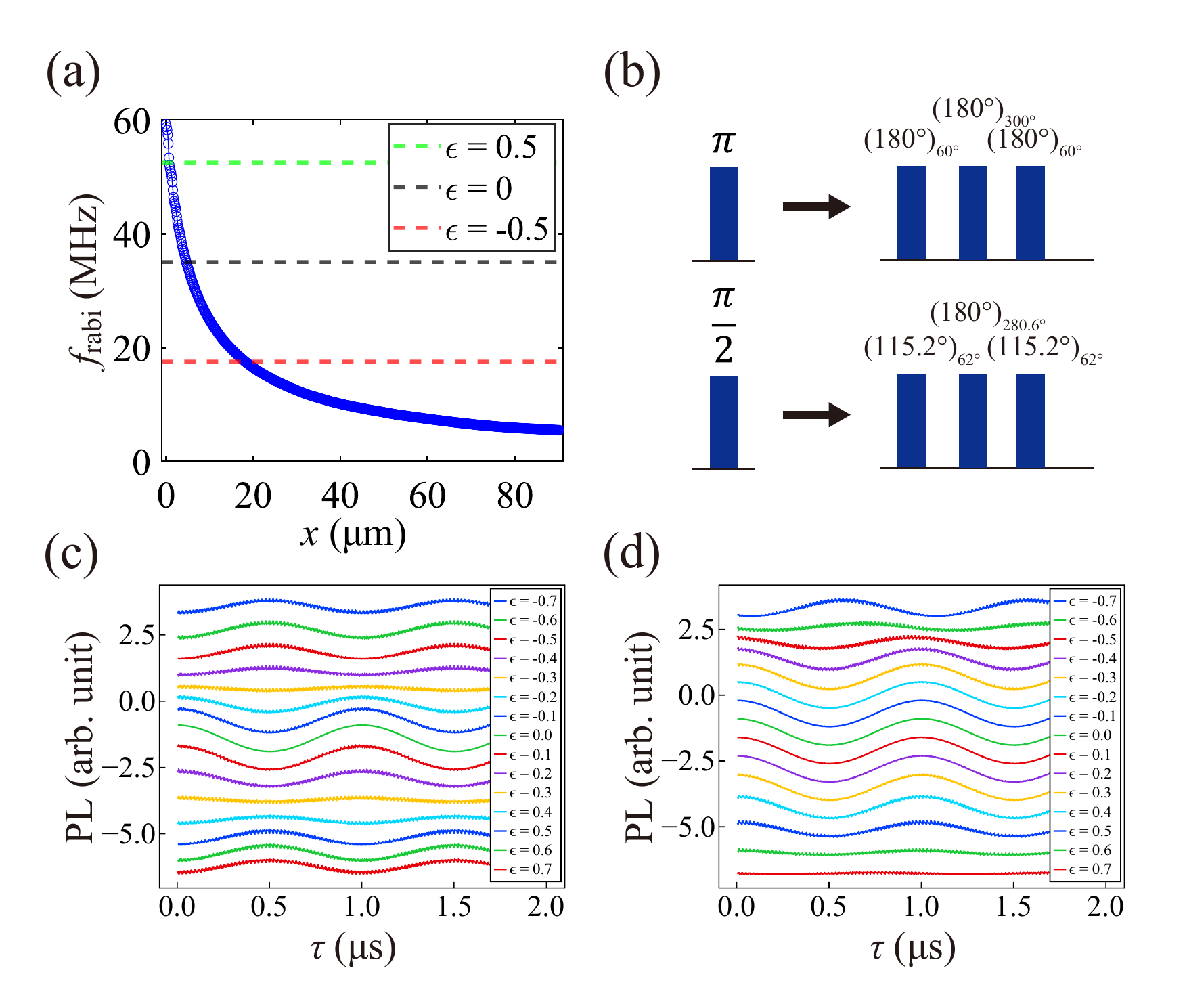}
    \caption{Details of the implementation of wideband spin-wave sensing based on the AC Zeeman effect. 
    (a) Rabi frequency distribution $f_{\mathrm{rabi}}(x)$ at a magnetic field $B_{0} = 18.37 \, \mathrm{mT}$, microwave frequency $f_{\mathrm{mw}} = 2356.7 \, \mathrm{MHz}$, and input microwave power $P_{\mathrm{in}} = 30.8 \, \mathrm{dBm}$.
    (b) SCROFULOUS composite pulses.
    (c) (d) Numerical simulation results of the protocol to detect the AC Zeeman effect under various pulse amplitude error rates $\epsilon$. Detuning of the signal microwave is set to $\Delta = 200 \, \mathrm{MHz}$ and the amplitude is $f_{\mathrm{rabi}} = 20 \, \mathrm{MHz}$ regarding Rabi frequency. 
    (c) is the result without composite pulses, and (d) corresponds to the result with composite pulses.
    }
    \label{fig:composite_pulse}
\end{figure}

\par
Figures~\ref{fig:composite_pulse}(c) and (d) show the results of numerical simulations to evaluate how much pulse amplitude errors can be compensated by utilizing composite pulses.
In the numerical simulations, we calculate the time evolution in the protocol shown in Fig.~\ref{fig:setup_principle}(d) in the main text. We set an external magnetic field $B_{0} = 18.37 \, \mathrm{mT}$, a detuning of the off-resonant signal microwave $\Delta = 200 \, \mathrm{MHz}$, and an amplitude $f_{\mathrm{rabi}} = 20 \, \mathrm{MHz}$ regarding Rabi frequency. 
Additionally, the resonant microwave amplitude for pulse operations $f_{\mathrm{pulse}}$ is set to $35 \, \mathrm{MHz}$, and this value is used to determine the pulse lengths of the $\frac{\pi}{2}$ pulse and the $\pi$ pulse.
In this situation, the NV spin can be regarded as a two-level system with $m_{s} = 0$ state and $m_{s} = -1$ state, and the rotating wave approximation is valid. Therefore, the time evolution of the spin is calculated using the exact expression for a two-level spin system under circularly polarized microwave application (details are described in the Supplementary Information of Ref.~\cite{ogawa2023demonstration}). We introduce an error rate $\epsilon$ as $f^{\prime}_{\mathrm{pulse}} = f_{\mathrm{pulse}} (1 + \epsilon)$ to investigate the effect of pulse amplitude errors. 
\par
Figure~\ref{fig:composite_pulse}(c) displays the simulation results of using conventional pulses under various pulse amplitude error rates, and Fig.~\ref{fig:composite_pulse}(d) illustrates the results of using the SCROFULOUS composite pulses. With conventional pulses, the signal oscillations disappear when the absolute error rate reaches approximately $0.3$. However, when using SCROFULOUS composite pulses, signal oscillations due to the AC Zeeman effect can still be observed even with an error rate of around $\pm \, 0.5$.
\par
Based on these simulation results, in our measurement, the resonant microwave amplitude for pulse operations is set to $35 \, \mathrm{MHz}$, and we focus on the region where the error rate $|\epsilon|$ is smaller than 0.5. The region that meets these conditions ranges from $x = 1.95 \, \mathrm{\mu m}$ to $x = 14.3 \, \mathrm{\mu m}$ [see Fig.~\ref{fig:composite_pulse}(a)].

\renewcommand{\thefigure}{H\arabic{figure}}
\renewcommand{\theequation}{H\arabic{equation}}
\setcounter{equation}{0}
\setcounter{figure}{0}

\section*{Appendix H: Numerical simulation of stripline microwave field}
\label{app:stripline_mw}
\begin{figure}[b]
    \centering
    \includegraphics[width=\linewidth]{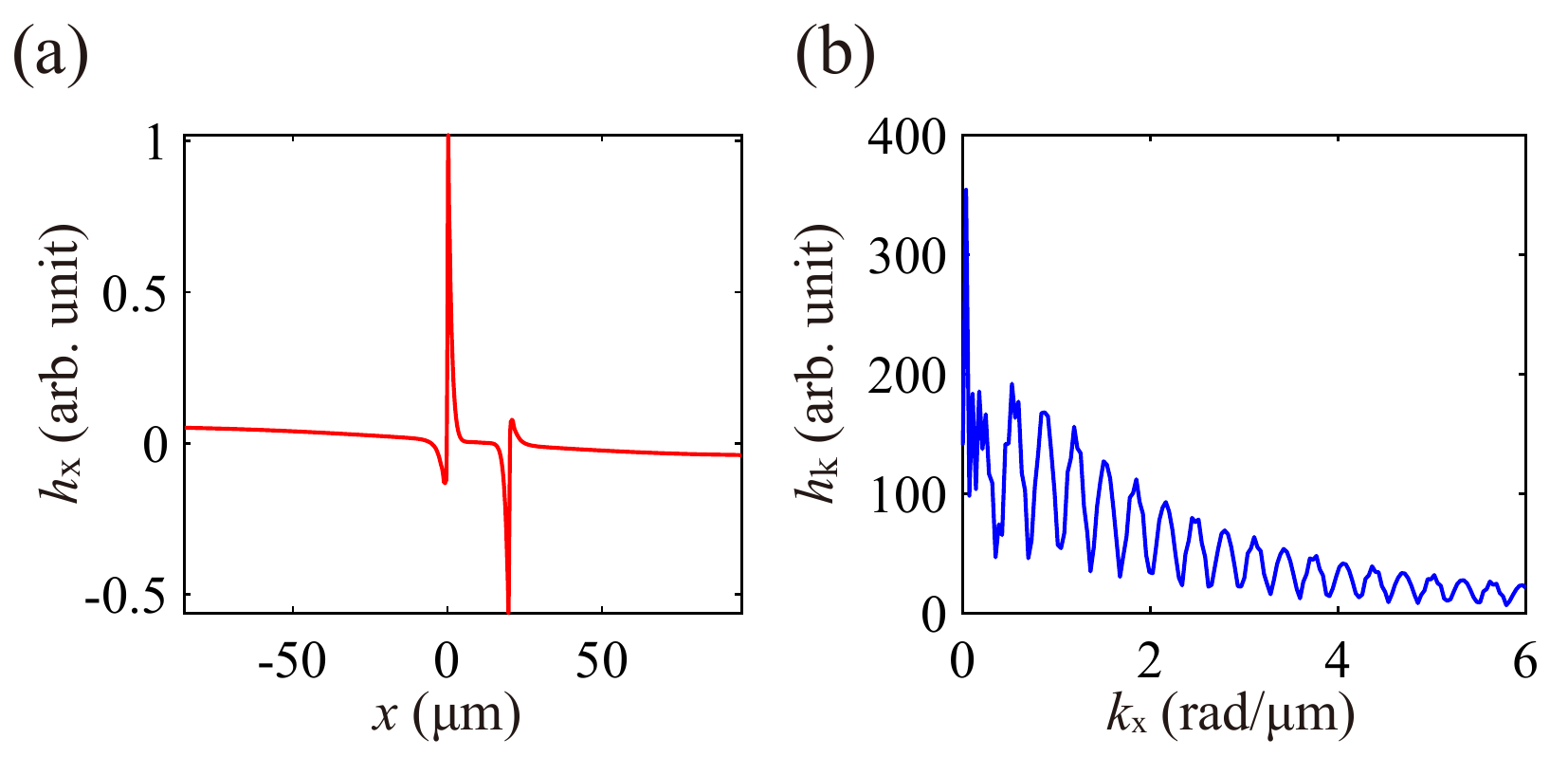}
    \caption{Numerical simulation results of microwave distribution from a stripline.
    (a) Distribution of the microwave amplitude in $x$-direction $h_{x}$ in the real space.
    (b) Distribution of the microwave amplitude in $x$-direction $h_{k}$ in the wavenumber space.
    }
    \label{fig:stripline_simulation}
\end{figure}

The amplitude of the spin waves excited by a stripline is proportional to the Fourier amplitude of the microwave field of the stripline at the wavenumber of the spin waves \cite{rezende2020fundamentals}.
In this section, we simulate the microwave distribution from the stripline to evaluate the wavenumber dependence of the spin-wave amplitude obtained in the wideband spin-wave sensing. For the simulation, we use three-dimensional electromagnetic simulation software CST MICROWAVE STUDIO$^{\textregistered}$.
For computational efficiency, we assume that the stripline consists only of $\mathrm{Cu} \, 200 \, \mathrm{nm}$, and the thickness of the YIG is set to $1 \, \mathrm{\mu m}$. Additionally, the length of the stripline is shortened to $400 \, \mathrm{\mu m}$ (actual length is $2 \, \mathrm{mm}$). 
\par
We simulate the distribution of microwave amplitude in the $x$-direction $h_{x}$ with the microwave frequency $f_{\mathrm{mw}} = 2165.8 \, \mathrm{MHz}$.
Figure~\ref{fig:stripline_simulation}(a) shows the simulation results of the microwave amplitude in real space and Fig.~\ref{fig:stripline_simulation}(b) presents the amplitude distribution in wavenumber space by performing a Fourier transformation of the distribution $h_{x}$ in Fig.~\ref{fig:stripline_simulation}(a).
From the result of the wavenumber dependence of the amplitude, it can be expected that the spin-wave amplitude decreases as the wavenumber increases and oscillates periodically.

\renewcommand{\thefigure}{I\arabic{figure}}
\renewcommand{\theequation}{I\arabic{equation}}
\setcounter{equation}{0}
\setcounter{figure}{0}

\section*{Appendix I: Evaluation of spin-wave amplitude in the wideband spin-wave sensing}
\label{app:sw_amplitude}
\begin{figure}[b]
    \centering
    \includegraphics[width=\linewidth]{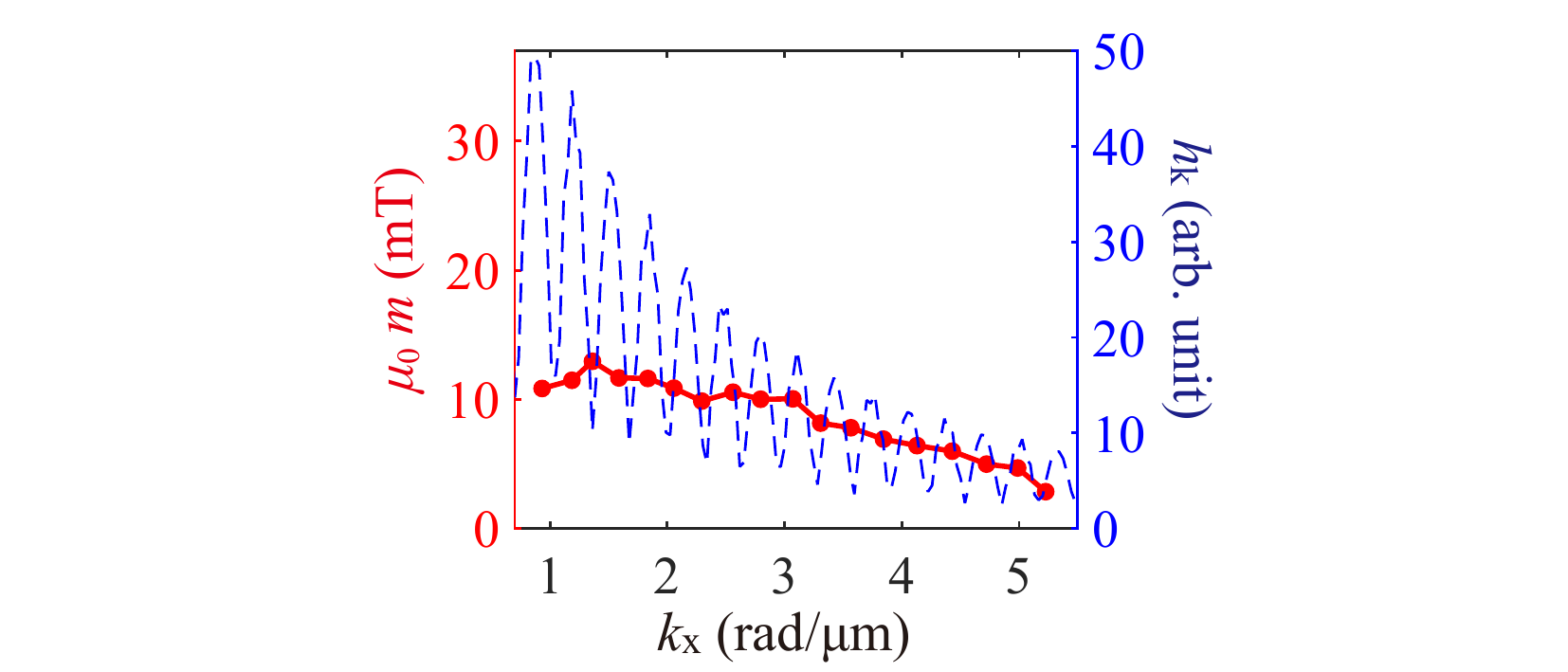}
    \caption{Wavenumber dependence of the spin-wave amplitude. (red) Experimentally obtained spin-wave amplitude multiplied by the permeability of vacuum $\mu_{0}$. (blue) Expected amplitude from the numerical simulation.
    }
    \label{fig:sw_amp_wide}
\end{figure}

This section compares the experimentally obtained spin-wave amplitudes in the wideband spin-wave sensing and the amplitudes expected from the numerical simulation results.
\par
Figure~\ref{fig:sw_amp_wide} shows the wavenumber dependence of the spin-wave amplitude.
The red points represent the wavenumber dependence of the spin-wave amplitude at the center position of the field of view ($x = 8.1 \, \mathrm{\mu m}$), estimated from the fitting results for the data in Fig.~\ref{fig:stripline_simulation}(d) in the main text.
The blue dotted line represents the Fourier amplitude of the stripline's microwave field in Fig.~\ref{fig:sw_amp_wide}(b), scaled by a constant factor.
While the experiment and simulation results show a similar trend in that they decrease as the wavenumber increases, the oscillations seen in the simulations are not visible in the experimental results.
Since the intervals of wavenumbers obtained in the experiment are close to those of the oscillations in the simulations, we conduct a measurement over finer frequency ranges for a more detailed comparison. 
\par
\begin{figure}[t]
    \centering
    \includegraphics[width=\linewidth]{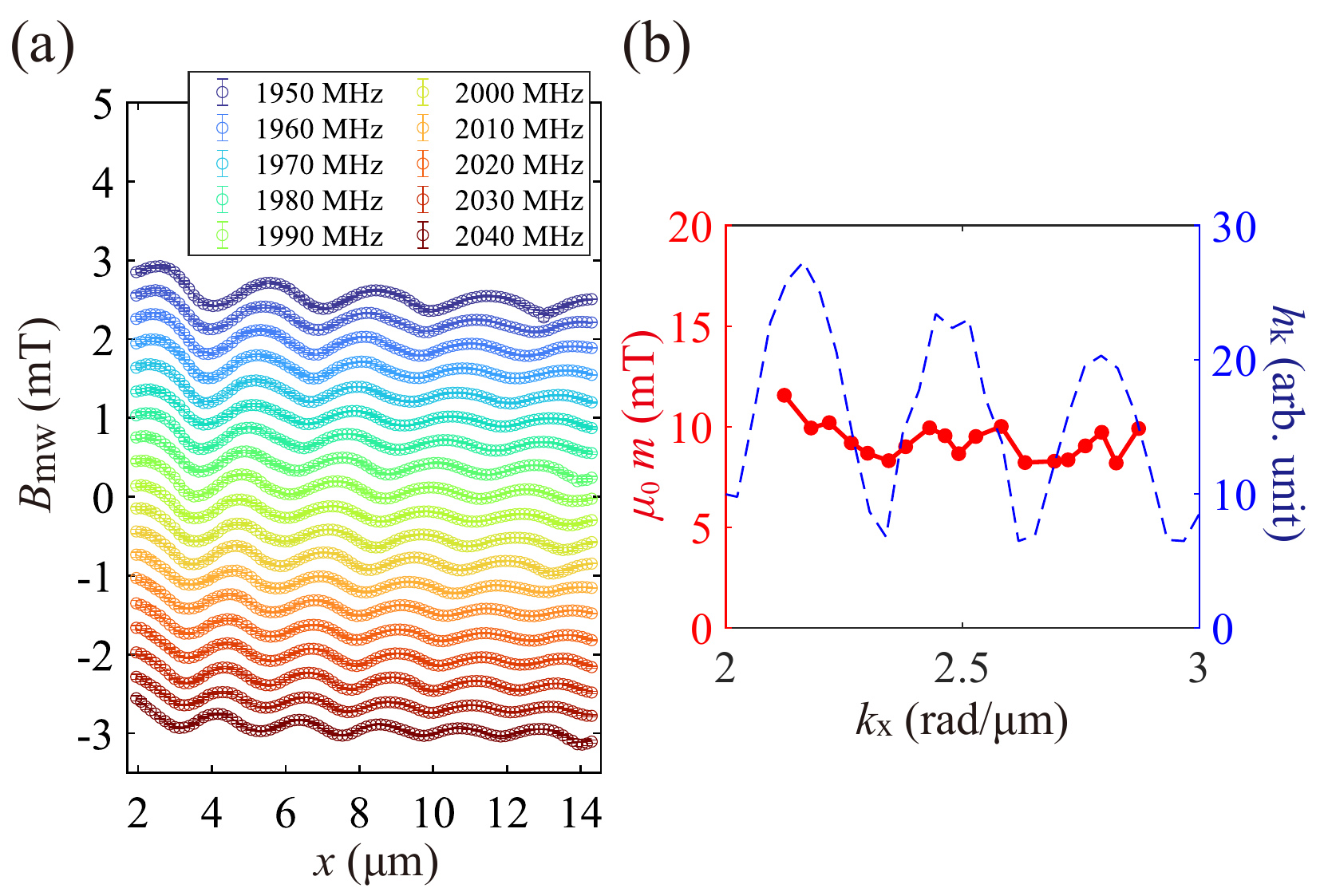}
    \caption{Wavenumber dependence of the spin-wave amplitude in a fine frequency range.
    (a) Microwave amplitude distributions across frequencies from $f_{\mathrm{sw}} = 1950 \, \mathrm{MHz}$ to $2040 \, \mathrm{MHz}$ at intervals of $5 \, \mathrm{MHz}$.
    (b) Wavenumber dependence of the spin-wave amplitude. (red) Experimentally obtained spin-wave amplitude multiplied by the permeability of vacuum $\mu_{0}$. (blue) Expected amplitude from the numerical simulation.
    }
    \label{fig:sw_amp_narrow}
\end{figure}

Figure~\ref{fig:sw_amp_narrow}(a) shows the microwave amplitude distributions obtained by sweeping the spin-wave frequency from $f_{\mathrm{sw}} = 1950 \, \mathrm{MHz}$ to $f_{\mathrm{sw}} = 2040 \, \mathrm{MHz}$ at intervals of $5 \, \mathrm{MHz}$. An offset of $0.3 \, \mathrm{mT}$ is added to each data.
In the figure, there seems to be no particular frequency where the oscillation amplitude of the microwave becomes significantly larger or smaller compared to other frequencies.
Figure~\ref{fig:sw_amp_narrow}(b) shows the wavenumber dependence of the spin-wave amplitude at the center of the field of view (red) and the expectation from the numerical simulation scaled by a constant factor (blue), similar to Fig.~\ref{fig:sw_amp_wide}. Here, too, no clear oscillations are observed.
One possible reason for the lack of oscillations in the spin-wave amplitude is that the decay length of the spin waves in the YIG sample used in the experiment is relatively short. The decay length of the spin waves in the wideband spin-wave sensing is typically estimated to be around $20 \, \mathrm{\mu m}$.
As indicated by Eq.~(\ref{eq:domega}), the decay of the spin waves is due to the linewidth of wavenumbers caused by damping. A decay length of $20 \, \mathrm{\mu m}$ corresponds to a linewidth $\Delta k = 0.05 \, \mathrm{rad}/\mathrm{\mu m}$ in wavenumber.
The stripline used in this study has a relatively wide width of $20 \, \mathrm{\mu m}$, resulting in small intervals of oscillations. Therefore, the spread of wavenumbers may average out the oscillations seen in the simulations, leading only to observable attenuation. For further investigation, experiments using a narrower stripline could be helpful.
Factors such as local inhomogeneities in the stripline could also contribute to the disappearance of sharp peaks predicted by simulations.

\renewcommand{\thefigure}{J\arabic{figure}}
\renewcommand{\theequation}{J\arabic{equation}}
\setcounter{equation}{0}
\setcounter{figure}{0}
\setcounter{subsection}{0}

\section*{Appendix J: Calculation of rigorous expression of the resonance frequency shift due to the AC Zeeman effect}
\label{app:AC_Zeeman_rigorous}
The resonance frequency shift caused by the AC Zeeman effect corresponds to a shift caused by off-resonant microwave irradiation in a rotating frame. 
In this section, we derive the expression of the shift in a situation where the detuning $\Delta$ is sufficiently large compared to the microwave amplitude $f_{\mathrm{rabi}}$.
In the derivation for the NV centers, previous studies Refs.~\cite{van2015nanometre} and \cite{li2019wideband} derived it by calculating an effective Hamiltonian using the Schrieffer-Wolff transformation. Here, we derive it based on the Floquet perturbation theory \cite{sambe1973steady, beloy2009theory}.
\par
\subsection{Floquet perturbation theory}
The time evolution of a quantum state $\ket{\psi(t)}$ is described by the Schr\"{o}dinger equation.
\begin{equation}
   \qty(\hat{H}(t) - i \hbar \frac{\partial }{\partial t}) \ket{\psi(t)} = 0,
   \label{schrodinger}
\end{equation}
where $\hat{H}(t)$ is the Hamiltonian of the system, which is assumed to have periodicity with a period $\tau$:
\begin{equation}
    \hat{H}(t+\tau) = \hat{H}(t).
\end{equation}
According to the Floquet theorem, the solution of the Schr\"{o}dinger equation can be expressed as
\begin{equation}
    \ket{\psi(t)} = e^{-\frac{i\mathcal{E}t}{\hbar}} \ket{\phi(t)},
\end{equation}
where $\mathcal{E}$ represents the quasi-energy and $\ket{\phi(t)}$ has the same periodicity as the Hamiltonian:
\begin{equation}
    \ket{\phi(t+\tau)} = \ket{\phi(t)}.
\end{equation}
By defining the Floquet Hamiltonian $\hat{\mathcal{H}}(t)$ as
\begin{equation}
    \hat{\mathcal{H}}(t) =  \hat{H}(t) - i \hbar \frac{\partial }{\partial t}, \\
\end{equation}
the time-dependent Schr\"{o}dinger equation can be expressed in the same form as the time-independent Schr\"{o}dinger equation:
\begin{equation}
    \hat{\mathcal{H}}(t) \ket{\phi(t)} = \mathcal{E} \ket{\phi(t)}.
\end{equation}
By applying a perturbation theory to this Floquet Hamiltonian, we calculate the resonance frequency shift due to an external oscillating field. \par
To apply a perturbation theory, we separate the Hamiltonian into the following unperturbed and perturbed terms:
\begin{equation}
    \hat{H}(t) = \hat{H}_{0} + V(t),
    \label{eq:total_floquet_hamiltonian}
\end{equation}
where $\hat{H}_{0}$ is the time-independent unperturbed Hamiltonian, and $V(t)$ is the perturbation term with a period $\tau$.
We assume that the eigenstates of the unperturbed Hamiltonian are labeled by a quantum number $n$:
\begin{equation}
    \hat{H}_{0} \ket{n} = E_{\mathrm{n}} \ket{n}.
\end{equation}
In this case, the eigenenergies and eigenstates of the unperturbed Floquet Hamiltonian 
\begin{equation}
    \hat{\mathcal{H}}_{0}(t) =  \hat{H}_{0}(t) - i \hbar \frac{\partial }{\partial t} \\
\end{equation}
can be expressed as
\begin{equation}
    \label{eq:up_floquet}
    \hat{\mathcal{H}_{0}}(t) |n,q \rangle \rangle = E_{n,q} |n,q \rangle \rangle, \\
\end{equation}
where 
\begin{equation}
    \begin{split}
        E_{n,q} &= E_{n} + q \omega, \\
        |n,q \rangle \rangle &= e^{i\omega q t} \ket{n}, \\
    \end{split}
\end{equation}
and $\omega = \frac{2\pi}{\tau}$. The state $|n,q \rangle \rangle$, where $q \in \mathbb{Z}$ represents the number of photons, form an extended Hilbert space. The inner product between two states $|u(t) \rangle \rangle$ and $|v(t) \rangle \rangle$ in this space is given by: 
\begin{equation}
    \langle \langle u(t) | v(t) \rangle \rangle = \frac{1}{\tau} \int_{-\frac{\tau}{2}}^{\frac{\tau}{2}} \langle \langle u(t)|v(t) \rangle \rangle dt.
\end{equation}
\par
We calculate the shift of the unperturbed eigenstate $|n,q \rangle \rangle$ and its eigenenergy $E_{n,q}$ due to the perturbation term. \\
We expand the eigenstates of the Hamiltonian $|\xi(t) \rangle \rangle$ with the unperturbed states as
\begin{equation}
    |\xi(t) \rangle \rangle = \sum_{m,p} c_{m,p} |m,p \rangle \rangle.
\end{equation}
The coefficients and energies can be expanded as
\begin{equation}
    \begin{split}
        c_{m,p} &= c^{(0)}_{m,p} \delta_{m,n} \delta_{p,q} +  c^{(1)}_{m,p} + \ldots, \\
        E &= E_{n,q} + E^{(1)} + E^{(2)} +  \ldots.
    \end{split}
\end{equation}
Since Eq.~(\ref{eq:up_floquet}) has the same form as the time-independent Schr\"{o}dinger equation, by using calculations similar to those in the time-independent perturbation theory, we can derive the perturbed energy up to the second order and coefficients to the first order as
\begin{equation}
    \label{eq:Ep}
    \begin{split}
        E^{(1)} &= \langle \langle n,q | V | n,q \rangle \rangle, \\
        E^{(2)} &= \sum_{m,p} {}^{\prime} \frac{\langle \langle n,q|V|m,p \rangle \rangle \langle \langle m,p|V|n,q \rangle \rangle}{E_{n,q} - E_{m,p}}, \\
    \end{split}
\end{equation}
and
\begin{equation}
    \begin{split}
        c_{m,p}^{(0)} &= 1, \\
        c_{m,p}^{(1)} &= \frac{\langle \langle m,p|V|n,q \rangle \rangle}{E_{n,q} - E_{m,p}} \, \qty((n,q) \neq (m,p)). \\
    \end{split}
\end{equation}
Here, $\sum_{m,p} {}^{\prime}$ means to sum over all $(m,p)$ except for $(n,q)$. 
\par
Next, we calculate the specific expression of the energy shift when the periodic perturbation term $V(t)$ can be decomposed as
\begin{equation}
    V(t) = V_{s} + V_{+}e^{i\omega t} + V_{-} e^{-i\omega t}.
    \label{eq:v}
\end{equation}
By substituting into Eq.~(\ref{eq:Ep}), the energy shift up to the second order can be calculated as
\begin{equation}
    \label{eq:eshift}
    \begin{split}
        E^{(1)} &= \braket{n| V_{s} | n}, \\
        E^{(2)} &= \sum_{m} {}^{\prime} \bigg( \frac{|\braket{n|V_{s}|m}|^{2}}{E_{n} - E_{m}} + \frac{\braket{n|V_{+}|m} \braket{m|V_{-}|n}}{E_{n} - E_{m} + \omega} \\
         &\hspace{2cm} + \frac{\braket{n|V_{-}|m} \braket{m|V_{+}|n}}{E_{n} - E_{m} - \omega}\bigg).
    \end{split}
\end{equation}
What is interesting here is that the energy shift does not depend on the number of photons $q$.
Therefore, the energy shift for the state $|n, q \rangle \rangle$ can be regarded as a steady energy shift of the time-independent eigenstate $\ket{n}$.

\subsection{Calculation of AC Zeeman shift of NV center under microwave irradiation}
We consider the following Hamiltonian of an NV spin with an external magnetic field and microwave field:
\begin{equation}
    \frac{\hat{H}}{\hbar} = D \hat{S^{2}_{z}} + \gamma^{\prime}_{e} B_{z} \hat{S_{z}} + \gamma^{\prime}_{e} (B_{-} e^{-i \omega t} + B_{+} e^{i \omega t}) \hat{S_{x}}. \\
\end{equation}
Hereafter, we will omit the Planck constant $\hbar$ for convenience.
By transitioning to the rotating frame using the unitary transformation
\begin{equation}
    \hat{U} = \exp(i \omega \hat{S_{z}}t),
\end{equation}
the effective Hamiltonian in the rotating frame $\hat{H}^{\prime}$ can be expressed as
\begin{equation}
    \begin{split}
        \hat{H}^{\prime} &= \hat{U} \hat{H} {\hat{U}}^{\dag} - i \hat{U} {\dot{U}}^{\dag} \\
                    &= \mqty(D + \gamma^{\prime}_{e} B_{z} + \omega & \Omega^{\prime}_{\mathrm{mw}} & 0 \\
                    \overline{\Omega^{\prime}_{\mathrm{mw}}} & 0 &  \Omega^{\prime}_{\mathrm{mw}} \\
                      0 & \overline{\Omega^{\prime}_{\mathrm{mw}}} & D - \gamma^{\prime}_{e} B_{z} - \omega), \\
    \end{split}
\end{equation}
where $\Omega^{\prime}_{\mathrm{mw}} = \frac{\gamma^{\prime}_{e}}{\sqrt{2}} (B_{-} + B_{+} e^{2 i \omega t})$. \\
With the angular frequency $2\omega$, this Hamiltonian can be related to the terms in Eqs.~(\ref{eq:total_floquet_hamiltonian}) and (\ref{eq:v}) as follows:
\begin{equation}
    \begin{split}
        \hat{H_{0}} &= \mqty(D + \gamma^{\prime}_{e} B_{z} + \omega & 0 & 0 \\
                           0 & 0 & 0 \\
                           0 & 0  & D - \gamma^{\prime}_{e} B_{z} - \omega), \\
        V_{s} &= \mqty(0 & \frac{\gamma^{\prime}_{e}B_{-}}{\sqrt{2}} & 0 \\
                 \frac{\gamma^{\prime}_{e}B_{-}}{\sqrt{2}} & 0 & \frac{\gamma^{\prime}_{e}B_{-}}{\sqrt{2}}  \\
                 0 & \frac{\gamma^{\prime}_{e}B_{-}}{\sqrt{2}} & 0), \\
        V_{+} &= \mqty(0 & \frac{\gamma^{\prime}_{e}B_{+}}{\sqrt{2}} & 0 \\
                 0 & 0 & \frac{\gamma^{\prime}_{e}B_{+}}{\sqrt{2}}  \\
                 0 & 0 & 0), \\
        V_{-} &= \mqty(0 & 0 & 0 \\
                 \frac{\gamma^{\prime}_{e}B_{+}}{\sqrt{2}} & 0 & 0  \\
                0 & \frac{\gamma^{\prime}_{e}B_{+}}{\sqrt{2}} & 0). \\
    \end{split}
\end{equation}
The eigenstates of the unperturbed Hamiltonian are the states obtained by a rotational transformation of the eigenstates of $\hat{S_{z}}$, labeled by $\ket{m_{\mathrm{S}}}$ (where $m_{\mathrm{S}} = 0, \, \pm \, 1$). We denote the eigenangular frequencies of the unperturbed Hamiltonian as $\omega_{m_{\mathrm{S}}}$.
The eigenenergy up to second-order terms for each state can be calculated using Eq.~($\ref{eq:eshift}$) as
\begin{equation}
    \label{eq:Ep_NV}
    \begin{split}
        E_{0} &= -\frac{(\gamma^{\prime}_{e} B_{-})^{2}}{2(\omega_{-} - \omega)} - \frac{(\gamma^{\prime}_{e} B_{-})^{2}}{2(\omega_{+} + \omega)} - \frac{(\gamma^{\prime}_{e} B_{+})^{2}}{2(\omega_{-} + \omega)} - \frac{(\gamma^{\prime}_{e} B_{+})^{2}}{2(\omega_{+} - \omega)},  \\
        E_{-} &= D - \gamma^{\prime}_{e} B_{z} - \omega + \frac{(\gamma^{\prime}_{e} B_{-})^{2}}{2(\omega_{-} - \omega)} + \frac{(\gamma^{\prime}_{e} B_{+})^{2}}{2(\omega_{-} + \omega)}, \\
        E_{+} &= D + \gamma^{\prime}_{e} B_{z} + \omega + \frac{(\gamma^{\prime}_{e} B_{-})^{2}}{2(\omega_{+} + \omega)} + \frac{(\gamma^{\prime}_{e} B_{+})^{2}}{2(\omega_{+} - \omega)}. 
    \end{split}
\end{equation}
In this study, we use the CP-2 sequence to detect the energy shift between $E_{0}$ and $E_{-}$ during off-resonant microwave irradiation.
We denote the corresponding angular frequency $\omega_{\mathrm{ACZ}}$ and it can be calculated by Eq.~(\ref{eq:Ep_NV}) as
\begin{equation}
    \omega_{\mathrm{ACZ}} = \frac{(\gamma^{\prime}_{e} B_{-})^{2}}{\omega_{-} - \omega} + \frac{(\gamma^{\prime}_{e} B_{-})^{2}}{2(\omega_{+} + \omega)} + \frac{(\gamma^{\prime}_{e} B_{+})^{2}}{\omega_{-} + \omega} + \frac{(\gamma^{\prime}_{e} B_{+})^{2}}{2(\omega_{+} - \omega)},
    \label{eq:ac_zeeman_all_supp}
\end{equation}
which is consistent with Refs.~\cite{van2015nanometre} and \cite{li2019wideband}.
Regarding terms after the second item in Eq.~(\ref{eq:ac_zeeman_all_supp}), those containing $\omega_{+}$ derive from the effect of $m_{\mathrm{S}} = +1$ state, and terms containing $B_{+}$ from the breakdown of the rotating wave approximation (RWA). In particular, the effects of these terms become pronounced when the detuning becomes comparable to the resonance frequencies (around zero-field splitting when the magnetic field is not large).
In the analysis of the experiments in the main text, the detuning is up to approximately $500 \, \mathrm{MHz}$, and since the contribution from the terms after the second term in Eq.~(\ref{eq:ac_zeeman_all_supp}) is small, we consider only the first term.

\subsection{Approximate expression of the resonance frequency shift by the AC Zeeman effect for a very large detuning}

\begin{figure}[b]
    \centering
    \includegraphics[width=\linewidth]{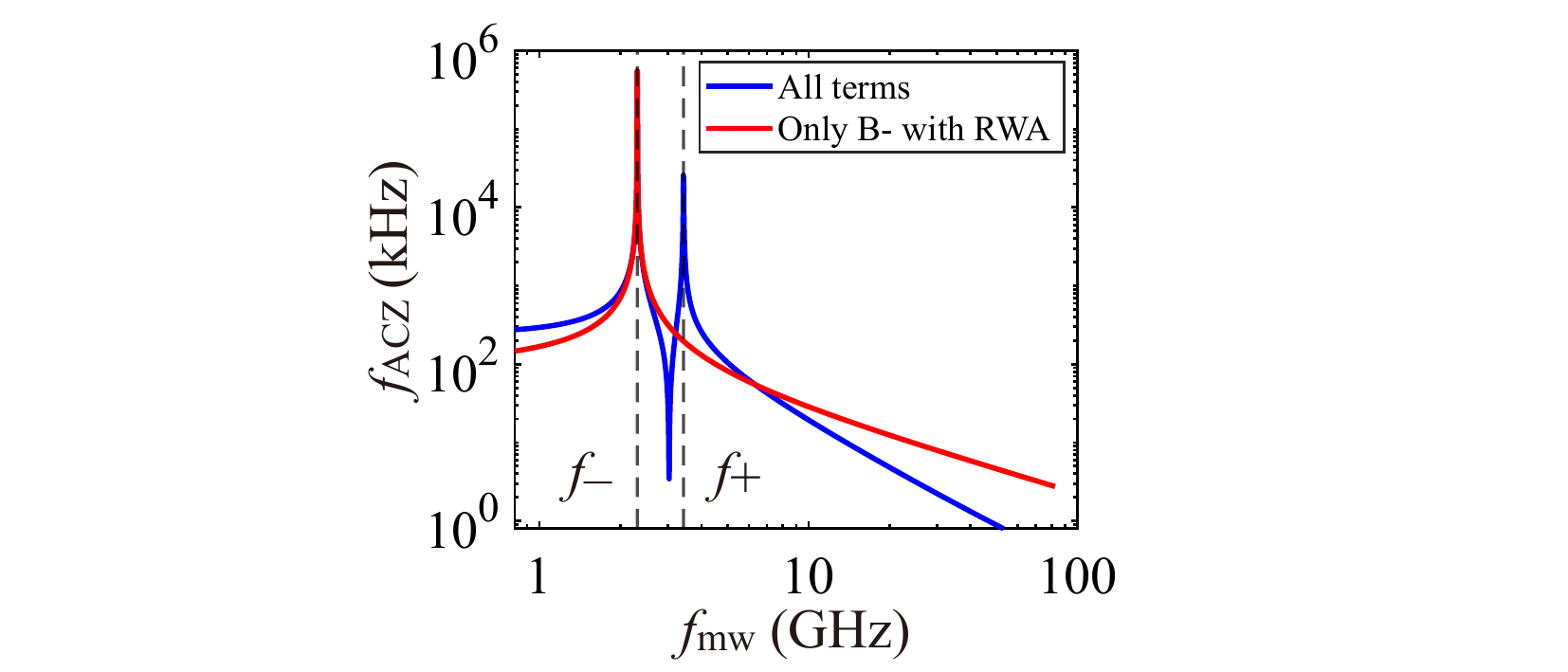}
    \caption{Numerical simulation results of the microwave frequency dependence of the frequency shift due to the AC Zeeman effect ($B_{-} = 0.53 \, \mathrm{mT}$, $B_{+} = 0.51 \, \mathrm{mT}$). (red) The results of considering only the first term in Eq.~(\ref{eq:ac_zeeman_all_supp}). (blue) Results of taking into account all terms in Eq.~(\ref{eq:ac_zeeman_all_supp}).}
    \label{fig:ac_zeeman_rigorous}
\end{figure}

Figure~\ref{fig:ac_zeeman_rigorous} shows the dependence of the resonance frequency shift by the AC Zeeman effect $f_{\mathrm{ACZ}} \, (= \, \frac{\omega_{\mathrm{ACZ}}}{2\pi})$ on the microwave frequency $f_{\mathrm{mw}}$. Here, the external magnetic field $B_{z}$ is set to $B_{z} = 20 \, \mathrm{mT}$, and the microwave amplitudes are set to $B_{-} = 0.53 \, \mathrm{mT}$ and $B_{+} = 0.51 \, \mathrm{mT}$, respectively.
The red line is the result of considering only the first term in Eq.~(\ref{eq:ac_zeeman_all_supp}), and the blue line represents considering all terms. $f_{\pm}$ correspond to the resonance frequencies between the $m_{\mathrm{S}} = 0$ and $m_{\mathrm{S}} = \pm 1$ states of the NV spin respectively.
For small detunings ($\Delta = |f_{\mathrm{mw}}-f_{\mathrm{-}}| < 500 \, \mathrm{MHz}$), the approximation considering only the first term is valid. However, as the microwave frequency increases further, careful analysis becomes necessary because singularities occur when the microwave frequency approaches the frequency between $m_{\mathrm{s}} = 0$ and $m_{\mathrm{s}} = +1$ at first, and all terms need to be considered for even larger detunings. \par
In particular, when the detuning is sufficiently large compared to the energy scales between the levels of the NV center ($\omega \gg \omega_{\pm}$), Eq.~(\ref{eq:ac_zeeman_all_supp}) can be approximated as
\begin{equation}
    \omega_{\mathrm{ACZ}} \simeq -\frac{\gamma^{\prime 2}_{e}}{2 \omega} (B_{-}^{2} - B_{+}^{2}).
    \label{eq:ac_zeeman_approx}
\end{equation}
In surface spin-wave interferometry measurements, the reference microwave can be assumed to be linearly polarized, while the microwave from the spin waves is circularly polarized. When the reference microwave amplitude $B_{\mathrm{ref}}$ is sufficiently larger than the microwave amplitude from the spin waves $B_{\mathrm{sw}}$, the microwave amplitudes of each polarization can be expressed as
\begin{equation}
    \label{eq:Bpm}
    \begin{split}
        B_{-} &= B_{\mathrm{ref}} + B_{\mathrm{sw}},  \\
        B_{+} &= B_{\mathrm{ref}}. \\
    \end{split}
\end{equation}
By substituting Eq.~(\ref{eq:Bpm}) into Eq.~(\ref{eq:ac_zeeman_approx}), we can obtain the approximated expression of the AC Zeeman shift as 
\begin{equation}
    \omega_{\mathrm{ACZ}} \simeq -\frac{\gamma^{\prime 2}_{e}}{\omega} B_{\mathrm{ref}} B_{\mathrm{sw}}.
\end{equation}
Therefore, it is possible to derive the microwave amplitude from the spin waves by calibrating the reference microwave amplitude $B_{\mathrm{ref}}$ in advance.

\renewcommand{\thefigure}{K\arabic{figure}}
\renewcommand{\theequation}{K\arabic{equation}}
\setcounter{equation}{0}
\setcounter{figure}{0}


\bibliography{citation}

\begin{thebibliography}{55}%
\makeatletter
\providecommand \@ifxundefined [1]{%
 \@ifx{#1\undefined}
}%
\providecommand \@ifnum [1]{%
 \ifnum #1\expandafter \@firstoftwo
 \else \expandafter \@secondoftwo
 \fi
}%
\providecommand \@ifx [1]{%
 \ifx #1\expandafter \@firstoftwo
 \else \expandafter \@secondoftwo
 \fi
}%
\providecommand \natexlab [1]{#1}%
\providecommand \enquote  [1]{``#1''}%
\providecommand \bibnamefont  [1]{#1}%
\providecommand \bibfnamefont [1]{#1}%
\providecommand \citenamefont [1]{#1}%
\providecommand \href@noop [0]{\@secondoftwo}%
\providecommand \href [0]{\begingroup \@sanitize@url \@href}%
\providecommand \@href[1]{\@@startlink{#1}\@@href}%
\providecommand \@@href[1]{\endgroup#1\@@endlink}%
\providecommand \@sanitize@url [0]{\catcode `\\12\catcode `\$12\catcode `\&12\catcode `\#12\catcode `\^12\catcode `\_12\catcode `\%12\relax}%
\providecommand \@@startlink[1]{}%
\providecommand \@@endlink[0]{}%
\providecommand \url  [0]{\begingroup\@sanitize@url \@url }%
\providecommand \@url [1]{\endgroup\@href {#1}{\urlprefix }}%
\providecommand \urlprefix  [0]{URL }%
\providecommand \Eprint [0]{\href }%
\providecommand \doibase [0]{https://doi.org/}%
\providecommand \selectlanguage [0]{\@gobble}%
\providecommand \bibinfo  [0]{\@secondoftwo}%
\providecommand \bibfield  [0]{\@secondoftwo}%
\providecommand \translation [1]{[#1]}%
\providecommand \BibitemOpen [0]{}%
\providecommand \bibitemStop [0]{}%
\providecommand \bibitemNoStop [0]{.\EOS\space}%
\providecommand \EOS [0]{\spacefactor3000\relax}%
\providecommand \BibitemShut  [1]{\csname bibitem#1\endcsname}%
\let\auto@bib@innerbib\@empty
\bibitem [{\citenamefont {Pirro}\ \emph {et~al.}(2021)\citenamefont {Pirro}, \citenamefont {Vasyuchka}, \citenamefont {Serga},\ and\ \citenamefont {Hillebrands}}]{Pirro2021}%
  \BibitemOpen
  \bibfield  {author} {\bibinfo {author} {\bibfnamefont {P.}~\bibnamefont {Pirro}}, \bibinfo {author} {\bibfnamefont {V.~I.}\ \bibnamefont {Vasyuchka}}, \bibinfo {author} {\bibfnamefont {A.~A.}\ \bibnamefont {Serga}},\ and\ \bibinfo {author} {\bibfnamefont {B.}~\bibnamefont {Hillebrands}},\ }\href {https://doi.org/10.1038/s41578-021-00332-w} {\bibfield  {journal} {\bibinfo  {journal} {Nat. Rev. Mater.}\ }\textbf {\bibinfo {volume} {6}},\ \bibinfo {pages} {1114} (\bibinfo {year} {2021})}\BibitemShut {NoStop}%
\bibitem [{\citenamefont {Bryant}\ \emph {et~al.}(1988)\citenamefont {Bryant}, \citenamefont {Jeffries},\ and\ \citenamefont {Nakamura}}]{bryant1988spin}%
  \BibitemOpen
  \bibfield  {author} {\bibinfo {author} {\bibfnamefont {P.}~\bibnamefont {Bryant}}, \bibinfo {author} {\bibfnamefont {C.}~\bibnamefont {Jeffries}},\ and\ \bibinfo {author} {\bibfnamefont {K.}~\bibnamefont {Nakamura}},\ }\href {https://doi.org/10.1103/PhysRevLett.60.1185} {\bibfield  {journal} {\bibinfo  {journal} {Phys. Rev. Lett.}\ }\textbf {\bibinfo {volume} {60}},\ \bibinfo {pages} {1185} (\bibinfo {year} {1988})}\BibitemShut {NoStop}%
\bibitem [{\citenamefont {Sulymenko}\ \emph {et~al.}(2018)\citenamefont {Sulymenko}, \citenamefont {Prokopenko}, \citenamefont {Tyberkevych}, \citenamefont {Slavin},\ and\ \citenamefont {Serga}}]{Sulymenko2018}%
  \BibitemOpen
  \bibfield  {author} {\bibinfo {author} {\bibfnamefont {O.~R.}\ \bibnamefont {Sulymenko}}, \bibinfo {author} {\bibfnamefont {O.~V.}\ \bibnamefont {Prokopenko}}, \bibinfo {author} {\bibfnamefont {V.~S.}\ \bibnamefont {Tyberkevych}}, \bibinfo {author} {\bibfnamefont {A.~N.}\ \bibnamefont {Slavin}},\ and\ \bibinfo {author} {\bibfnamefont {A.~A.}\ \bibnamefont {Serga}},\ }\href {https://doi.org/10.1063/1.5041426} {\bibfield  {journal} {\bibinfo  {journal} {Low Temp. Phys.}\ }\textbf {\bibinfo {volume} {44}},\ \bibinfo {pages} {775} (\bibinfo {year} {2018})}\BibitemShut {NoStop}%
\bibitem [{\citenamefont {Demokritov}\ \emph {et~al.}(2006)\citenamefont {Demokritov}, \citenamefont {Demidov}, \citenamefont {Dzyapko}, \citenamefont {Melkov}, \citenamefont {Serga}, \citenamefont {Hillebrands},\ and\ \citenamefont {Slavin}}]{Demokritov2006}%
  \BibitemOpen
  \bibfield  {author} {\bibinfo {author} {\bibfnamefont {S.~O.}\ \bibnamefont {Demokritov}}, \bibinfo {author} {\bibfnamefont {V.~E.}\ \bibnamefont {Demidov}}, \bibinfo {author} {\bibfnamefont {O.}~\bibnamefont {Dzyapko}}, \bibinfo {author} {\bibfnamefont {G.~A.}\ \bibnamefont {Melkov}}, \bibinfo {author} {\bibfnamefont {A.~A.}\ \bibnamefont {Serga}}, \bibinfo {author} {\bibfnamefont {B.}~\bibnamefont {Hillebrands}},\ and\ \bibinfo {author} {\bibfnamefont {A.~N.}\ \bibnamefont {Slavin}},\ }\href {https://doi.org/10.1038/nature05117} {\bibfield  {journal} {\bibinfo  {journal} {Nature}\ }\textbf {\bibinfo {volume} {443}},\ \bibinfo {pages} {430} (\bibinfo {year} {2006})}\BibitemShut {NoStop}%
\bibitem [{\citenamefont {Flebus}\ \emph {et~al.}(2024)\citenamefont {Flebus}, \citenamefont {Grundler}, \citenamefont {Rana}, \citenamefont {Otani}, \citenamefont {Barsukov}, \citenamefont {Barman}, \citenamefont {Gubbiotti}, \citenamefont {Landeros}, \citenamefont {Akerman}, \citenamefont {Ebels}, \citenamefont {Pirro}, \citenamefont {Demidov}, \citenamefont {Schultheiss}, \citenamefont {Csaba}, \citenamefont {Wang}, \citenamefont {Ciubotaru}, \citenamefont {Nikonov}, \citenamefont {Che}, \citenamefont {Hertel}, \citenamefont {Ono}, \citenamefont {Afanasiev}, \citenamefont {Mentink}, \citenamefont {Rasing}, \citenamefont {Hillebrands}, \citenamefont {Kusminskiy}, \citenamefont {Zhang}, \citenamefont {Du}, \citenamefont {Finco}, \citenamefont {van~der Sar}, \citenamefont {Luo}, \citenamefont {Shiota}, \citenamefont {Sklenar}, \citenamefont {Yu},\ and\ \citenamefont {Rao}}]{Flebus2024}%
  \BibitemOpen
  \bibfield  {author} {\bibinfo {author} {\bibfnamefont {B.}~\bibnamefont {Flebus}}, \bibinfo {author} {\bibfnamefont {D.}~\bibnamefont {Grundler}}, \bibinfo {author} {\bibfnamefont {B.}~\bibnamefont {Rana}}, \bibinfo {author} {\bibfnamefont {Y.~C.}\ \bibnamefont {Otani}}, \bibinfo {author} {\bibfnamefont {I.}~\bibnamefont {Barsukov}}, \bibinfo {author} {\bibfnamefont {A.}~\bibnamefont {Barman}}, \bibinfo {author} {\bibfnamefont {G.}~\bibnamefont {Gubbiotti}}, \bibinfo {author} {\bibfnamefont {P.}~\bibnamefont {Landeros}}, \bibinfo {author} {\bibfnamefont {J.}~\bibnamefont {Akerman}}, \bibinfo {author} {\bibfnamefont {U.}~\bibnamefont {Ebels}}, \bibinfo {author} {\bibfnamefont {P.}~\bibnamefont {Pirro}}, \bibinfo {author} {\bibfnamefont {V.~E.}\ \bibnamefont {Demidov}}, \bibinfo {author} {\bibfnamefont {K.}~\bibnamefont {Schultheiss}}, \bibinfo {author} {\bibfnamefont {G.}~\bibnamefont {Csaba}}, \bibinfo {author} {\bibfnamefont {Q.}~\bibnamefont {Wang}}, \bibinfo {author} {\bibfnamefont {F.}~\bibnamefont
  {Ciubotaru}}, \bibinfo {author} {\bibfnamefont {D.~E.}\ \bibnamefont {Nikonov}}, \bibinfo {author} {\bibfnamefont {P.}~\bibnamefont {Che}}, \bibinfo {author} {\bibfnamefont {R.}~\bibnamefont {Hertel}}, \bibinfo {author} {\bibfnamefont {T.}~\bibnamefont {Ono}}, \bibinfo {author} {\bibfnamefont {D.}~\bibnamefont {Afanasiev}}, \bibinfo {author} {\bibfnamefont {J.}~\bibnamefont {Mentink}}, \bibinfo {author} {\bibfnamefont {T.}~\bibnamefont {Rasing}}, \bibinfo {author} {\bibfnamefont {B.}~\bibnamefont {Hillebrands}}, \bibinfo {author} {\bibfnamefont {S.~V.}\ \bibnamefont {Kusminskiy}}, \bibinfo {author} {\bibfnamefont {W.}~\bibnamefont {Zhang}}, \bibinfo {author} {\bibfnamefont {C.~R.}\ \bibnamefont {Du}}, \bibinfo {author} {\bibfnamefont {A.}~\bibnamefont {Finco}}, \bibinfo {author} {\bibfnamefont {T.}~\bibnamefont {van~der Sar}}, \bibinfo {author} {\bibfnamefont {Y.~K.}\ \bibnamefont {Luo}}, \bibinfo {author} {\bibfnamefont {Y.}~\bibnamefont {Shiota}}, \bibinfo {author} {\bibfnamefont {J.}~\bibnamefont
  {Sklenar}}, \bibinfo {author} {\bibfnamefont {T.}~\bibnamefont {Yu}},\ and\ \bibinfo {author} {\bibfnamefont {J.}~\bibnamefont {Rao}},\ }\href {https://doi.org/10.1088/1361-648X/ad399c} {\bibfield  {journal} {\bibinfo  {journal} {J. Condens. Matter Phys.}\ }\textbf {\bibinfo {volume} {36}},\ \bibinfo {pages} {36350} (\bibinfo {year} {2024})}\BibitemShut {NoStop}%
\bibitem [{\citenamefont {Doherty}\ \emph {et~al.}(2013)\citenamefont {Doherty}, \citenamefont {Manson}, \citenamefont {Delaney}, \citenamefont {Jelezko}, \citenamefont {Wrachtrup},\ and\ \citenamefont {Hollenberg}}]{doherty2013nitrogen}%
  \BibitemOpen
  \bibfield  {author} {\bibinfo {author} {\bibfnamefont {M.~W.}\ \bibnamefont {Doherty}}, \bibinfo {author} {\bibfnamefont {N.~B.}\ \bibnamefont {Manson}}, \bibinfo {author} {\bibfnamefont {P.}~\bibnamefont {Delaney}}, \bibinfo {author} {\bibfnamefont {F.}~\bibnamefont {Jelezko}}, \bibinfo {author} {\bibfnamefont {J.}~\bibnamefont {Wrachtrup}},\ and\ \bibinfo {author} {\bibfnamefont {L.~C.}\ \bibnamefont {Hollenberg}},\ }\href {https://doi.org/https://doi.org/10.1016/j.physrep.2013.02.001} {\bibfield  {journal} {\bibinfo  {journal} {Phys. Rep.}\ }\textbf {\bibinfo {volume} {528}},\ \bibinfo {pages} {1} (\bibinfo {year} {2013})}\BibitemShut {NoStop}%
\bibitem [{\citenamefont {Wang}\ \emph {et~al.}(2015)\citenamefont {Wang}, \citenamefont {Yuan}, \citenamefont {Huang}, \citenamefont {Rong}, \citenamefont {Wang}, \citenamefont {Xu}, \citenamefont {Duan}, \citenamefont {Ju}, \citenamefont {Shi},\ and\ \citenamefont {Du}}]{Wang2015}%
  \BibitemOpen
  \bibfield  {author} {\bibinfo {author} {\bibfnamefont {P.}~\bibnamefont {Wang}}, \bibinfo {author} {\bibfnamefont {Z.}~\bibnamefont {Yuan}}, \bibinfo {author} {\bibfnamefont {P.}~\bibnamefont {Huang}}, \bibinfo {author} {\bibfnamefont {X.}~\bibnamefont {Rong}}, \bibinfo {author} {\bibfnamefont {M.}~\bibnamefont {Wang}}, \bibinfo {author} {\bibfnamefont {X.}~\bibnamefont {Xu}}, \bibinfo {author} {\bibfnamefont {C.}~\bibnamefont {Duan}}, \bibinfo {author} {\bibfnamefont {C.}~\bibnamefont {Ju}}, \bibinfo {author} {\bibfnamefont {F.}~\bibnamefont {Shi}},\ and\ \bibinfo {author} {\bibfnamefont {J.}~\bibnamefont {Du}},\ }\href {https://doi.org/10.1038/ncomms7631} {\bibfield  {journal} {\bibinfo  {journal} {Nat. Commun.}\ }\textbf {\bibinfo {volume} {6}},\ \bibinfo {pages} {6631} (\bibinfo {year} {2015})}\BibitemShut {NoStop}%
\bibitem [{\citenamefont {Horsley}\ \emph {et~al.}(2018)\citenamefont {Horsley}, \citenamefont {Appel}, \citenamefont {Wolters}, \citenamefont {Achard}, \citenamefont {Tallaire}, \citenamefont {Maletinsky},\ and\ \citenamefont {Treutlein}}]{Horsley2018}%
  \BibitemOpen
  \bibfield  {author} {\bibinfo {author} {\bibfnamefont {A.}~\bibnamefont {Horsley}}, \bibinfo {author} {\bibfnamefont {P.}~\bibnamefont {Appel}}, \bibinfo {author} {\bibfnamefont {J.}~\bibnamefont {Wolters}}, \bibinfo {author} {\bibfnamefont {J.}~\bibnamefont {Achard}}, \bibinfo {author} {\bibfnamefont {A.}~\bibnamefont {Tallaire}}, \bibinfo {author} {\bibfnamefont {P.}~\bibnamefont {Maletinsky}},\ and\ \bibinfo {author} {\bibfnamefont {P.}~\bibnamefont {Treutlein}},\ }\href {https://doi.org/10.1103/physrevapplied.10.044039} {\bibfield  {journal} {\bibinfo  {journal} {Phys. Rev. Appl.}\ }\textbf {\bibinfo {volume} {10}},\ \bibinfo {pages} {044039} (\bibinfo {year} {2018})}\BibitemShut {NoStop}%
\bibitem [{\citenamefont {Van~der Sar}\ \emph {et~al.}(2015)\citenamefont {Van~der Sar}, \citenamefont {Casola}, \citenamefont {Walsworth},\ and\ \citenamefont {Yacoby}}]{van2015nanometre}%
  \BibitemOpen
  \bibfield  {author} {\bibinfo {author} {\bibfnamefont {T.}~\bibnamefont {Van~der Sar}}, \bibinfo {author} {\bibfnamefont {F.}~\bibnamefont {Casola}}, \bibinfo {author} {\bibfnamefont {R.}~\bibnamefont {Walsworth}},\ and\ \bibinfo {author} {\bibfnamefont {A.}~\bibnamefont {Yacoby}},\ }\href {https://doi.org/10.1038/ncomms8886} {\bibfield  {journal} {\bibinfo  {journal} {Nat. Commun.}\ }\textbf {\bibinfo {volume} {6}},\ \bibinfo {pages} {7886} (\bibinfo {year} {2015})}\BibitemShut {NoStop}%
\bibitem [{\citenamefont {Andrich}\ \emph {et~al.}(2017)\citenamefont {Andrich}, \citenamefont {de~las Casas}, \citenamefont {Liu}, \citenamefont {Bretscher}, \citenamefont {Berman}, \citenamefont {Heremans}, \citenamefont {Nealey},\ and\ \citenamefont {Awschalom}}]{Andrich2017}%
  \BibitemOpen
  \bibfield  {author} {\bibinfo {author} {\bibfnamefont {P.}~\bibnamefont {Andrich}}, \bibinfo {author} {\bibfnamefont {C.~F.}\ \bibnamefont {de~las Casas}}, \bibinfo {author} {\bibfnamefont {X.}~\bibnamefont {Liu}}, \bibinfo {author} {\bibfnamefont {H.~L.}\ \bibnamefont {Bretscher}}, \bibinfo {author} {\bibfnamefont {J.~R.}\ \bibnamefont {Berman}}, \bibinfo {author} {\bibfnamefont {F.~J.}\ \bibnamefont {Heremans}}, \bibinfo {author} {\bibfnamefont {P.~F.}\ \bibnamefont {Nealey}},\ and\ \bibinfo {author} {\bibfnamefont {D.~D.}\ \bibnamefont {Awschalom}},\ }\href {https://doi.org/10.1038/s41534-017-0029-z} {\bibfield  {journal} {\bibinfo  {journal} {npj Quantum Inf.}\ }\textbf {\bibinfo {volume} {3}},\ \bibinfo {pages} {28} (\bibinfo {year} {2017})}\BibitemShut {NoStop}%
\bibitem [{\citenamefont {Kikuchi}\ \emph {et~al.}(2017)\citenamefont {Kikuchi}, \citenamefont {Prananto}, \citenamefont {Hayashi}, \citenamefont {Laraoui}, \citenamefont {Mizuochi}, \citenamefont {Hatano}, \citenamefont {Saitoh}, \citenamefont {Kim}, \citenamefont {Meriles},\ and\ \citenamefont {An}}]{kikuchi2017long}%
  \BibitemOpen
  \bibfield  {author} {\bibinfo {author} {\bibfnamefont {D.}~\bibnamefont {Kikuchi}}, \bibinfo {author} {\bibfnamefont {D.}~\bibnamefont {Prananto}}, \bibinfo {author} {\bibfnamefont {K.}~\bibnamefont {Hayashi}}, \bibinfo {author} {\bibfnamefont {A.}~\bibnamefont {Laraoui}}, \bibinfo {author} {\bibfnamefont {N.}~\bibnamefont {Mizuochi}}, \bibinfo {author} {\bibfnamefont {M.}~\bibnamefont {Hatano}}, \bibinfo {author} {\bibfnamefont {E.}~\bibnamefont {Saitoh}}, \bibinfo {author} {\bibfnamefont {Y.}~\bibnamefont {Kim}}, \bibinfo {author} {\bibfnamefont {C.~A.}\ \bibnamefont {Meriles}},\ and\ \bibinfo {author} {\bibfnamefont {T.}~\bibnamefont {An}},\ }\href {https://doi.org/10.7567/APEX.10.103004} {\bibfield  {journal} {\bibinfo  {journal} {Appl. Phys. Express}\ }\textbf {\bibinfo {volume} {10}},\ \bibinfo {pages} {103004} (\bibinfo {year} {2017})}\BibitemShut {NoStop}%
\bibitem [{\citenamefont {Bertelli}\ \emph {et~al.}(2020)\citenamefont {Bertelli}, \citenamefont {Carmiggelt}, \citenamefont {Yu}, \citenamefont {Simon}, \citenamefont {Pothoven}, \citenamefont {Bauer}, \citenamefont {Blanter}, \citenamefont {Aarts},\ and\ \citenamefont {Van Der~Sar}}]{bertelli2020magnetic}%
  \BibitemOpen
  \bibfield  {author} {\bibinfo {author} {\bibfnamefont {I.}~\bibnamefont {Bertelli}}, \bibinfo {author} {\bibfnamefont {J.~J.}\ \bibnamefont {Carmiggelt}}, \bibinfo {author} {\bibfnamefont {T.}~\bibnamefont {Yu}}, \bibinfo {author} {\bibfnamefont {B.~G.}\ \bibnamefont {Simon}}, \bibinfo {author} {\bibfnamefont {C.~C.}\ \bibnamefont {Pothoven}}, \bibinfo {author} {\bibfnamefont {G.~E.}\ \bibnamefont {Bauer}}, \bibinfo {author} {\bibfnamefont {Y.~M.}\ \bibnamefont {Blanter}}, \bibinfo {author} {\bibfnamefont {J.}~\bibnamefont {Aarts}},\ and\ \bibinfo {author} {\bibfnamefont {T.}~\bibnamefont {Van Der~Sar}},\ }\href {https://doi.org/10.1126/sciadv.abd3556} {\bibfield  {journal} {\bibinfo  {journal} {Sci. Adv.}\ }\textbf {\bibinfo {volume} {6}},\ \bibinfo {pages} {eabd3556} (\bibinfo {year} {2020})}\BibitemShut {NoStop}%
\bibitem [{\citenamefont {Zhou}\ \emph {et~al.}(2021)\citenamefont {Zhou}, \citenamefont {Carmiggelt}, \citenamefont {G{\"a}chter}, \citenamefont {Esterlis}, \citenamefont {Sels}, \citenamefont {St^^c3^^b6hr}, \citenamefont {Du}, \citenamefont {Fernandez}, \citenamefont {Rodriguez-Nieva}, \citenamefont {B{\"u}ttner}, \citenamefont {Demler},\ and\ \citenamefont {Yacoby}}]{zhou2021magnon}%
  \BibitemOpen
  \bibfield  {author} {\bibinfo {author} {\bibfnamefont {T.~X.}\ \bibnamefont {Zhou}}, \bibinfo {author} {\bibfnamefont {J.~J.}\ \bibnamefont {Carmiggelt}}, \bibinfo {author} {\bibfnamefont {L.~M.}\ \bibnamefont {G{\"a}chter}}, \bibinfo {author} {\bibfnamefont {I.}~\bibnamefont {Esterlis}}, \bibinfo {author} {\bibfnamefont {D.}~\bibnamefont {Sels}}, \bibinfo {author} {\bibfnamefont {R.~J.}\ \bibnamefont {St^^c3^^b6hr}}, \bibinfo {author} {\bibfnamefont {C.}~\bibnamefont {Du}}, \bibinfo {author} {\bibfnamefont {D.}~\bibnamefont {Fernandez}}, \bibinfo {author} {\bibfnamefont {J.~F.}\ \bibnamefont {Rodriguez-Nieva}}, \bibinfo {author} {\bibfnamefont {F.}~\bibnamefont {B{\"u}ttner}}, \bibinfo {author} {\bibfnamefont {E.}~\bibnamefont {Demler}},\ and\ \bibinfo {author} {\bibfnamefont {A.}~\bibnamefont {Yacoby}},\ }\href {https://doi.org/10.1073/pnas.2019473118} {\bibfield  {journal} {\bibinfo  {journal} {Proc. Natl. Acad. Sci. USA.}\ }\textbf {\bibinfo {volume} {118}},\ \bibinfo {pages} {e2019473118} (\bibinfo
  {year} {2021})}\BibitemShut {NoStop}%
\bibitem [{\citenamefont {Jelezko}\ \emph {et~al.}(2004)\citenamefont {Jelezko}, \citenamefont {Gaebel}, \citenamefont {Popa}, \citenamefont {Gruber},\ and\ \citenamefont {Wrachtrup}}]{Jelezko2004}%
  \BibitemOpen
  \bibfield  {author} {\bibinfo {author} {\bibfnamefont {F.}~\bibnamefont {Jelezko}}, \bibinfo {author} {\bibfnamefont {T.}~\bibnamefont {Gaebel}}, \bibinfo {author} {\bibfnamefont {I.}~\bibnamefont {Popa}}, \bibinfo {author} {\bibfnamefont {A.}~\bibnamefont {Gruber}},\ and\ \bibinfo {author} {\bibfnamefont {J.}~\bibnamefont {Wrachtrup}},\ }\href {https://doi.org/10.1103/physrevlett.92.076401} {\bibfield  {journal} {\bibinfo  {journal} {Phys. Rev. Lett.}\ }\textbf {\bibinfo {volume} {92}},\ \bibinfo {pages} {076401} (\bibinfo {year} {2004})}\BibitemShut {NoStop}%
\bibitem [{\citenamefont {Carmiggelt}\ \emph {et~al.}(2023)\citenamefont {Carmiggelt}, \citenamefont {Bertelli}, \citenamefont {Mulder}, \citenamefont {Teepe}, \citenamefont {Elyasi}, \citenamefont {Simon}, \citenamefont {Bauer}, \citenamefont {Blanter},\ and\ \citenamefont {van~der Sar}}]{Carmiggelt2023}%
  \BibitemOpen
  \bibfield  {author} {\bibinfo {author} {\bibfnamefont {J.~J.}\ \bibnamefont {Carmiggelt}}, \bibinfo {author} {\bibfnamefont {I.}~\bibnamefont {Bertelli}}, \bibinfo {author} {\bibfnamefont {R.~W.}\ \bibnamefont {Mulder}}, \bibinfo {author} {\bibfnamefont {A.}~\bibnamefont {Teepe}}, \bibinfo {author} {\bibfnamefont {M.}~\bibnamefont {Elyasi}}, \bibinfo {author} {\bibfnamefont {B.~G.}\ \bibnamefont {Simon}}, \bibinfo {author} {\bibfnamefont {G.~E.~W.}\ \bibnamefont {Bauer}}, \bibinfo {author} {\bibfnamefont {Y.~M.}\ \bibnamefont {Blanter}},\ and\ \bibinfo {author} {\bibfnamefont {T.}~\bibnamefont {van~der Sar}},\ }\href {https://doi.org/10.1038/s41467-023-36146-3} {\bibfield  {journal} {\bibinfo  {journal} {Nat. Commun.}\ }\textbf {\bibinfo {volume} {14}},\ \bibinfo {pages} {490} (\bibinfo {year} {2023})}\BibitemShut {NoStop}%
\bibitem [{\citenamefont {Wu}\ \emph {et~al.}(2024)\citenamefont {Wu}, \citenamefont {Liu}, \citenamefont {Ren}, \citenamefont {Leung}, \citenamefont {Leung}, \citenamefont {Ho}, \citenamefont {Wang}, \citenamefont {Shao},\ and\ \citenamefont {Yang}}]{Wu2024}%
  \BibitemOpen
  \bibfield  {author} {\bibinfo {author} {\bibfnamefont {J.}~\bibnamefont {Wu}}, \bibinfo {author} {\bibfnamefont {J.}~\bibnamefont {Liu}}, \bibinfo {author} {\bibfnamefont {Z.}~\bibnamefont {Ren}}, \bibinfo {author} {\bibfnamefont {M.~Y.}\ \bibnamefont {Leung}}, \bibinfo {author} {\bibfnamefont {W.~K.}\ \bibnamefont {Leung}}, \bibinfo {author} {\bibfnamefont {K.~O.}\ \bibnamefont {Ho}}, \bibinfo {author} {\bibfnamefont {X.}~\bibnamefont {Wang}}, \bibinfo {author} {\bibfnamefont {Q.}~\bibnamefont {Shao}},\ and\ \bibinfo {author} {\bibfnamefont {S.}~\bibnamefont {Yang}},\ }\href {https://doi.org/10.1038/s44306-024-00035-2} {\bibfield  {journal} {\bibinfo  {journal} {npj Spintronics}\ }\textbf {\bibinfo {volume} {2}},\ \bibinfo {pages} {30} (\bibinfo {year} {2024})}\BibitemShut {NoStop}%
\bibitem [{\citenamefont {Autler}\ and\ \citenamefont {Townes}(1955)}]{autler1955stark}%
  \BibitemOpen
  \bibfield  {author} {\bibinfo {author} {\bibfnamefont {S.~H.}\ \bibnamefont {Autler}}\ and\ \bibinfo {author} {\bibfnamefont {C.~H.}\ \bibnamefont {Townes}},\ }\href {https://doi.org/10.1103/PhysRev.100.703} {\bibfield  {journal} {\bibinfo  {journal} {Phys. Rev.}\ }\textbf {\bibinfo {volume} {100}},\ \bibinfo {pages} {703} (\bibinfo {year} {1955})}\BibitemShut {NoStop}%
\bibitem [{\citenamefont {Ramsey}(1955)}]{ramsey1955resonance}%
  \BibitemOpen
  \bibfield  {author} {\bibinfo {author} {\bibfnamefont {N.~F.}\ \bibnamefont {Ramsey}},\ }\href {https://doi.org/10.1103/PhysRev.100.1191} {\bibfield  {journal} {\bibinfo  {journal} {Phys. Rev.}\ }\textbf {\bibinfo {volume} {100}},\ \bibinfo {pages} {1191} (\bibinfo {year} {1955})}\BibitemShut {NoStop}%
\bibitem [{\citenamefont {Wei}\ \emph {et~al.}(1997)\citenamefont {Wei}, \citenamefont {Windsor},\ and\ \citenamefont {Manson}}]{wei1997strongly}%
  \BibitemOpen
  \bibfield  {author} {\bibinfo {author} {\bibfnamefont {C.}~\bibnamefont {Wei}}, \bibinfo {author} {\bibfnamefont {A.~S.}\ \bibnamefont {Windsor}},\ and\ \bibinfo {author} {\bibfnamefont {N.~B.}\ \bibnamefont {Manson}},\ }\href {https://doi.org/10.1088/0953-4075/30/21/022} {\bibfield  {journal} {\bibinfo  {journal} {Journal of Physics B: Atomic, Molecular and Optical Physics}\ }\textbf {\bibinfo {volume} {30}},\ \bibinfo {pages} {4877} (\bibinfo {year} {1997})}\BibitemShut {NoStop}%
\bibitem [{\citenamefont {Grimm}\ \emph {et~al.}(2000)\citenamefont {Grimm}, \citenamefont {Weidem{\"u}ller},\ and\ \citenamefont {Ovchinnikov}}]{grimm2000optical}%
  \BibitemOpen
  \bibfield  {author} {\bibinfo {author} {\bibfnamefont {R.}~\bibnamefont {Grimm}}, \bibinfo {author} {\bibfnamefont {M.}~\bibnamefont {Weidem{\"u}ller}},\ and\ \bibinfo {author} {\bibfnamefont {Y.~B.}\ \bibnamefont {Ovchinnikov}},\ }in\ \href {https://doi.org/https://doi.org/10.1016/S1049-250X(08)60186-X} {\emph {\bibinfo {booktitle} {Adv. At. Mol. Opt. Phys.}}},\ Vol.~\bibinfo {volume} {42}\ (\bibinfo  {publisher} {Elsevier},\ \bibinfo {year} {2000})\ pp.\ \bibinfo {pages} {95--170}\BibitemShut {NoStop}%
\bibitem [{\citenamefont {Meyer}\ \emph {et~al.}(2020)\citenamefont {Meyer}, \citenamefont {Castillo}, \citenamefont {Cox},\ and\ \citenamefont {Kunz}}]{meyer2020assessment}%
  \BibitemOpen
  \bibfield  {author} {\bibinfo {author} {\bibfnamefont {D.~H.}\ \bibnamefont {Meyer}}, \bibinfo {author} {\bibfnamefont {Z.~A.}\ \bibnamefont {Castillo}}, \bibinfo {author} {\bibfnamefont {K.~C.}\ \bibnamefont {Cox}},\ and\ \bibinfo {author} {\bibfnamefont {P.~D.}\ \bibnamefont {Kunz}},\ }\href {https://doi.org/10.1088/1361-6455/ab6051} {\bibfield  {journal} {\bibinfo  {journal} {J. Phys. B: At. Mol. Opt. Phys.}\ }\textbf {\bibinfo {volume} {53}},\ \bibinfo {pages} {034001} (\bibinfo {year} {2020})}\BibitemShut {NoStop}%
\bibitem [{\citenamefont {Simons}\ \emph {et~al.}(2016{\natexlab{a}})\citenamefont {Simons}, \citenamefont {Gordon},\ and\ \citenamefont {Holloway}}]{simons2016simultaneous}%
  \BibitemOpen
  \bibfield  {author} {\bibinfo {author} {\bibfnamefont {M.~T.}\ \bibnamefont {Simons}}, \bibinfo {author} {\bibfnamefont {J.~A.}\ \bibnamefont {Gordon}},\ and\ \bibinfo {author} {\bibfnamefont {C.~L.}\ \bibnamefont {Holloway}},\ }\href {https://doi.org/10.1063/1.4963106} {\bibfield  {journal} {\bibinfo  {journal} {J. Appl. Phys.}\ }\textbf {\bibinfo {volume} {120}},\ \bibinfo {pages} {123103} (\bibinfo {year} {2016}{\natexlab{a}})}\BibitemShut {NoStop}%
\bibitem [{\citenamefont {Simons}\ \emph {et~al.}(2016{\natexlab{b}})\citenamefont {Simons}, \citenamefont {Gordon}, \citenamefont {Holloway}, \citenamefont {Anderson}, \citenamefont {Miller},\ and\ \citenamefont {Raithel}}]{simons2016using}%
  \BibitemOpen
  \bibfield  {author} {\bibinfo {author} {\bibfnamefont {M.~T.}\ \bibnamefont {Simons}}, \bibinfo {author} {\bibfnamefont {J.~A.}\ \bibnamefont {Gordon}}, \bibinfo {author} {\bibfnamefont {C.~L.}\ \bibnamefont {Holloway}}, \bibinfo {author} {\bibfnamefont {D.~A.}\ \bibnamefont {Anderson}}, \bibinfo {author} {\bibfnamefont {S.~A.}\ \bibnamefont {Miller}},\ and\ \bibinfo {author} {\bibfnamefont {G.}~\bibnamefont {Raithel}},\ }\href {https://doi.org/10.1063/1.4947231} {\bibfield  {journal} {\bibinfo  {journal} {Appl. Phys. Lett.}\ }\textbf {\bibinfo {volume} {108}},\ \bibinfo {pages} {174101} (\bibinfo {year} {2016}{\natexlab{b}})}\BibitemShut {NoStop}%
\bibitem [{\citenamefont {Li}\ \emph {et~al.}(2019)\citenamefont {Li}, \citenamefont {Wang}, \citenamefont {Cheng}, \citenamefont {Wang}, \citenamefont {Wang}, \citenamefont {Duan}, \citenamefont {Liu}, \citenamefont {Shi},\ and\ \citenamefont {Du}}]{li2019wideband}%
  \BibitemOpen
  \bibfield  {author} {\bibinfo {author} {\bibfnamefont {R.}~\bibnamefont {Li}}, \bibinfo {author} {\bibfnamefont {C.-J.}\ \bibnamefont {Wang}}, \bibinfo {author} {\bibfnamefont {Z.}~\bibnamefont {Cheng}}, \bibinfo {author} {\bibfnamefont {P.}~\bibnamefont {Wang}}, \bibinfo {author} {\bibfnamefont {Y.}~\bibnamefont {Wang}}, \bibinfo {author} {\bibfnamefont {C.}~\bibnamefont {Duan}}, \bibinfo {author} {\bibfnamefont {H.}~\bibnamefont {Liu}}, \bibinfo {author} {\bibfnamefont {F.}~\bibnamefont {Shi}},\ and\ \bibinfo {author} {\bibfnamefont {J.}~\bibnamefont {Du}},\ }\href {https://doi.org/10.1103/PhysRevA.99.062328} {\bibfield  {journal} {\bibinfo  {journal} {Phys. Rev. A}\ }\textbf {\bibinfo {volume} {99}},\ \bibinfo {pages} {062328} (\bibinfo {year} {2019})}\BibitemShut {NoStop}%
\bibitem [{\citenamefont {Ogawa}\ \emph {et~al.}(2023)\citenamefont {Ogawa}, \citenamefont {Nishimura}, \citenamefont {Sasaki},\ and\ \citenamefont {Kobayashi}}]{ogawa2023demonstration}%
  \BibitemOpen
  \bibfield  {author} {\bibinfo {author} {\bibfnamefont {K.}~\bibnamefont {Ogawa}}, \bibinfo {author} {\bibfnamefont {S.}~\bibnamefont {Nishimura}}, \bibinfo {author} {\bibfnamefont {K.}~\bibnamefont {Sasaki}},\ and\ \bibinfo {author} {\bibfnamefont {K.}~\bibnamefont {Kobayashi}},\ }\href {https://doi.org/10.1063/5.0175456} {\bibfield  {journal} {\bibinfo  {journal} {Appl. Phys. Lett.}\ }\textbf {\bibinfo {volume} {123}},\ \bibinfo {pages} {214002} (\bibinfo {year} {2023})}\BibitemShut {NoStop}%
\bibitem [{\citenamefont {Tetienne}\ \emph {et~al.}(2018)\citenamefont {Tetienne}, \citenamefont {de~Gille}, \citenamefont {Broadway}, \citenamefont {Teraji}, \citenamefont {Lillie}, \citenamefont {McCoey}, \citenamefont {Dontschuk}, \citenamefont {Hall}, \citenamefont {Stacey}, \citenamefont {Simpson},\ and\ \citenamefont {Hollenberg}}]{tetienne2018spin}%
  \BibitemOpen
  \bibfield  {author} {\bibinfo {author} {\bibfnamefont {J.-P.}\ \bibnamefont {Tetienne}}, \bibinfo {author} {\bibfnamefont {R.~W.}\ \bibnamefont {de~Gille}}, \bibinfo {author} {\bibfnamefont {D.~A.}\ \bibnamefont {Broadway}}, \bibinfo {author} {\bibfnamefont {T.}~\bibnamefont {Teraji}}, \bibinfo {author} {\bibfnamefont {S.~E.}\ \bibnamefont {Lillie}}, \bibinfo {author} {\bibfnamefont {J.~M.}\ \bibnamefont {McCoey}}, \bibinfo {author} {\bibfnamefont {N.}~\bibnamefont {Dontschuk}}, \bibinfo {author} {\bibfnamefont {L.~T.}\ \bibnamefont {Hall}}, \bibinfo {author} {\bibfnamefont {A.}~\bibnamefont {Stacey}}, \bibinfo {author} {\bibfnamefont {D.~A.}\ \bibnamefont {Simpson}},\ and\ \bibinfo {author} {\bibfnamefont {L.~C.~L.}\ \bibnamefont {Hollenberg}},\ }\href {https://doi.org/10.1103/PhysRevB.97.085402} {\bibfield  {journal} {\bibinfo  {journal} {Phys. Rev. B}\ }\textbf {\bibinfo {volume} {97}},\ \bibinfo {pages} {085402} (\bibinfo {year} {2018})}\BibitemShut {NoStop}%
\bibitem [{\citenamefont {Fukushima}\ \emph {et~al.}(2022)\citenamefont {Fukushima}, \citenamefont {Ueda}, \citenamefont {Moriuchi}, \citenamefont {Kida}, \citenamefont {Hagiwara},\ and\ \citenamefont {Matsuno}}]{fukushima2022spin}%
  \BibitemOpen
  \bibfield  {author} {\bibinfo {author} {\bibfnamefont {K.}~\bibnamefont {Fukushima}}, \bibinfo {author} {\bibfnamefont {K.}~\bibnamefont {Ueda}}, \bibinfo {author} {\bibfnamefont {N.}~\bibnamefont {Moriuchi}}, \bibinfo {author} {\bibfnamefont {T.}~\bibnamefont {Kida}}, \bibinfo {author} {\bibfnamefont {M.}~\bibnamefont {Hagiwara}},\ and\ \bibinfo {author} {\bibfnamefont {J.}~\bibnamefont {Matsuno}},\ }\href {https://doi.org/10.1063/5.0124583} {\bibfield  {journal} {\bibinfo  {journal} {Appl. Phys. Lett.}\ }\textbf {\bibinfo {volume} {121}},\ \bibinfo {pages} {232403} (\bibinfo {year} {2022})}\BibitemShut {NoStop}%
\bibitem [{\citenamefont {Serga}\ \emph {et~al.}(2010)\citenamefont {Serga}, \citenamefont {Chumak},\ and\ \citenamefont {Hillebrands}}]{Serga2010}%
  \BibitemOpen
  \bibfield  {author} {\bibinfo {author} {\bibfnamefont {A.~A.}\ \bibnamefont {Serga}}, \bibinfo {author} {\bibfnamefont {A.~V.}\ \bibnamefont {Chumak}},\ and\ \bibinfo {author} {\bibfnamefont {B.}~\bibnamefont {Hillebrands}},\ }\href {https://doi.org/10.1088/0022-3727/43/26/264002} {\bibfield  {journal} {\bibinfo  {journal} {J. Phys. D}\ }\textbf {\bibinfo {volume} {43}},\ \bibinfo {pages} {264002} (\bibinfo {year} {2010})}\BibitemShut {NoStop}%
\bibitem [{\citenamefont {Sasaki}\ \emph {et~al.}(2016)\citenamefont {Sasaki}, \citenamefont {Monnai}, \citenamefont {Saijo}, \citenamefont {Fujita}, \citenamefont {Watanabe}, \citenamefont {Ishi-Hayase}, \citenamefont {Itoh},\ and\ \citenamefont {Abe}}]{Sasaki2016}%
  \BibitemOpen
  \bibfield  {author} {\bibinfo {author} {\bibfnamefont {K.}~\bibnamefont {Sasaki}}, \bibinfo {author} {\bibfnamefont {Y.}~\bibnamefont {Monnai}}, \bibinfo {author} {\bibfnamefont {S.}~\bibnamefont {Saijo}}, \bibinfo {author} {\bibfnamefont {R.}~\bibnamefont {Fujita}}, \bibinfo {author} {\bibfnamefont {H.}~\bibnamefont {Watanabe}}, \bibinfo {author} {\bibfnamefont {J.}~\bibnamefont {Ishi-Hayase}}, \bibinfo {author} {\bibfnamefont {K.~M.}\ \bibnamefont {Itoh}},\ and\ \bibinfo {author} {\bibfnamefont {E.}~\bibnamefont {Abe}},\ }\href {https://doi.org/10.1063/1.4952418} {\bibfield  {journal} {\bibinfo  {journal} {Rev. Sci. Instrum.}\ }\textbf {\bibinfo {volume} {87}},\ \bibinfo {pages} {053904} (\bibinfo {year} {2016})}\BibitemShut {NoStop}%
\bibitem [{\citenamefont {Mariani}\ \emph {et~al.}(2020)\citenamefont {Mariani}, \citenamefont {Nomoto}, \citenamefont {Kashiwaya},\ and\ \citenamefont {Nomura}}]{Mariani2020}%
  \BibitemOpen
  \bibfield  {author} {\bibinfo {author} {\bibfnamefont {G.}~\bibnamefont {Mariani}}, \bibinfo {author} {\bibfnamefont {S.}~\bibnamefont {Nomoto}}, \bibinfo {author} {\bibfnamefont {S.}~\bibnamefont {Kashiwaya}},\ and\ \bibinfo {author} {\bibfnamefont {S.}~\bibnamefont {Nomura}},\ }\href {https://doi.org/10.1038/s41598-020-61669-w} {\bibfield  {journal} {\bibinfo  {journal} {Sci, Rep.}\ }\textbf {\bibinfo {volume} {10}},\ \bibinfo {pages} {4813} (\bibinfo {year} {2020})}\BibitemShut {NoStop}%
\bibitem [{\citenamefont {Alsid}\ \emph {et~al.}(2023)\citenamefont {Alsid}, \citenamefont {Schloss}, \citenamefont {Steinecker}, \citenamefont {Barry}, \citenamefont {Maccabe}, \citenamefont {Wang}, \citenamefont {Cappellaro},\ and\ \citenamefont {Braje}}]{alsid2023solid}%
  \BibitemOpen
  \bibfield  {author} {\bibinfo {author} {\bibfnamefont {S.~T.}\ \bibnamefont {Alsid}}, \bibinfo {author} {\bibfnamefont {J.~M.}\ \bibnamefont {Schloss}}, \bibinfo {author} {\bibfnamefont {M.~H.}\ \bibnamefont {Steinecker}}, \bibinfo {author} {\bibfnamefont {J.~F.}\ \bibnamefont {Barry}}, \bibinfo {author} {\bibfnamefont {A.~C.}\ \bibnamefont {Maccabe}}, \bibinfo {author} {\bibfnamefont {G.}~\bibnamefont {Wang}}, \bibinfo {author} {\bibfnamefont {P.}~\bibnamefont {Cappellaro}},\ and\ \bibinfo {author} {\bibfnamefont {D.~A.}\ \bibnamefont {Braje}},\ }\href {https://doi.org/10.1103/PhysRevApplied.19.054095} {\bibfield  {journal} {\bibinfo  {journal} {Phys. Rev. Appl.}\ }\textbf {\bibinfo {volume} {19}},\ \bibinfo {pages} {054095} (\bibinfo {year} {2023})}\BibitemShut {NoStop}%
\bibitem [{\citenamefont {Schneider}\ \emph {et~al.}(2010)\citenamefont {Schneider}, \citenamefont {Serga}, \citenamefont {Chumak}, \citenamefont {Sandweg}, \citenamefont {Trudel}, \citenamefont {Wolff}, \citenamefont {Kostylev}, \citenamefont {Tiberkevich}, \citenamefont {Slavin},\ and\ \citenamefont {Hillebrands}}]{Schneider2010}%
  \BibitemOpen
  \bibfield  {author} {\bibinfo {author} {\bibfnamefont {T.}~\bibnamefont {Schneider}}, \bibinfo {author} {\bibfnamefont {A.~A.}\ \bibnamefont {Serga}}, \bibinfo {author} {\bibfnamefont {A.~V.}\ \bibnamefont {Chumak}}, \bibinfo {author} {\bibfnamefont {C.~W.}\ \bibnamefont {Sandweg}}, \bibinfo {author} {\bibfnamefont {S.}~\bibnamefont {Trudel}}, \bibinfo {author} {\bibfnamefont {S.}~\bibnamefont {Wolff}}, \bibinfo {author} {\bibfnamefont {M.~P.}\ \bibnamefont {Kostylev}}, \bibinfo {author} {\bibfnamefont {V.~S.}\ \bibnamefont {Tiberkevich}}, \bibinfo {author} {\bibfnamefont {A.~N.}\ \bibnamefont {Slavin}},\ and\ \bibinfo {author} {\bibfnamefont {B.}~\bibnamefont {Hillebrands}},\ }\href {https://doi.org/10.1103/PhysRevLett.104.197203} {\bibfield  {journal} {\bibinfo  {journal} {Phys. Rev. Lett.}\ }\textbf {\bibinfo {volume} {104}},\ \bibinfo {pages} {197203} (\bibinfo {year} {2010})}\BibitemShut {NoStop}%
\bibitem [{\citenamefont {Hurben}\ and\ \citenamefont {Patton}(1998)}]{hurben1998}%
  \BibitemOpen
  \bibfield  {author} {\bibinfo {author} {\bibfnamefont {M.~J.}\ \bibnamefont {Hurben}}\ and\ \bibinfo {author} {\bibfnamefont {C.~E.}\ \bibnamefont {Patton}},\ }\href {https://doi.org/10.1063/1.367194} {\bibfield  {journal} {\bibinfo  {journal} {J. Appl. Phys.}\ }\textbf {\bibinfo {volume} {83}},\ \bibinfo {pages} {4344} (\bibinfo {year} {1998})}\BibitemShut {NoStop}%
\bibitem [{\citenamefont {Bertelli}\ \emph {et~al.}(2021)\citenamefont {Bertelli}, \citenamefont {Simon}, \citenamefont {Yu}, \citenamefont {Aarts}, \citenamefont {Bauer}, \citenamefont {Blanter},\ and\ \citenamefont {van~der Sar}}]{Bertelli2021}%
  \BibitemOpen
  \bibfield  {author} {\bibinfo {author} {\bibfnamefont {I.}~\bibnamefont {Bertelli}}, \bibinfo {author} {\bibfnamefont {B.~G.}\ \bibnamefont {Simon}}, \bibinfo {author} {\bibfnamefont {T.}~\bibnamefont {Yu}}, \bibinfo {author} {\bibfnamefont {J.}~\bibnamefont {Aarts}}, \bibinfo {author} {\bibfnamefont {G.~E.~W.}\ \bibnamefont {Bauer}}, \bibinfo {author} {\bibfnamefont {Y.~M.}\ \bibnamefont {Blanter}},\ and\ \bibinfo {author} {\bibfnamefont {T.}~\bibnamefont {van~der Sar}},\ }\href {https://doi.org/https://doi.org/10.1002/qute.202100094} {\bibfield  {journal} {\bibinfo  {journal} {Adv. Quantum Technol.}\ }\textbf {\bibinfo {volume} {4}},\ \bibinfo {pages} {2100094} (\bibinfo {year} {2021})}\BibitemShut {NoStop}%
\bibitem [{\citenamefont {Chang}\ \emph {et~al.}(2014)\citenamefont {Chang}, \citenamefont {Li}, \citenamefont {Zhang}, \citenamefont {Liu}, \citenamefont {Hoffmann}, \citenamefont {Deng},\ and\ \citenamefont {Wu}}]{chang2014nano}%
  \BibitemOpen
  \bibfield  {author} {\bibinfo {author} {\bibfnamefont {H.}~\bibnamefont {Chang}}, \bibinfo {author} {\bibfnamefont {P.}~\bibnamefont {Li}}, \bibinfo {author} {\bibfnamefont {W.}~\bibnamefont {Zhang}}, \bibinfo {author} {\bibfnamefont {T.}~\bibnamefont {Liu}}, \bibinfo {author} {\bibfnamefont {A.}~\bibnamefont {Hoffmann}}, \bibinfo {author} {\bibfnamefont {L.}~\bibnamefont {Deng}},\ and\ \bibinfo {author} {\bibfnamefont {M.}~\bibnamefont {Wu}},\ }\href {https://doi.org/10.1109/LMAG.2014.2350958} {\bibfield  {journal} {\bibinfo  {journal} {IEEE Magn. Lett.}\ }\textbf {\bibinfo {volume} {5}},\ \bibinfo {pages} {1} (\bibinfo {year} {2014})}\BibitemShut {NoStop}%
\bibitem [{\citenamefont {Qin}\ \emph {et~al.}(2018)\citenamefont {Qin}, \citenamefont {H^^c3^^a4m^^c3^^a4l^^c3^^a4inen}, \citenamefont {Arjas}, \citenamefont {Witteveen},\ and\ \citenamefont {Dijken}}]{Qin2018}%
  \BibitemOpen
  \bibfield  {author} {\bibinfo {author} {\bibfnamefont {H.}~\bibnamefont {Qin}}, \bibinfo {author} {\bibfnamefont {S.~J.}\ \bibnamefont {H^^c3^^a4m^^c3^^a4l^^c3^^a4inen}}, \bibinfo {author} {\bibfnamefont {K.}~\bibnamefont {Arjas}}, \bibinfo {author} {\bibfnamefont {J.}~\bibnamefont {Witteveen}},\ and\ \bibinfo {author} {\bibfnamefont {S.~V.}\ \bibnamefont {Dijken}},\ }\href {https://doi.org/10.1103/PhysRevB.98.224422} {\bibfield  {journal} {\bibinfo  {journal} {Phys. Rev. B}\ }\textbf {\bibinfo {volume} {98}},\ \bibinfo {pages} {1} (\bibinfo {year} {2018})}\BibitemShut {NoStop}%
\bibitem [{\citenamefont {Cummins}\ \emph {et~al.}(2003)\citenamefont {Cummins}, \citenamefont {Llewellyn},\ and\ \citenamefont {Jones}}]{cummins2003tackling}%
  \BibitemOpen
  \bibfield  {author} {\bibinfo {author} {\bibfnamefont {H.~K.}\ \bibnamefont {Cummins}}, \bibinfo {author} {\bibfnamefont {G.}~\bibnamefont {Llewellyn}},\ and\ \bibinfo {author} {\bibfnamefont {J.~A.}\ \bibnamefont {Jones}},\ }\href {https://doi.org/10.1103/PhysRevA.67.042308} {\bibfield  {journal} {\bibinfo  {journal} {Phys. Rev. A}\ }\textbf {\bibinfo {volume} {67}},\ \bibinfo {pages} {042308} (\bibinfo {year} {2003})}\BibitemShut {NoStop}%
\bibitem [{\citenamefont {Nomura}\ \emph {et~al.}(2021)\citenamefont {Nomura}, \citenamefont {Kaida}, \citenamefont {Watanabe},\ and\ \citenamefont {Kashiwaya}}]{nomura2021composite}%
  \BibitemOpen
  \bibfield  {author} {\bibinfo {author} {\bibfnamefont {S.}~\bibnamefont {Nomura}}, \bibinfo {author} {\bibfnamefont {K.}~\bibnamefont {Kaida}}, \bibinfo {author} {\bibfnamefont {H.}~\bibnamefont {Watanabe}},\ and\ \bibinfo {author} {\bibfnamefont {S.}~\bibnamefont {Kashiwaya}},\ }\href {https://doi.org/10.1063/5.0052161} {\bibfield  {journal} {\bibinfo  {journal} {J. Appl. Phys.}\ }\textbf {\bibinfo {volume} {130}},\ \bibinfo {pages} {024503} (\bibinfo {year} {2021})}\BibitemShut {NoStop}%
\bibitem [{\citenamefont {Rietwyk}\ \emph {et~al.}(2024)\citenamefont {Rietwyk}, \citenamefont {Shaji}, \citenamefont {Robertson}, \citenamefont {Healey}, \citenamefont {Singh}, \citenamefont {Scholten}, \citenamefont {Reineck}, \citenamefont {Broadway},\ and\ \citenamefont {Tetienne}}]{Kevin2024practical}%
  \BibitemOpen
  \bibfield  {author} {\bibinfo {author} {\bibfnamefont {K.~J.}\ \bibnamefont {Rietwyk}}, \bibinfo {author} {\bibfnamefont {A.}~\bibnamefont {Shaji}}, \bibinfo {author} {\bibfnamefont {I.~O.}\ \bibnamefont {Robertson}}, \bibinfo {author} {\bibfnamefont {A.~J.}\ \bibnamefont {Healey}}, \bibinfo {author} {\bibfnamefont {P.}~\bibnamefont {Singh}}, \bibinfo {author} {\bibfnamefont {S.~C.}\ \bibnamefont {Scholten}}, \bibinfo {author} {\bibfnamefont {P.}~\bibnamefont {Reineck}}, \bibinfo {author} {\bibfnamefont {D.~A.}\ \bibnamefont {Broadway}},\ and\ \bibinfo {author} {\bibfnamefont {J.-P.}\ \bibnamefont {Tetienne}},\ }\href {https://doi.org/10.1116/5.0230098} {\bibfield  {journal} {\bibinfo  {journal} {AVS Quantum Science}\ }\textbf {\bibinfo {volume} {6}},\ \bibinfo {pages} {044402} (\bibinfo {year} {2024})}\BibitemShut {NoStop}%
\bibitem [{\citenamefont {Nishimura}\ \emph {et~al.}(2024)\citenamefont {Nishimura}, \citenamefont {Tsukamoto}, \citenamefont {Sasaki},\ and\ \citenamefont {Kobayashi}}]{nishimura2024investigations}%
  \BibitemOpen
  \bibfield  {author} {\bibinfo {author} {\bibfnamefont {S.}~\bibnamefont {Nishimura}}, \bibinfo {author} {\bibfnamefont {M.}~\bibnamefont {Tsukamoto}}, \bibinfo {author} {\bibfnamefont {K.}~\bibnamefont {Sasaki}},\ and\ \bibinfo {author} {\bibfnamefont {K.}~\bibnamefont {Kobayashi}},\ }\href {https://doi.org/10.48550/arXiv.2402.1442} {\bibfield  {journal} {\bibinfo  {journal} {arXiv:2402.14422}\ } (\bibinfo {year} {2024})}\BibitemShut {NoStop}%
\bibitem [{\citenamefont {Scott}\ \emph {et~al.}(2004)\citenamefont {Scott}, \citenamefont {Patton}, \citenamefont {Kostylev},\ and\ \citenamefont {Kalinikos}}]{Scott2004}%
  \BibitemOpen
  \bibfield  {author} {\bibinfo {author} {\bibfnamefont {M.~M.}\ \bibnamefont {Scott}}, \bibinfo {author} {\bibfnamefont {C.~E.}\ \bibnamefont {Patton}}, \bibinfo {author} {\bibfnamefont {M.~P.}\ \bibnamefont {Kostylev}},\ and\ \bibinfo {author} {\bibfnamefont {B.~A.}\ \bibnamefont {Kalinikos}},\ }\href {https://doi.org/10.1063/1.1699503} {\bibfield  {journal} {\bibinfo  {journal} {J. Appl. Phys.}\ }\textbf {\bibinfo {volume} {95}},\ \bibinfo {pages} {6294} (\bibinfo {year} {2004})}\BibitemShut {NoStop}%
\bibitem [{\citenamefont {Hula}\ \emph {et~al.}(2020)\citenamefont {Hula}, \citenamefont {Schultheiss}, \citenamefont {Buzdakov}, \citenamefont {K^^c3^^b6rber}, \citenamefont {Bejarano}, \citenamefont {Flacke}, \citenamefont {Liensberger}, \citenamefont {Weiler}, \citenamefont {Shaw}, \citenamefont {Nembach}, \citenamefont {Fassbender},\ and\ \citenamefont {Schultheiss}}]{Hula2020}%
  \BibitemOpen
  \bibfield  {author} {\bibinfo {author} {\bibfnamefont {T.}~\bibnamefont {Hula}}, \bibinfo {author} {\bibfnamefont {K.}~\bibnamefont {Schultheiss}}, \bibinfo {author} {\bibfnamefont {A.}~\bibnamefont {Buzdakov}}, \bibinfo {author} {\bibfnamefont {L.}~\bibnamefont {K^^c3^^b6rber}}, \bibinfo {author} {\bibfnamefont {M.}~\bibnamefont {Bejarano}}, \bibinfo {author} {\bibfnamefont {L.}~\bibnamefont {Flacke}}, \bibinfo {author} {\bibfnamefont {L.}~\bibnamefont {Liensberger}}, \bibinfo {author} {\bibfnamefont {M.}~\bibnamefont {Weiler}}, \bibinfo {author} {\bibfnamefont {J.~M.}\ \bibnamefont {Shaw}}, \bibinfo {author} {\bibfnamefont {H.~T.}\ \bibnamefont {Nembach}}, \bibinfo {author} {\bibfnamefont {J.}~\bibnamefont {Fassbender}},\ and\ \bibinfo {author} {\bibfnamefont {H.}~\bibnamefont {Schultheiss}},\ }\href {https://doi.org/10.1063/5.0015269} {\bibfield  {journal} {\bibinfo  {journal} {Appl. Phys. Lett.}\ }\textbf {\bibinfo {volume} {117}},\ \bibinfo {pages} {042404} (\bibinfo {year} {2020})}\BibitemShut
  {NoStop}%
\bibitem [{\citenamefont {Lake}\ \emph {et~al.}(2022)\citenamefont {Lake}, \citenamefont {Divinskiy}, \citenamefont {Schmidt}, \citenamefont {Demokritov},\ and\ \citenamefont {Demidov}}]{lake2022interplay}%
  \BibitemOpen
  \bibfield  {author} {\bibinfo {author} {\bibfnamefont {S.}~\bibnamefont {Lake}}, \bibinfo {author} {\bibfnamefont {B.}~\bibnamefont {Divinskiy}}, \bibinfo {author} {\bibfnamefont {G.}~\bibnamefont {Schmidt}}, \bibinfo {author} {\bibfnamefont {S.}~\bibnamefont {Demokritov}},\ and\ \bibinfo {author} {\bibfnamefont {V.}~\bibnamefont {Demidov}},\ }\href {https://doi.org/10.1103/PhysRevApplied.17.034010} {\bibfield  {journal} {\bibinfo  {journal} {Phys. Rev. Appl.}\ }\textbf {\bibinfo {volume} {17}},\ \bibinfo {pages} {034010} (\bibinfo {year} {2022})}\BibitemShut {NoStop}%
\bibitem [{\citenamefont {Sambe}(1973)}]{sambe1973steady}%
  \BibitemOpen
  \bibfield  {author} {\bibinfo {author} {\bibfnamefont {H.}~\bibnamefont {Sambe}},\ }\href {https://doi.org/10.1103/PhysRevA.7.2203} {\bibfield  {journal} {\bibinfo  {journal} {Phys. Rev. A}\ }\textbf {\bibinfo {volume} {7}},\ \bibinfo {pages} {2203} (\bibinfo {year} {1973})}\BibitemShut {NoStop}%
\bibitem [{\citenamefont {Beloy}(2009)}]{beloy2009theory}%
  \BibitemOpen
  \bibfield  {author} {\bibinfo {author} {\bibfnamefont {K.}~\bibnamefont {Beloy}},\ }\emph {\bibinfo {title} {Theory of the ac stark effect on the atomic hyperfine structure and applications to microwave atomic clocks}},\ \href@noop {} {Ph.D. thesis},\ \bibinfo  {school} {University of Nevada} (\bibinfo {year} {2009})\BibitemShut {NoStop}%
\bibitem [{\citenamefont {Healey}\ \emph {et~al.}(2020)\citenamefont {Healey}, \citenamefont {Stacey}, \citenamefont {Johnson}, \citenamefont {Broadway}, \citenamefont {Teraji}, \citenamefont {Simpson}, \citenamefont {Tetienne},\ and\ \citenamefont {Hollenberg}}]{healey2020comparison}%
  \BibitemOpen
  \bibfield  {author} {\bibinfo {author} {\bibfnamefont {A.}~\bibnamefont {Healey}}, \bibinfo {author} {\bibfnamefont {A.}~\bibnamefont {Stacey}}, \bibinfo {author} {\bibfnamefont {B.}~\bibnamefont {Johnson}}, \bibinfo {author} {\bibfnamefont {D.}~\bibnamefont {Broadway}}, \bibinfo {author} {\bibfnamefont {T.}~\bibnamefont {Teraji}}, \bibinfo {author} {\bibfnamefont {D.}~\bibnamefont {Simpson}}, \bibinfo {author} {\bibfnamefont {J.-P.}\ \bibnamefont {Tetienne}},\ and\ \bibinfo {author} {\bibfnamefont {L.}~\bibnamefont {Hollenberg}},\ }\href {https://doi.org/10.1103/PhysRevMaterials.4.104605} {\bibfield  {journal} {\bibinfo  {journal} {Phys. Rev. Mater.}\ }\textbf {\bibinfo {volume} {4}},\ \bibinfo {pages} {104605} (\bibinfo {year} {2020})}\BibitemShut {NoStop}%
\bibitem [{\citenamefont {Stigloher}\ \emph {et~al.}(2016)\citenamefont {Stigloher}, \citenamefont {Decker}, \citenamefont {K\"orner}, \citenamefont {Tanabe}, \citenamefont {Moriyama}, \citenamefont {Taniguchi}, \citenamefont {Hata}, \citenamefont {Madami}, \citenamefont {Gubbiotti}, \citenamefont {Kobayashi}, \citenamefont {Ono},\ and\ \citenamefont {Back}}]{Stigloher2016}%
  \BibitemOpen
  \bibfield  {author} {\bibinfo {author} {\bibfnamefont {J.}~\bibnamefont {Stigloher}}, \bibinfo {author} {\bibfnamefont {M.}~\bibnamefont {Decker}}, \bibinfo {author} {\bibfnamefont {H.~S.}\ \bibnamefont {K\"orner}}, \bibinfo {author} {\bibfnamefont {K.}~\bibnamefont {Tanabe}}, \bibinfo {author} {\bibfnamefont {T.}~\bibnamefont {Moriyama}}, \bibinfo {author} {\bibfnamefont {T.}~\bibnamefont {Taniguchi}}, \bibinfo {author} {\bibfnamefont {H.}~\bibnamefont {Hata}}, \bibinfo {author} {\bibfnamefont {M.}~\bibnamefont {Madami}}, \bibinfo {author} {\bibfnamefont {G.}~\bibnamefont {Gubbiotti}}, \bibinfo {author} {\bibfnamefont {K.}~\bibnamefont {Kobayashi}}, \bibinfo {author} {\bibfnamefont {T.}~\bibnamefont {Ono}},\ and\ \bibinfo {author} {\bibfnamefont {C.~H.}\ \bibnamefont {Back}},\ }\href {https://doi.org/10.1103/PhysRevLett.117.037204} {\bibfield  {journal} {\bibinfo  {journal} {Phys. Rev. Lett.}\ }\textbf {\bibinfo {volume} {117}},\ \bibinfo {pages} {037204} (\bibinfo {year} {2016})}\BibitemShut {NoStop}%
\bibitem [{\citenamefont {Heussner}\ \emph {et~al.}(2020)\citenamefont {Heussner}, \citenamefont {Talmelli}, \citenamefont {Geilen}, \citenamefont {Heinz}, \citenamefont {Br^^c3^^a4cher}, \citenamefont {Meyer}, \citenamefont {Ciubotaru}, \citenamefont {Adelmann}, \citenamefont {Yamamoto}, \citenamefont {Serga}, \citenamefont {Hillebrands},\ and\ \citenamefont {Pirro}}]{Heussner2020}%
  \BibitemOpen
  \bibfield  {author} {\bibinfo {author} {\bibfnamefont {F.}~\bibnamefont {Heussner}}, \bibinfo {author} {\bibfnamefont {G.}~\bibnamefont {Talmelli}}, \bibinfo {author} {\bibfnamefont {M.}~\bibnamefont {Geilen}}, \bibinfo {author} {\bibfnamefont {B.}~\bibnamefont {Heinz}}, \bibinfo {author} {\bibfnamefont {T.}~\bibnamefont {Br^^c3^^a4cher}}, \bibinfo {author} {\bibfnamefont {T.}~\bibnamefont {Meyer}}, \bibinfo {author} {\bibfnamefont {F.}~\bibnamefont {Ciubotaru}}, \bibinfo {author} {\bibfnamefont {C.}~\bibnamefont {Adelmann}}, \bibinfo {author} {\bibfnamefont {K.}~\bibnamefont {Yamamoto}}, \bibinfo {author} {\bibfnamefont {A.~A.}\ \bibnamefont {Serga}}, \bibinfo {author} {\bibfnamefont {B.}~\bibnamefont {Hillebrands}},\ and\ \bibinfo {author} {\bibfnamefont {P.}~\bibnamefont {Pirro}},\ }\href {https://doi.org/https://doi.org/10.1002/pssr.201900695} {\bibfield  {journal} {\bibinfo  {journal} {Phys. Status Solidi RRL}\ }\textbf {\bibinfo {volume} {14}},\ \bibinfo {pages} {1900695} (\bibinfo {year}
  {2020})}\BibitemShut {NoStop}%
\bibitem [{\citenamefont {Schulz}\ \emph {et~al.}(2023)\citenamefont {Schulz}, \citenamefont {Litzius}, \citenamefont {Powalla}, \citenamefont {Birch}, \citenamefont {Gallardo}, \citenamefont {Satheesh}, \citenamefont {Weigand}, \citenamefont {Scholz}, \citenamefont {Lotsch}, \citenamefont {Sch^^c3^^bctz}, \citenamefont {Burghard},\ and\ \citenamefont {Wintz}}]{Schulz2023}%
  \BibitemOpen
  \bibfield  {author} {\bibinfo {author} {\bibfnamefont {F.}~\bibnamefont {Schulz}}, \bibinfo {author} {\bibfnamefont {K.}~\bibnamefont {Litzius}}, \bibinfo {author} {\bibfnamefont {L.}~\bibnamefont {Powalla}}, \bibinfo {author} {\bibfnamefont {M.~T.}\ \bibnamefont {Birch}}, \bibinfo {author} {\bibfnamefont {R.~A.}\ \bibnamefont {Gallardo}}, \bibinfo {author} {\bibfnamefont {S.}~\bibnamefont {Satheesh}}, \bibinfo {author} {\bibfnamefont {M.}~\bibnamefont {Weigand}}, \bibinfo {author} {\bibfnamefont {T.}~\bibnamefont {Scholz}}, \bibinfo {author} {\bibfnamefont {B.~V.}\ \bibnamefont {Lotsch}}, \bibinfo {author} {\bibfnamefont {G.}~\bibnamefont {Sch^^c3^^bctz}}, \bibinfo {author} {\bibfnamefont {M.}~\bibnamefont {Burghard}},\ and\ \bibinfo {author} {\bibfnamefont {S.}~\bibnamefont {Wintz}},\ }\href {https://doi.org/10.1021/acs.nanolett.3c02212} {\bibfield  {journal} {\bibinfo  {journal} {Nano Lett.}\ }\textbf {\bibinfo {volume} {23}},\ \bibinfo {pages} {10126} (\bibinfo {year} {2023})}\BibitemShut {NoStop}%
\bibitem [{\citenamefont {Xu}\ \emph {et~al.}(2023)\citenamefont {Xu}, \citenamefont {Jia}, \citenamefont {Huang}, \citenamefont {Meng}, \citenamefont {Zhang}, \citenamefont {Zhang}, \citenamefont {Cheng}, \citenamefont {Lan}, \citenamefont {Dong}, \citenamefont {Wei}, \citenamefont {Feng}, \citenamefont {He}, \citenamefont {Yuan}, \citenamefont {Zhu}, \citenamefont {He}, \citenamefont {Wan}, \citenamefont {Wei}, \citenamefont {Wang}, \citenamefont {Shao}, \citenamefont {Gu}, \citenamefont {Coey}, \citenamefont {Shi}, \citenamefont {Zhang}, \citenamefont {Han},\ and\ \citenamefont {Yu}}]{Xu2023}%
  \BibitemOpen
  \bibfield  {author} {\bibinfo {author} {\bibfnamefont {H.}~\bibnamefont {Xu}}, \bibinfo {author} {\bibfnamefont {K.}~\bibnamefont {Jia}}, \bibinfo {author} {\bibfnamefont {Y.}~\bibnamefont {Huang}}, \bibinfo {author} {\bibfnamefont {F.}~\bibnamefont {Meng}}, \bibinfo {author} {\bibfnamefont {Q.}~\bibnamefont {Zhang}}, \bibinfo {author} {\bibfnamefont {Y.}~\bibnamefont {Zhang}}, \bibinfo {author} {\bibfnamefont {C.}~\bibnamefont {Cheng}}, \bibinfo {author} {\bibfnamefont {G.}~\bibnamefont {Lan}}, \bibinfo {author} {\bibfnamefont {J.}~\bibnamefont {Dong}}, \bibinfo {author} {\bibfnamefont {J.}~\bibnamefont {Wei}}, \bibinfo {author} {\bibfnamefont {J.}~\bibnamefont {Feng}}, \bibinfo {author} {\bibfnamefont {C.}~\bibnamefont {He}}, \bibinfo {author} {\bibfnamefont {Z.}~\bibnamefont {Yuan}}, \bibinfo {author} {\bibfnamefont {M.}~\bibnamefont {Zhu}}, \bibinfo {author} {\bibfnamefont {W.}~\bibnamefont {He}}, \bibinfo {author} {\bibfnamefont {C.}~\bibnamefont {Wan}}, \bibinfo {author} {\bibfnamefont
  {H.}~\bibnamefont {Wei}}, \bibinfo {author} {\bibfnamefont {S.}~\bibnamefont {Wang}}, \bibinfo {author} {\bibfnamefont {Q.}~\bibnamefont {Shao}}, \bibinfo {author} {\bibfnamefont {L.}~\bibnamefont {Gu}}, \bibinfo {author} {\bibfnamefont {M.}~\bibnamefont {Coey}}, \bibinfo {author} {\bibfnamefont {Y.}~\bibnamefont {Shi}}, \bibinfo {author} {\bibfnamefont {G.}~\bibnamefont {Zhang}}, \bibinfo {author} {\bibfnamefont {X.}~\bibnamefont {Han}},\ and\ \bibinfo {author} {\bibfnamefont {G.}~\bibnamefont {Yu}},\ }\href {https://doi.org/10.1038/s41467-023-39529-8} {\bibfield  {journal} {\bibinfo  {journal} {Nat. Commun.}\ }\textbf {\bibinfo {volume} {14}},\ \bibinfo {pages} {3824} (\bibinfo {year} {2023})}\BibitemShut {NoStop}%
\bibitem [{\citenamefont {Kalinikos}\ and\ \citenamefont {Slavin}(1986)}]{kalinikos1986theory}%
  \BibitemOpen
  \bibfield  {author} {\bibinfo {author} {\bibfnamefont {B.~A.}\ \bibnamefont {Kalinikos}}\ and\ \bibinfo {author} {\bibfnamefont {A.~N.}\ \bibnamefont {Slavin}},\ }\href {https://doi.org/10.1088/0022-3719/19/35/014} {\bibfield  {journal} {\bibinfo  {journal} {J. Phys. C: Solid State Phys.}\ }\textbf {\bibinfo {volume} {19}},\ \bibinfo {pages} {7013} (\bibinfo {year} {1986})}\BibitemShut {NoStop}%
\bibitem [{\citenamefont {Yu}\ \emph {et~al.}(2019)\citenamefont {Yu}, \citenamefont {Liu}, \citenamefont {Yu}, \citenamefont {Blanter},\ and\ \citenamefont {Bauer}}]{Yu2019}%
  \BibitemOpen
  \bibfield  {author} {\bibinfo {author} {\bibfnamefont {T.}~\bibnamefont {Yu}}, \bibinfo {author} {\bibfnamefont {C.}~\bibnamefont {Liu}}, \bibinfo {author} {\bibfnamefont {H.}~\bibnamefont {Yu}}, \bibinfo {author} {\bibfnamefont {Y.~M.}\ \bibnamefont {Blanter}},\ and\ \bibinfo {author} {\bibfnamefont {G.~E.~W.}\ \bibnamefont {Bauer}},\ }\href {https://doi.org/10.1103/PhysRevB.99.134424} {\bibfield  {journal} {\bibinfo  {journal} {Phys. Rev. B}\ }\textbf {\bibinfo {volume} {99}},\ \bibinfo {pages} {134424} (\bibinfo {year} {2019})}\BibitemShut {NoStop}%
\bibitem [{\citenamefont {Arias}\ and\ \citenamefont {Mills}(1999)}]{Rodrigo1999Extrinsic}%
  \BibitemOpen
  \bibfield  {author} {\bibinfo {author} {\bibfnamefont {R.}~\bibnamefont {Arias}}\ and\ \bibinfo {author} {\bibfnamefont {D.~L.}\ \bibnamefont {Mills}},\ }\href {https://doi.org/10.1103/PhysRevB.60.7395} {\bibfield  {journal} {\bibinfo  {journal} {Phys. Rev. B}\ }\textbf {\bibinfo {volume} {60}},\ \bibinfo {pages} {7395} (\bibinfo {year} {1999})}\BibitemShut {NoStop}%
\bibitem [{\citenamefont {Tsukamoto}\ \emph {et~al.}(2021)\citenamefont {Tsukamoto}, \citenamefont {Ogawa}, \citenamefont {Ozawa}, \citenamefont {Iwasaki}, \citenamefont {Hatano}, \citenamefont {Sasaki},\ and\ \citenamefont {Kobayashi}}]{Tsukamoto2021}%
  \BibitemOpen
  \bibfield  {author} {\bibinfo {author} {\bibfnamefont {M.}~\bibnamefont {Tsukamoto}}, \bibinfo {author} {\bibfnamefont {K.}~\bibnamefont {Ogawa}}, \bibinfo {author} {\bibfnamefont {H.}~\bibnamefont {Ozawa}}, \bibinfo {author} {\bibfnamefont {T.}~\bibnamefont {Iwasaki}}, \bibinfo {author} {\bibfnamefont {M.}~\bibnamefont {Hatano}}, \bibinfo {author} {\bibfnamefont {K.}~\bibnamefont {Sasaki}},\ and\ \bibinfo {author} {\bibfnamefont {K.}~\bibnamefont {Kobayashi}},\ }\href {https://doi.org/10.1063/5.0054809} {\bibfield  {journal} {\bibinfo  {journal} {Appl. Phys. Lett.}\ }\textbf {\bibinfo {volume} {118}},\ \bibinfo {pages} {264002} (\bibinfo {year} {2021})}\BibitemShut {NoStop}%
\bibitem [{\citenamefont {Rezende}(2020)}]{rezende2020fundamentals}%
  \BibitemOpen
  \bibfield  {author} {\bibinfo {author} {\bibfnamefont {S.~M.}\ \bibnamefont {Rezende}},\ }\href@noop {} {\emph {\bibinfo {title} {Fundamentals of magnonics}}},\ Vol.\ \bibinfo {volume} {969}\ (\bibinfo  {publisher} {Springer},\ \bibinfo {year} {2020})\BibitemShut {NoStop}%
\end{thebibliography}%

\end{document}